\documentclass[twocolappendix, twocolumn, 12pt]{aastex63}
\usepackage{amsmath, amssymb, bm}
\usepackage[utf8]{inputenc}
\usepackage{booktabs}
\usepackage{hyperref}
\usepackage{multirow}
\usepackage{enumitem}

\newcommand{\package}[1]{\texttt{#1}}

\definecolor{myblue}{HTML}{1F77B4}
\definecolor{mygreen}{HTML}{2CA02C}
\definecolor{myred}{HTML}{D62728}
\definecolor{mymagenta}{HTML}{D33682}
\definecolor{codepurple}{HTML}{C42043}

\hypersetup{
bookmarks=true,         
unicode=true,           
colorlinks=true,        
linkcolor=myred,        
citecolor=myblue,       
filecolor=magenta,      
urlcolor=mygreen        
}

\newcommand{\galex}{\textit{GALEX}}
\newcommand{\hst}{\textit{HST}}

\newcommand{\wise}{\textit{WISE}}

\newcommand{\galexfuv}{$FUV$}
\newcommand{\galexnuv}{$NUV$}

\newcommand{\sdssu}{$u'$}
\newcommand{\sdssg}{$g'$}
\newcommand{\sdssr}{$r'$}
\newcommand{\sdssi}{$i'$}
\newcommand{\sdssz}{$z'$}

\newcommand{\psg}{$g_{\rm PS1}$}
\newcommand{\psr}{$r_{\rm PS1}$}
\renewcommand{\psi}{$i_{\rm PS1}$}
\newcommand{\psz}{$z_{\rm PS1}$}
\newcommand{\psy}{$y_{\rm PS1}$}

\newcommand{\lsg}{$g$}
\newcommand{\lsr}{$r$}
\newcommand{\lsz}{$z$}

\newcommand{\twomassj}{$J$}
\newcommand{\twomassh}{$H$}
\newcommand{\twomassk}{$K_{\rm s}$}

\newcommand{\wiseone}{$W1$}
\newcommand{\wisetwo}{$W2$}

\received{August 13, 2020}

\submitjournal{ApJS}

\shorttitle{The PTF CCSN Host-Galaxy Sample}
\shortauthors{Schulze et al.}

\newcommand{\berkeley}{Department of Astronomy, University of California, Berkeley, CA 94720-3411, USA}

\newcommand{\bgu}{Department of Mechanical Engineering, Ben-Gurion University of the Negev, 8410501 Beer-Sheva, Israel}

\newcommand{\caltechastro}{Cahill Center for Astrophysics, California Institute of Technology, 1200 E. California Blvd. Pasadena, CA 91125, USA}

\newcommand{\caltechastrotwo}{Division of Physics, Mathematics, and Astronomy, California Institute of Technology, Pasadena, CA 91125, USA}

\newcommand{\caltechobs}{The Caltech Optical Observatories, California Institute of Technology, Pasadena, CA 91125, USA}

\newcommand{\cfa}{Center for Astrophysics \textbar{} Harvard \& Smithsonian, 60 Garden Street, Cambridge, MA 02138-1516, USA}

\newcommand{\cifar}{CIFAR Azrieli Global Scholars program, CIFAR, Toronto, Canada}

\newcommand{\Delaware}{Department of Physics and Astronomy, University of Delaware, Newark, DE, 19716, USA}

\newcommand{\Delawaretwo}{Joseph R. Biden, Jr. School of Public Policy and Administration, University of Delaware, Newark, DE, 19716, USA}

\newcommand{\dtuspace}{DTU Space, National Space Institute, Technical University of Denmark, Elektrovej 327, DK-2800 Kgs. Lyngby, Denmark
}

\newcommand{\esogarching}{European Southern Observatory, Karl-Schwarzschild-Str 2, 85748 Garching, Germany}

\newcommand{\huji}{Racah Institute of Physics, The Hebrew University of Jerusalem, 91904 Jerusalem, Israel}

\newcommand{\inaf}{INAF - Osservatorio Astrofisico di Catania, Via Santa Sofia 78, 95123, Catania, Italy}

\newcommand{\kavli}{Kavli Institute for the Physics and Mathematics of the Universe (WPI), The University of Tokyo Institutes for Advanced Study, The University of Tokyo, Kashiwa, Chiba 277-8583, Japan}

\newcommand{\lancaster}{Department of Physics, Lancaster University, Lancaster, LA1 4YB, U.K.}

\newcommand{\lascumbres}{Las Cumbres Observatory, 6740 Cortona Dr. Suite 102, Goleta, CA 93117, USA}

\newcommand{\lbnl}{Lawrence Berkeley National Laboratory, 1 Cyclotron Road, MS 50B-4206, Berkeley, CA 94720, USA}

\newcommand{\ljmu}{Astrophysics Research Institute, Liverpool John Moores University, 146 Brownlow Hill, Liverpool, L3 5RF, UK}

\newcommand{\miller}{Miller Senior Fellow, Miller Institute for Basic Research in Science, University of California, Berkeley, CA 94720, USA}

\newcommand{\ncu}{Graduate Institute of Astronomy, National Central University, 300 Jhongda Road, Zhongli, Taoyuan 32001, Taiwan}

\newcommand{\newark}{Data Science Institute, University of Delaware, Newark, DE, 19716, USA}

\newcommand{\nsf}{NSF's National Optical-Infrared Astronomy Research Laboratory, 950 North Cherry Avenue, Tucson, AZ 85719, USA}

\newcommand{\nthu}{Institute of Astronomy, National Tsing Hua University, Hsinchu 30013, Taiwan}

\newcommand{\nyu}{Center for Cosmology and Particle Physics, New York University, NY 10003, USA}

\newcommand{\nyutwo}{Center for Urban Science and Progress, New York University, 370 Jay St, Brooklyn, NY 11201, USA}

\newcommand{\okcastro}{Department of Astronomy, The Oskar Klein Centre, Stockholm University, AlbaNova, 10691 Stockholm, Sweden}

\newcommand{\okcphysics}{Department of Physics, The Oskar Klein Centre, Stockholm University, AlbaNova, 10691 Stockholm, Sweden}

\newcommand{\purdue}{Department of Physics and Astronomy, Purdue University, 525 Northwestern Avenue, West Lafayette, IN 47907, USA}

\newcommand{\ruhrunibochum}{Ruhr-Universit\"at Bochum, Astronomisches Institut, German Centre for Cosmological Lensing, Universit\"atsstr. 150, 44801 Bochum, Germany}

\newcommand{\sandiego}{Department of Astronomy / Mount Laguna Observatory, San Diego State University, 5500 Campanile Drive, San Diego, CA 92812-1221, USA}

\newcommand{\santacruz}{Department of Astronomy and Astrophysics, University of California, Santa Cruz, CA 95064, USA}

\newcommand{\serbia}{Astronomical Observatory, Volgina 7, 11060 Belgrade, Republic of Serbia}

\newcommand{\slovenia}{Centre for Astrophysics and Cosmology, University of Nova Gorica, Vipavska 11c, 5270 Ajdov\v{s}\u{c}ina, Slovenia}

\newcommand{\sotonpa}{School of Physics and Astronomy, University of Southampton, Southampton, SO17 1BJ, UK}

\newcommand{\tauisrael}{School of Physics and Astronomy, Tel Aviv University, Tel Aviv, 69978, Israel}

\newcommand{\trinity}{School of Physics, Trinity College Dublin, The University of Dublin, College Green, Dublin 2, Ireland}

\newcommand{\turku}{Department of Physics and Astronomy, University of Turku, 20014 Turku, Finland}

\newcommand{\ucd}{School of Physics, O'Brien Centre for Science North, University College Dublin, Belfield, Dublin 4, Ireland}

\newcommand{\UCSB}{University of California, Santa Barbara, Department of Physics, Santa Barbara, CA, 93111, USA}

\newcommand{\umd}{Department of Astronomy, University of Maryland, College Park, MD 20742, USA}

\newcommand{\uom}{Minnesota Institute for Astrophysics, University of Minnesota, 116 Church Street SE, Minneapolis, MN 55455, USA}

\newcommand{\wisastro}{Department of Particle Physics and Astrophysics, Weizmann Institute of Science, 234 Herzl St, 76100 Rehovot, Israel}

\newcommand{\wiscomplexsystems}{Department of Physics of Complex Systems, Weizmann Institute of Science, 234 Herzl St, 76100 Rehovot, Israel}

\begin{document}

\title{The Palomar Transient Factory Core-Collapse Supernova Host-Galaxy Sample.\\I. Host-Galaxy Distribution Functions and Environment-Dependence of CCSNe}

\correspondingauthor{Steve Schulze}
\email{steve.schulze@weizmann.ac.il}

\author[0000-0001-6797-1888]{Steve Schulze}
\affiliation{\wisastro}

\author{Ofer Yaron}
\affiliation{\wisastro}

\author[0000-0003-1546-6615]{Jesper Sollerman}
\affiliation{\okcastro}

\author[0000-0002-8597-0756]{Giorgos Leloudas}
\affil{\dtuspace}

\author[0000-0002-4223-103X]{Amit Gal}
\affil{\wiscomplexsystems}

\author[0000-0001-7363-7932]{Angus~H. Wright}
\affiliation{\ruhrunibochum}

\author[0000-0001-9454-4639]{Ragnhild Lunnan}
\affil{\okcastro}

\author[0000-0002-3653-5598]{Avishay Gal-Yam}
\affiliation{\wisastro}

\author{Eran O. Ofek}
\affiliation{\wisastro}

\author[0000-0001-8472-1996]{Daniel~A. Perley}
\affiliation{\ljmu}

\author[0000-0003-3460-0103]{Alexei~V. Filippenko}
\affiliation{\berkeley}
\affiliation{\miller}

\author[0000-0002-5619-4938]{Mansi~M. Kasliwal}
\affiliation{\caltechastrotwo}

\author{Shri R. Kulkarni}
\affiliation{\caltechastro}

\author[0000-0002-3389-0586]{Peter E. Nugent}
\affiliation{\lbnl}
\affiliation{\berkeley}

\author[0000-0001-9171-5236]{Robert~M. Quimby}
\affiliation{\sandiego}
\affiliation{\kavli}

\author[0000-0001-9053-4820]{Mark Sullivan}
\affiliation{\sotonpa}

\author[0000-0002-4667-6730]{Nora~Linn Strothjohann}
\affiliation{\wisastro}

\author[0000-0001-7090-4898]{Iair Arcavi}
\affiliation{\tauisrael}
\affiliation{\cifar}

\author[0000-0001-6760-3074]{Sagi Ben-Ami}
\affiliation{\wisastro}

\author{Federica Bianco}
\affiliation{\Delaware}
\affiliation{\Delawaretwo}
\affiliation{\newark}
\affiliation{\nyutwo}

\author[0000-0002-7777-216X]{Joshua~S. Bloom}
	\affiliation{\berkeley}
	\affiliation{\lbnl}

\author{Kishalay De}
\affiliation{\caltechastrotwo}

\author[0000-0003-2191-1674]{Morgan Fraser}
\affiliation{\ucd}

\author[0000-0002-4223-103X]{Christoffer~U. Fremling}
\affiliation{\caltechastrotwo}

\author[0000-0002-5936-1156]{Assaf Horesh}
\affiliation{\huji}

\author[0000-0001-5975-290X]{Joel Johansson}
\affiliation{\okcphysics}

\author[0000-0003-3142-997X]{Patrick~L. Kelly}
\affiliation{\uom}

\author[0000-0003-1416-8069]{Sladjana Kne\v{z}evi\'c}
\affil{\serbia}

\author[0000-0002-9770-3508]{Kate Maguire}
\affiliation{\trinity}

\author{Anders Nyholm}
\affil{\okcastro}

\author[0000-0003-0783-3323]{Seméli Papadogiannakis}
\affiliation{\okcphysics}

\author[0000-0003-4743-1679]{Tanja Petrushevska}
\affiliation{\slovenia}

\author[0000-0003-4557-0632]{Adam Rubin}
\affiliation{\esogarching}

\author[0000-0003-1710-9339]{Lin Yan}
\affil{\caltechobs}

\author{Yi Yang}
\affiliation{\wisastro}

\author{Scott~M. Adams}
\affil{\caltechastro}

\author[0000-0002-3429-2481]{Filomena Bufano}
\affiliation{\inaf}

\author{Kelsey~I. Clubb}
\affiliation{\berkeley}

\author{Ryan~J. Foley}
\affiliation{\santacruz}

\author[0000-0002-0809-6575]{Yoav Green}
\affiliation{\bgu}

\author[0000-0001-8242-4090]{Jussi Harmanen}
\affiliation{\turku}

\author[0000-0002-9017-3567]{Anna~Y.~Q. Ho}
\affiliation{\caltechastro}

\author[0000-0002-2960-978X]{Isobel~M. Hook}
\affiliation{\lancaster}

\author[0000-0002-0832-2974]{Griffin Hosseinzadeh}
\affiliation{\cfa}

\author{D.~Andrew Howell}
\affiliation{\lascumbres}
\affiliation{\UCSB}

\author[0000-0002-5105-344X]{Albert~K.~H. Kong}
\affiliation{\nthu}

\author[0000-0001-5455-3653]{Rubina Kotak}
\affiliation{\turku}

\author[0000-0001-6685-0479]{Thomas Matheson}
\affiliation{\nsf}

\author[0000-0001-5807-7893]{Curtis McCully}
\affiliation{\lascumbres}
\affiliation{\UCSB}

\author[0000-0002-0763-3885]{Dan Milisavljevic}
\affil{\purdue}

\author[0000-0001-8415-6720]{Yen-Chen Pan}
\affiliation{\ncu}

\author[0000-0003-1470-7173]{Dovi Poznanski}
\affiliation{\tauisrael}

\author{Isaac Shivvers}
\affiliation{\berkeley}

\author[0000-0002-3859-8074]{Sjoert van Velzen}
\affiliation{\umd}
\affiliation{\nyu}

\begin{abstract}
Several thousand core-collapse supernovae (CCSNe) of different flavors have been discovered so far. However, identifying their progenitors has remained an outstanding open question in astrophysics. Studies of SN host galaxies have proven to be powerful in providing constraints on the progenitor populations. In this paper, we present all CCSNe detected between 2009 and 2017 by the Palomar Transient Factory.  This sample includes 888 SNe of 12 distinct classes out to redshift $z\approx1$. We present the photometric properties of their host galaxies from the far-ultraviolet to the mid-infrared and model the host-galaxy spectral energy distributions to derive physical properties. The galaxy mass functions of Type Ic, Ib, IIb, II, and IIn SNe ranges from $10^{5}$ to $10^{11.5}~M_\odot$, probing the entire mass range of star-forming galaxies down to the least-massive star-forming galaxies known. Moreover, the galaxy mass distributions are consistent with models of star-formation-weighted mass functions. Regular CCSNe are hence direct tracers of star formation. Small but notable differences exist between some of the SN classes. Type Ib/c SNe prefer galaxies with slightly higher masses (i.e., higher metallicities) and star-formation rates than Type IIb and II SNe. These differences are less pronounced than previously thought. H-poor SLSNe and SNe~Ic-BL are scarce in galaxies above $10^{10}~M_\odot$. Their progenitors require environments with metallicities of $<0.4$ and $<1$ solar, respectively. In addition, the hosts of H-poor SLSNe are dominated by a younger stellar population than all other classes of CCSNe. Our findings corroborate the notion that low-metallicity \textit{and} young age play an important role in the formation of SLSN progenitors.
\end{abstract}


\section{Introduction}

Stars with zero-age-main-sequence (ZAMS) masses of at least $8~M_\odot$ can presumably explode as core-collapse supernovae (CCSNe; for a review, see, e.g., \citealt{Smartt2009a} and references therein) which can be detected out to large cosmological distances \citep{Howell2013a,Pan2017a,Smith2018a,Moriya2019a,Curtin2019a}. Their large distances allow examining a wide range of stellar populations and even star-formation environments that do not exist in the Milky Way. CCSNe are divided into three families (H-poor, H-rich, and interaction-powered SNe) and more than a dozen classes and subclasses based on the absence or presence of particular absorption and emission lines, line widths, and SN peak luminosities \citep[e.g.,][]{Filippenko1997a, GalYam2017a}. How the different SN types are related to each other, what their progenitors are, and what the mapping between SN and progenitor properties is have remained outstanding problems in contemporary astrophysics.

Given the typical SN distances, a direct search of their progenitors is unfeasible in most cases \citep[e.g.][]{Smartt2009a, vanDyk2017a}. Studies of their host galaxies have proven to be powerful to indirectly provide constraints on the progenitor populations. These studies have examined
(i) the host morphologies and locations of SNe in their hosts and large-scale structures \citep[e.g.,][]{vandenBergh1997a, Tsvetkov2004a, Hakobyan2008a, Habergham2010a, Habergham2012a, Hakobyan2014a, Hakobyan2016a},
(ii) how SNe trace the light and star formation in their hosts \citep[e.g.,][]{Fruchter2006a, Leloudas2010a, Svensson2010a, Anderson2012a, Kelly2012a, Lunnan2015a, Kangas2017a},
(iii) the galaxy masses, star-formation rates, and metallicities of the hosts \citep[e.g.,][]{Prieto2008a, Neill2011a, Kelly2012a, Stoll2013a, Taddia2013a, Lunnan2014a, Leloudas2015a,  Angus2016a, Perley2016a, Schulze2018a, Angus2019a, Hosseinzadeh2019a, Modjaz2019a, Taggart2019a},
(iv) the metallicities and stellar populations at the explosion sites \citep[e.g.,][]{Modjaz2008a, Leloudas2011a, Kelly2012a, Sanders2012a, Kuncarayakti2013a, Kuncarayakti2013b, Taddia2015a, Thoene2015a,  Galbany2016a, Chen2017b, Thoene2017a, Galbany2018a, Izzo2018a, Kuncarayakti2018a}, and
(v) the frequency ratio between different SN types as a function of galaxy properties \citep[e.g.,][]{Prantzos2003a, Boissier2009a, Prieto2008a, Arcavi2010a, Graur2017b, Graur2017a}.

In addition, these studies revealed the commonalities and diversities of various SN classes as well as the interplay of SN and host-galaxy properties. Despite the success of these studies, most of them were based on small or statistically biased samples. Deep wide-field synoptic surveys led to overcoming both limitations and paved the path for the next milestones in SN science. The Palomar Transient Factory \citep[PTF;][]{Law2009a, Rau2009a, Kulkarni2013a} is an example of such a survey. It used the 1.22~m (48 inch) Oschin Schmidt telescope (P48) at Mount Palomar (USA) and was operated between 2009 and mid-2017. PTF was an untargeted transient survey designed to systematically map out existing gaps in the transient phase-space and search for theoretically predicted but not yet detected phenomena. The crucial advantages of the PTF survey were its large field of view of 7.2 square degrees, to monitor a large area of the night sky, and its well-matched spectroscopic resources, to routinely obtain spectra of even the faintest transients detected with the P48. Between 2009 and mid-2017, PTF discovered over 3000 SNe ($\gtrsim2100$ SNe~Ia and $\gtrsim900$ CCSNe).

In this paper, we present the CCSN sample and its host-galaxy properties. Our scope is to (i) deduce the distribution functions of the host-galaxy properties such as mass and star-formation rate, (ii) quantify the environment dependence for the production efficiency for the main SN classes, and (iii) indirectly constrain the progenitor properties for the largest CCSN classes.

We assume $\Lambda$CDM cosmology with $H_0=67.8~{\rm km\,s}^{-1}\,{\rm Mpc}^{-1}$, $\Omega_M=0.308$, and $\Omega_\Lambda=0.692$ \citep{Planck2016a}. We report uncertainties at $1\sigma$ confidence. All magnitudes are reported in the AB system \citep{Oke1983a}. A machine-readable catalog of the PTF CCSN sample is available at \href{http://www.github.com/steveschulze/PTF}{http://www.github.com/steveschulze/PTF}. SN classification spectra will be publicly available on WISeREP\footnote{\href{https://wiserep.weizmann.ac.il/}{https://wiserep.weizmann.ac.il}} \citep{Yaron2012a}. Further data products, such as tabulated versions of the probability distribution functions, detailed outputs from the galaxy spectral energy distribution modeling, and the host identification, will be released after acceptance of the paper.

\section{Data}
\subsection{Supernova Data}

\subsubsection{Photometry}\label{data:sn_images}

We retrieved fully-reduced SN images obtained with P48 and the 1.5~m (60~inch) P60 telescopes from the NASA/IPAC Infrared Science Archive\footnote{\href{https://irsa.ipac.caltech.edu/Missions/ptf.html}{https://irsa.ipac.caltech.edu/Missions/ptf.html}} and the P60 archive\footnote{ \href{http://eakins.ipac.caltech.edu:8000/cgi-bin/P60/nph-p60login}{http://eakins.ipac.caltech.edu:8000/cgi-bin/P60/nph-p60login}}, respectively. The P48 image reduction is described by \citet{Laher2014a}, while the PTF photometric calibration and the photometric system are discussed by \citet{Ofek2012a}. We used these data only for the host-galaxy identification. The SN light curves and their analysis are beyond the scope of this paper.

\subsubsection{Spectroscopy}\label{data:sn_spectra}

Supernova spectra were obtained primarily with the low-resolution Double Beam Spectrograph (DBSP; \citealt{Oke1982a}) on the 5~m (200~inch) Hale telescope (P200) at Palomar Observatory (USA), the Low-Resolution Imaging Spectrometer (LRIS; \citealt{Oke1995a}) on the 10~m Keck~I telescope on Maunakea (USA), and the Kast double spectrograph\footnote{\href{https://mthamilton.ucolick.org/techdocs/instruments/kast/Tech\%20Report\%2066\%20KAST\%20Miller\%20Stone.pdf}{https://mthamilton.ucolick.org/techdocs/instruments/kast/\\Tech\%20Report\%2066\%20KAST\%20Miller\%20Stone.pdf}} on the 3~m Shane telescope at Lick Observatory on Mount Hamilton (USA).

\begin{table*}
\caption{Properties of the different imaging surveys}
\label{tab:surveys}
\footnotesize
\hspace{-1.5cm}
\begin{tabular}{lcllccc}
\toprule
Survey          & Spectral      & Filters                                       & Depth                         & Pixel scale       & FWHM(PSF)     & Ref.\\
                & range (\AA)   &                                               & (mag)                         & ($''/{\rm px}$)   & ($''$)        &           \\
\midrule
\galex          & 1542-2274     & \galexfuv, \galexnuv                      & 20, 21                            & 1.5               & 5--6          & 1\\
SDSS            & 3595-8897     & \sdssu, \sdssg, \sdssr, \sdssi, \sdssz    & 22.2, 23.1, 22.7, 22.2, 20.7      & 0.396             & 1.3--1.5      & 2\\
PS1             & 4776-9603     & \psg, \psr, \psi, \psz, \psy              & 23.3, 23.2, 23.1, 22.3, 21.3      & 0.258             & 1.0--1.5      & 3\\
Legacy Surveys  & 4635-9216     & \lsg, \lsr, \lsz                          & 24.0\&24.7, 23.5\&23.9, 22.5\&23.0   & 0.262             & 1.1--1.3      & 4\\
2MASS           & 12350-21590   & \twomassj, \twomassh, \twomassk           & 17.5, 17.2, 16.9                  & 2                 & 2.5--3.0      & 5\\
unWISE         & 33526-46028  & \wiseone, \wisetwo                        & 20.4, 19.9                        & 2.75              & 6             & 6\\
\bottomrule
\end{tabular}

\tablecomments{The depths of the images are reported for point sources at the $5\sigma$ confidence level. For the DESI Legacy Imaging Surveys, we report the depths of DR5\&DR7. The spectral ranges were taken from \href{http://svo.cab.inta-csic.es/main/index.php}{http://svo.cab.inta-csic.es/main/index.php}.
The column ``FWHM'' reports the span of the median FWHMs of all bands.
}
\tablerefs{1) \citet{Bianchi2014a}; 2) \href{https://www.sdss.org/dr14/imaging/other_info}{https://www.sdss.org/dr14/imaging/other\_info}; 3) \citet{Chambers2016a}; 4)  \citet{Dey2018a}; 5) \citet{Skrutskie2006a}; 6) \citet{Meisner2017a}
}
\end{table*}

We augmented this dataset with observations obtained with the low-resolution Gemini Multi-Object Spectrograph \citep[GMOS;][]{Hook2004a} on the 8.1~m Gemini telescopes on Hawaii (USA) and Cerro Pach\'on (Chile), Auxiliary-port CAMera \citep[ACAM;][]{Benn2008a} on the 4.2~m William Herschel Telescope (WHT) on the Canary Islands (Spain), the Ritchey-Chretien Focus Spectrograph\footnote{\href{https://www.noao.edu/kpno/manuals/l2mspect/node1.html}{https://www.noao.edu/kpno/manuals/l2mspect/node1.html}} (RC) on the 4~m Kitt Peak National Observatory (KPNO) telescope, and Alhambra Faint Object Spectrograph and Camera\footnote{\href{http://www.not.iac.es/instruments/alfosc}{http://www.not.iac.es/instruments/alfosc}} (ALFOSC) on the 2.56~m Nordic Optical Telescope (NOT) on the Canary Islands. We also obtained intermediate-resolution spectra of several objects with the optical-to-near-infrared echellete spectrograph X-shooter \citep{Vernet2011a} on the ESO 8.2~m Very Large Telescope (VLT) on Cerro Paranal, Chile, and with the optical DEep Imaging Multi-Object Spectrograph \citep[DEIMOS;][]{Faber2003a} on the 10~m Keck~II telescope on Maunakea. For two objects, we utilized publicly available spectra. A log of the spectroscopic observations, including references to the archival data, is presented in Appendix \ref{appendix:spec}.

\subsection{Host-Galaxy Data}
\subsubsection{Photometry}\label{data:host_photometry}

We retrieved science-ready coadded images from the \textit{Galaxy Evolution Explorer} (\galex) general release 6/7 \citep{Martin2005a}, the Sloan Digital Sky Survey data release 9 (SDSS DR 9; \citealt{Ahn2012a}), the Panoramic Survey Telescope and Rapid Response System (Pan-STARRS, PS1) DR1 \citep{Chambers2016a}, the Two Micron All Sky Survey \citep[2MASS;][]{Skrutskie2006a}, and preprocessed \wise\ images \citep{Wright2010a} from the unWISE archive \citep{Lang2014a}\footnote{\href{http://unwise.me}{http://unwise.me}}. The unWISE images are based on the public \wise\ data and include images from the ongoing NEOWISE-Reactivation mission R3 \citep{Mainzer2014a, Meisner2017a}.

Several fields were observed more than once with \galex. We considered only those images where the respective SN was within $0\fdg5$ from the center of a \galex\ pointing to avoid regions where the \galex\ zeropoint and astrometry begin to vary with distance from the pointing center \citep{Bianchi2017a}. If a field was observed multiple times in a given filter, we chose the deepest observation. For the unWISE images, we only considered observations where a given SN was $>50$ pixels from the chip edge.

The hosts of $\sim10\%$ of our SN sample were not detected by these surveys. For those objects, we retrieved deeper optical images from the DESI Legacy Imaging Surveys \citep[Legacy Surveys, LS;][]{Dey2018a} DR5-7, from the data archive of the 3.6~m Canada-France-Hawaii Telescope (USA), and images from the \textit{Hubble Space Telescope} (\textit{HST}). These data were provided as science-ready by the respective data archives. Furthermore, we augmented our dataset with photometry presented by \citet{Perley2016a}, \citet{De2018a} and \citet{Schulze2018a}.

The vital properties of the \galex, LS, PS1, SDSS, 2MASS, and unWISE data are summarized in Table~\ref{tab:surveys}.

\section{Methods}
\subsection{Host-Galaxy Photometry}\label{method:photometry}

\begin{figure}
\centering
\includegraphics[width=1\columnwidth]{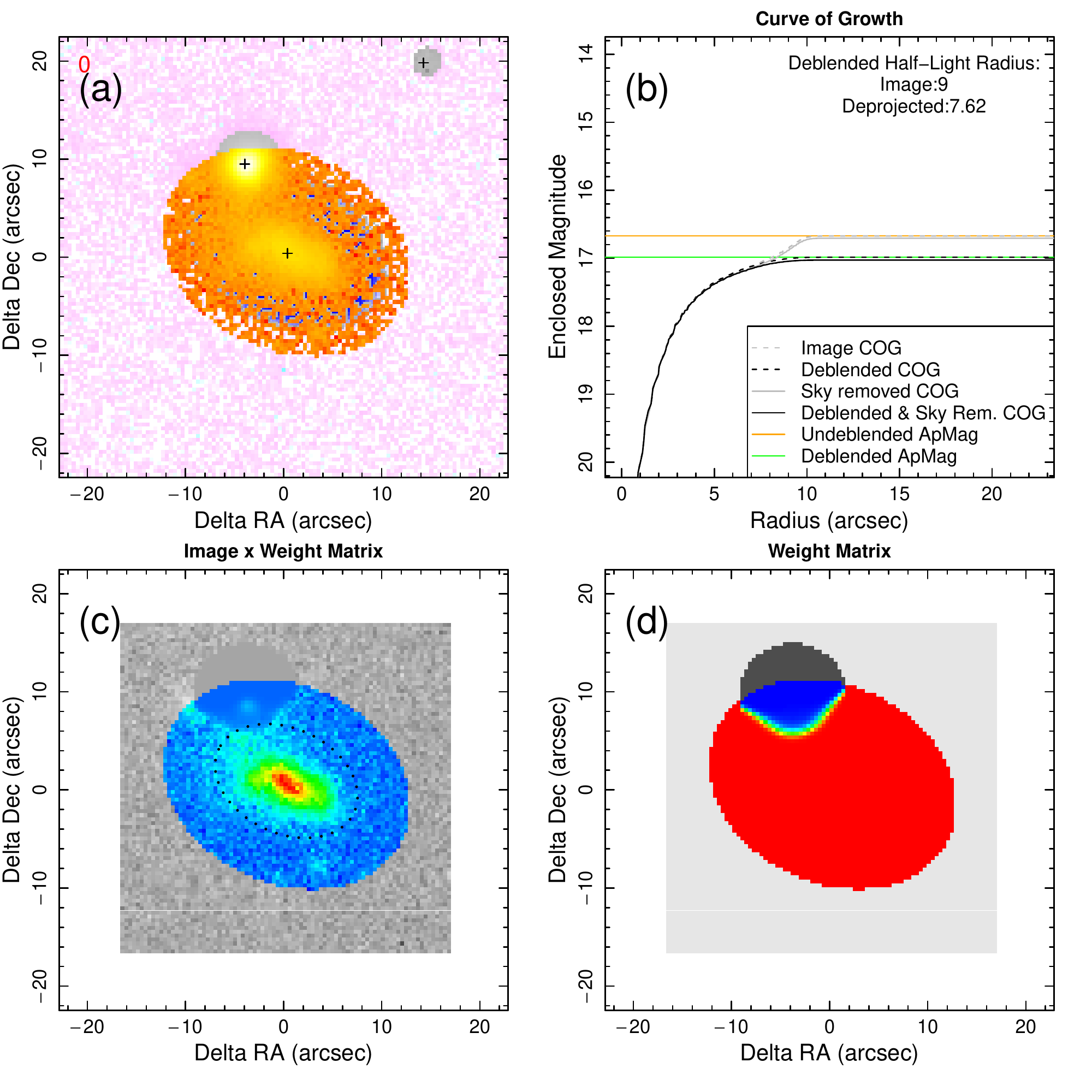}
\caption{Example of the photometry measurement with \package{LAMBDAR} and the importance of deblending techniques to recover the flux of the host galaxy of PTF10fqg. Panel $a$ shows the input SDSS $r'$-band image. Positive fluxes within the measurement aperture are shown in yellow. Pixels deemed to be part of the ``sky'' are indicated in pink. Panel $b$ illustrates the curve of growth (CoG) of the object photometry. The gray lines indicate the radial integral of Panel $a$ (before deblending). The black lines show the radial integral of Panel $c$ (after deblending). Horizontal orange and green lines mark the measured aperture magnitude for the object before and after deblending, respectively. The text in that panel describes the circular and deprojected half-light radii, in arcseconds, with the deprojection being based on the input aperture (prior to convolution). Panel $c$ shows the image stamp after deblending. The black dotted line marks the measured deblended and deprojected half-light radius. Panel $d$ shows the deblend weights of the host galaxy. Colored and grayscale pixels in Panels $c$ and $d$ mark those within and beyond the aperture, respectively. Figure adapted from \citet{Wright2016a}.
}
\label{fig:lambdar}
\end{figure}

The survey images are characterized by different pixel (px) scales and point-spread functions (PSFs): the pixel scales vary from $0.26 ''/{\rm px}$ (LS and PS1) to $2.75''/{\rm px}$ (unWISE), and the full width at half-maximum intensity (FWHM) of the PSFs varies from $1''$ (LS and PS1) to $6''$ (unWISE; Table \ref{tab:surveys}). These differences need to be taken into account to measure the total flux of galaxies and preserve intrinsic galaxy colors. The software package \package{LAMBDAR}\footnote{\href{https://github.com/AngusWright/LAMBDAR}{https://github.com/AngusWright/LAMBDAR}} \citep[Lambda Adaptive Multi-Band Deblending Algorithm in R;][]{Wright2016a} was designed for this task, to conduct consistent photometry on images that are neither pixel nor seeing matched while performing deblending, sky subtraction, and error estimation. \package{LAMBDAR} has three critical input parameters for each image: the PSF, the zeropoint, and the source catalog of the field (see below). With this information in hand, \package{LAMBDAR} convolves each aperture with the PSF of a given image, re-projects the PSF-matched aperture on the new pixel scale, and iteratively removes neighboring objects that are blended with the host galaxy. An example of this technique is shown in Fig.~\ref{fig:lambdar}.

\paragraph{PSF} The \galex\ PSFs are provided by the \galex\ Technical Documentation\footnote{\href{http://www.galex.caltech.edu/wiki/Public:Documentation}{http://www.galex.caltech.edu/wiki/Public:Documentation}}. We built the PSF of the unWISE images using the parameterization of \citet{Lang2014a}. For 2MASS, PS1, and SDSS images, we measured the median FWHM of point sources in each image and assumed that a PSF could be approximated by a Gaussian profile.

\paragraph{Zeropoint} We used tabulated zeropoints for \galex, PanSTARRS, SDSS, and unWISE images. Specifically, we set the zeropoints to
\begin{itemize}[leftmargin=*, noitemsep]
  \item  18.82 and 20.08~mag for \galex\ $FUV$ and $NUV$ data \citep{Morrissey2007a}, respectively;
  \item 22.5~mag for SDSS images\footnote{\href{https://www.sdss.org/dr12/algorithms/magnitudes}{https://www.sdss.org/dr12/algorithms/magnitudes}};
  \item $25 + 2.5 \times \log ({\rm exposure~time})$~mag for PS1 images\footnote{\href{https://outerspace.stsci.edu/display/PANSTARRS/PS1+FAQ+-+Frequently+asked+questions}{https://outerspace.stsci.edu/display/PANSTARRS/PS1+FAQ+-+Frequently+asked+questions}}, where the exposure time is given in seconds; and
  \item 22.5~mag for the unWISE $W1$ and $W2$ images \citep{Lang2014a}.
\end{itemize}
\noindent To extract the zeropoints of the 2MASS images, we identified stars and compared their instrumental magnitudes to the tabulated magnitudes in the 2MASS Point Source Catalog \citep{Skrutskie2006a}. The 2MASS and unWISE zeropoints were converted from the Vega system to the AB system using the offsets reported by \citet{Blanton2007a} and \citet[][their Table 3 in Section 4.4h]{Cutri2013a}.

\paragraph{Source catalog} The source catalog of each field contains the positions and aperture properties (minor and major semi-axis and the orientation of the ellipse) of the host galaxy and contaminant sources (stars, galaxies, active galactic nuclei). We obtained these properties with \package{Source Extractor}\footnote{\href{https://www.astromatic.net/software/sextractor}{https://www.astromatic.net/software/sextractor}} version 2.19.5 \citep{Bertin1996a} and, where needed, adjusted the size of the elliptical apertures to measure the total flux.

The photometry of the LS and CFHT images was extracted with the aperture-photometry tool presented by \citet{Schulze2018a}\footnote{\href{https://github.com/steveschulze/Photometry}{https://github.com/steveschulze/Photometry}}. The photometry of this software and \package{LAMBDAR} agree well for galaxies that are not very extended or blended with other objects.

All measurements are presented in Appendix~\ref{appendix:phot}. Measurements were corrected for Galactic extinction using the Python package \package{sfdmap}\footnote{\href{https://github.com/kbarbary/sfdmap}{https://github.com/kbarbary/sfdmap}} version 0.1.1 that makes use of the \citet{Schlafly2011a} extinction maps.

\subsection{Supernova Spectroscopy}\label{method:sn_spectra}

We used dedicated pipelines to reduce data of different instruments: Keck/LRIS data were processed with \package{LPIPE} \citep{Perley2019a}, VLT/X-shooter data with the ESO X-shooter instrument pipeline\footnote{\href{https://www.eso.org/sci/software/pipelines}{https://www.eso.org/sci/software/pipelines}}, and Gemini data with the \package{Gemini IRAF package}\footnote{\href{http://www.gemini.edu/sciops/data-and-results/processing-software}{http://www.gemini.edu/sciops/data-and-results/processing-software}}. Data from other telescopes were reduced with the software package \package{IRAF}\footnote{\href{https://iraf-community.github.io}{https://iraf-community.github.io}} \citep{Tody1986a}.

The typical steps of all software packages are bias subtraction, flat fielding, source extraction in a statistically optimal way \citep{Horne1986a}, wavelength calibration (in the air reference system), and flux calibration with spectrophotometric standard stars. In most cases, a telluric correction was attempted, too. The wavelength calibration of instruments with multiple arms, such as LRIS and Kast, can be nontrivial. This can lead to velocity offsets of $\sim200~{\rm km~s}^{-1}$ between different arms of the instruments.

\subsection{Supernova Classification}\label{method:classification}

All SNe were spectroscopically classified with the software packages \package{Supernova Identification}\footnote{\href{https://people.lam.fr/blondin.stephane/software/snid}{https://people.lam.fr/blondin.stephane/software/snid}} version 5.0 \citep[\package{SNID};][]{Blondin2007a} or  \package{Superfit}\footnote{\href{https://github.com/dahowell/superfit}{https://github.com/dahowell/superfit}} version 3.5 \citep{Howell2005a}, if the host contamination was significant. For both packages, we generated template libraries that include spectra from \citet{Silverman2012a}, \citet{Modjaz2014a}, \citet{Liu2016a}, \citet{Modjaz2016a}, and a private library built by S. Ben-Ami and extended by G. Leloudas to include superluminous supernovae (SLSNe) and other rare transients. We preferentially used spectra obtained around the time of maximum light. If not available or to break ambiguity in the classification, we used late-time data.

A minority of PTF CCSNe were classified by other teams, and we used these data if they were publicly available. In addition, a number of PTF CCSNe were previously discussed in the literature \citep[e.g.,][]{Fremling2018a, Quimby2018a, Taddia2019a, Modjaz2019a}. For full consistency, we repeated classification of these objects using our tools.

\subsection{Host-Galaxy Identification}\label{method:host_id}

To identify the hosts, we retrieved two P48 or P60 images for each SN field: one image around the time of SN brightness maximum (PTF+SN) and one without SN contribution (PTF-SN). The PTF-SN image was chosen to have seeing FWHM similar to the PTF+SN image and taken during dark/grey time. With both images in hand, we built the difference image (i.e., SN detection image). First, we aligned the images on the pixel level using \package{alipy}\footnote{\href{https://obswww.unige.ch/~tewes/alipy}{https://obswww.unige.ch/\~{}tewes/alipy}}, and then subtracted the images from each other using the software \package{High Order Transform of Psf ANd Template Subtraction}\footnote{\href{https://github.com/acbecker/hotpants}{https://github.com/acbecker/hotpants}} \citep[\package{HOTPANTS};][]{Becker2015a} version 5.1.11.

After that, we aligned the world-coordinate system of the PTF-SN image (usually in $r_{\rm PTF}$) to that of a host image (usually an $r'$-band image from LS, PS1 or SDSS) using the software \package{Software for Calibrating AstroMetry and Photometry}\footnote{\href{https://www.astromatic.net/software/scamp}{https://www.astromatic.net/software/scamp}} \citep[\package{SCAMP};][]{Bertin2006a} version 2.0.4 and applied the calibration of the world coordinate system on the difference image. In the final step, we built source catalogs of the host and difference images with the software \package{Source Extractor}. The closest object to the SN, within a reasonable distance, was declared as the host galaxy.

\subsection{Spectral Energy Distribution Fitting}\label{method:sed}
\subsubsection{Description of the Method}

To extract physical and phenomenological parameters of the host galaxies, we modeled the spectral energy distributions (SEDs) with the software package \package{Prospector}\footnote{\href{https://github.com/bd-j/prospector}{https://github.com/bd-j/prospector}} version 0.3 \citep{Leja2017a}. \package{Prospector} uses the \package{Flexible Stellar Population Synthesis}\footnote{\href{https://github.com/cconroy20/fsps}{https://github.com/cconroy20/fsps}} (\package{FSPS}) code \citep{Conroy2009a} to generate the underlying physical model and \package{python-fsps}\footnote{\href{http://dfm.io/python-fsps/current/}{http://dfm.io/python-fsps/current/\#}} \citep{Foreman-Mackey2014a} to interface with in Python.

The SED model required assumptions for the star-formation history (SFH), initial mass function (IMF), attenuation, and whether a contribution from the ionized gas should be taken into account. We used a linear-exponential SFH [functional form $t \times \exp\left(-t/\tau\right)$, where $t$ is the age of the SFH episode and $\tau$ is the $e$-folding timescale], the \citet{Chabrier2003a} IMF, the \citet{Calzetti2000a} attenuation model, and the \citet{Byler2017a} model for the ionized gas contribution.

The priors were set as distribution functions with broad ranges as specified in Table~\ref{tab:priors}. The physical parameters mass, star-formation rate (SFR), age, $e$-folding timescale of the SFH, extinction, and metallicity were inferred in a Bayesian way by sampling the posterior probability functions with the dynamic nested sampling package \package{dynesty}\footnote{\href{https://github.com/joshspeagle/dynesty}{https://github.com/joshspeagle/dynesty}} \citep{Speagle2020a}. For each model parameter, we report the median values of the marginalized posterior probability functions and their 1$\sigma$ confidence intervals.

\begin{table}
\caption{Model parameters and their priors of the galaxy SED modeling}
\hspace{-1cm}
\label{tab:priors}
\begin{tabular}{lll}
\toprule
Property									& Type		& Range		\\
\midrule
Galaxy mass ($\log\,M_\star/M_\odot)$		& Uniform	& 5--13		\\
$V$-band optical depth ($\tau_V$)$^\dagger$	& Uniform	& 0--8		\\
Stellar metallicity ($\log Z/Z_\odot$)	    & Uniform		& $-2$--0.5	\\
Age of the SF episode ($t_{\rm age}$/Gyr) 	& LogUniform		& 0.001-13.8\\
e-folding time-scale of the star-           & LogUniform	& 0.1--100  \\
formation episode	($\tau$/Gyr)	\\
\bottomrule
\end{tabular}
\tablenotetext{\dagger}{The optical depth in $V$ band was converted to the selective-to-total extinction via $E(B-V)=1.086\times\tau_V/\kappa(V)$, where $\kappa(V)$ is the $V$-band opacity of the \citet{Calzetti2000a} attenuation model.}
\end{table}

\subsubsection{Quality of the SED Modelling}\label{res:sed2}

\begin{figure}
\centering
\includegraphics[width=0.49\columnwidth]{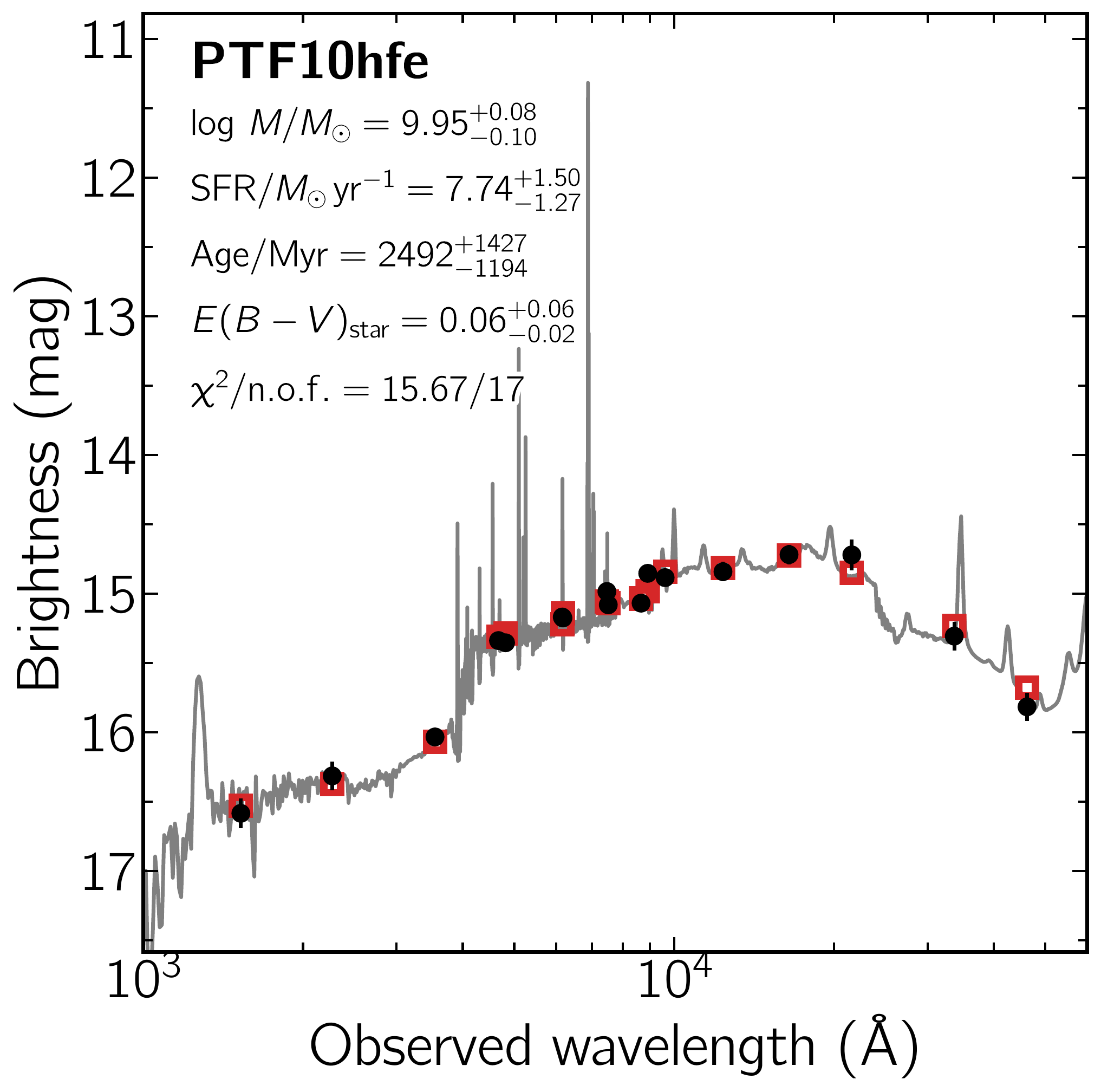}
\includegraphics[width=0.49\columnwidth]{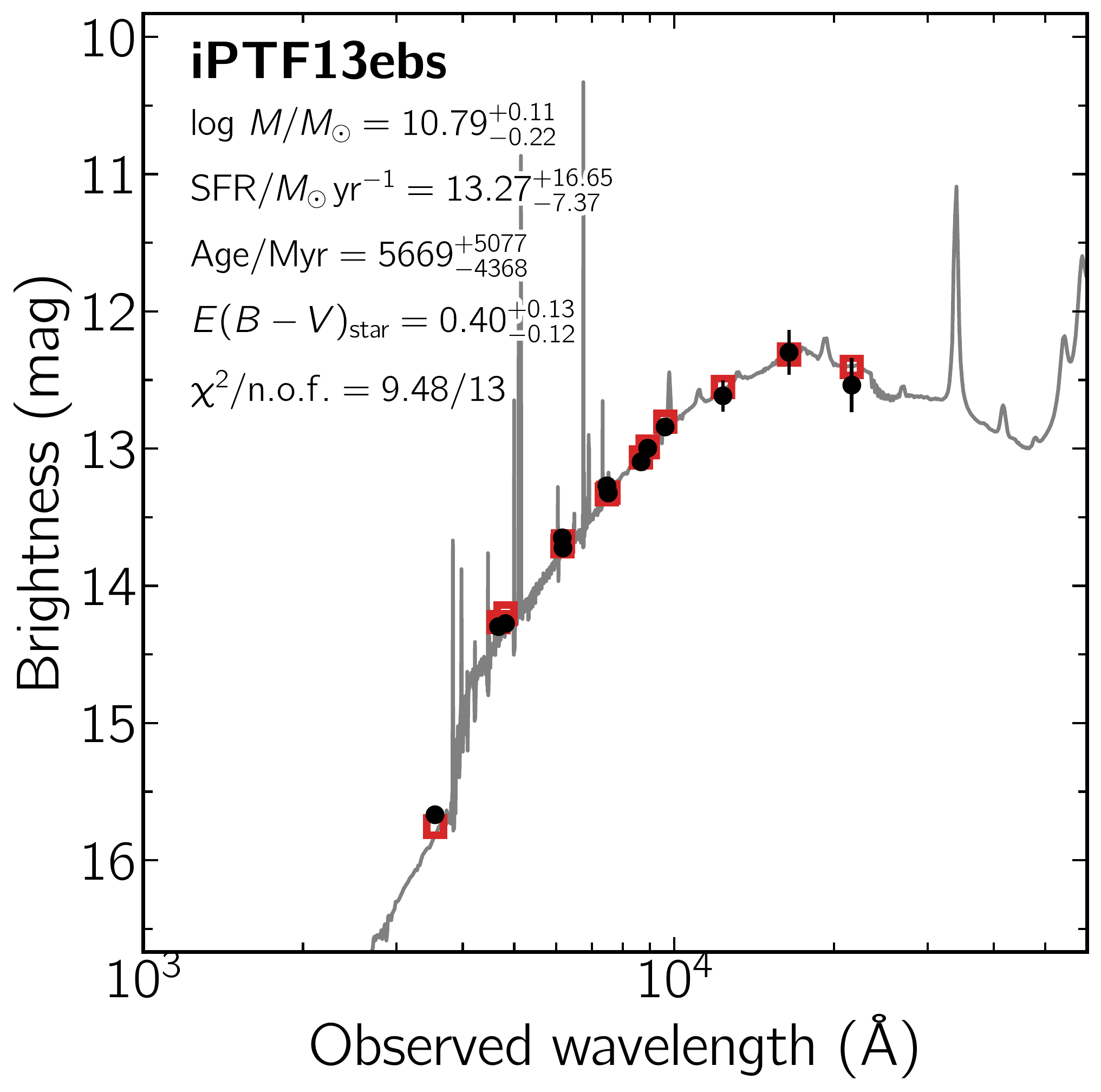}
\includegraphics[width=0.49\columnwidth]{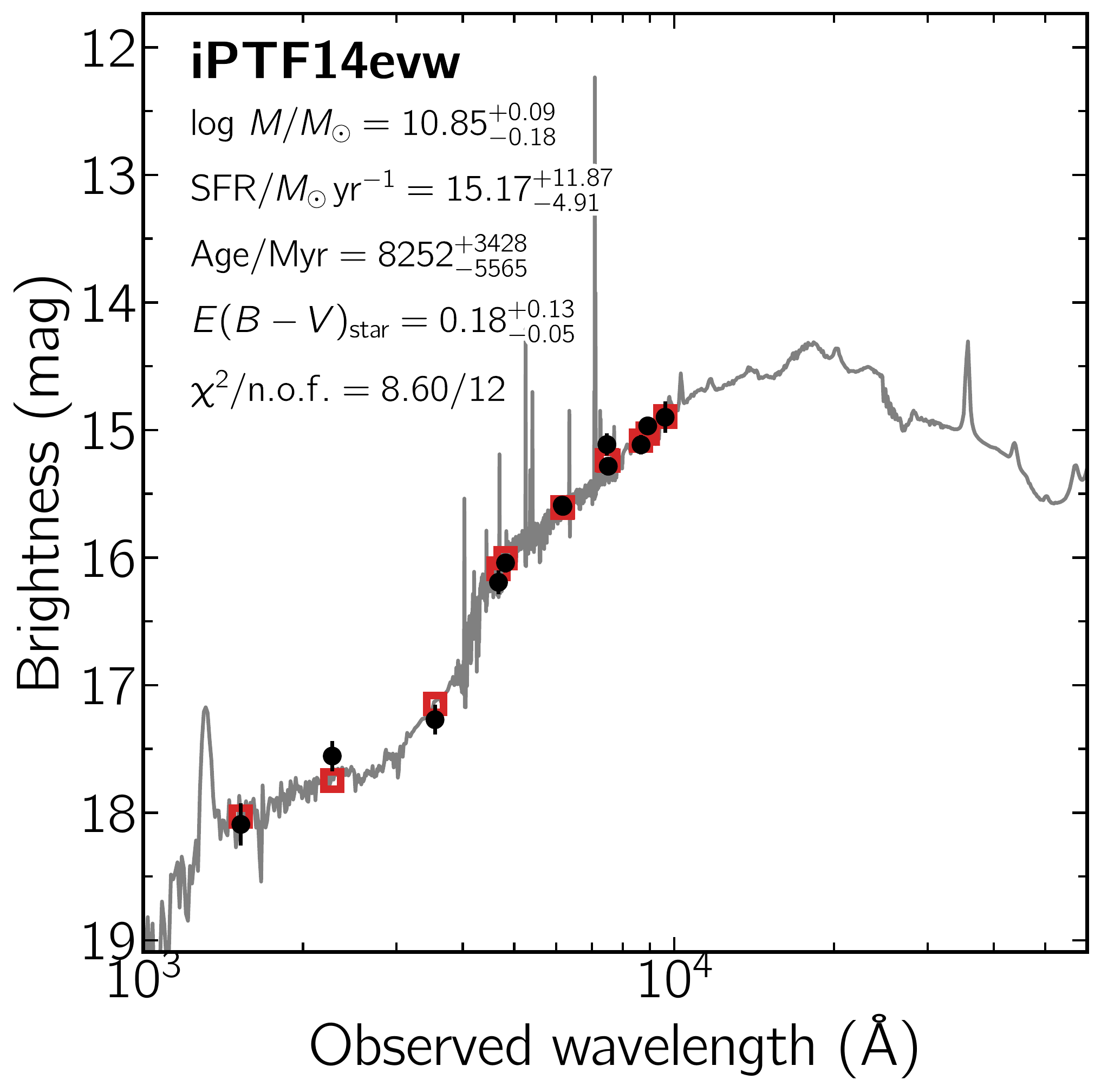}
\includegraphics[width=0.49\columnwidth]{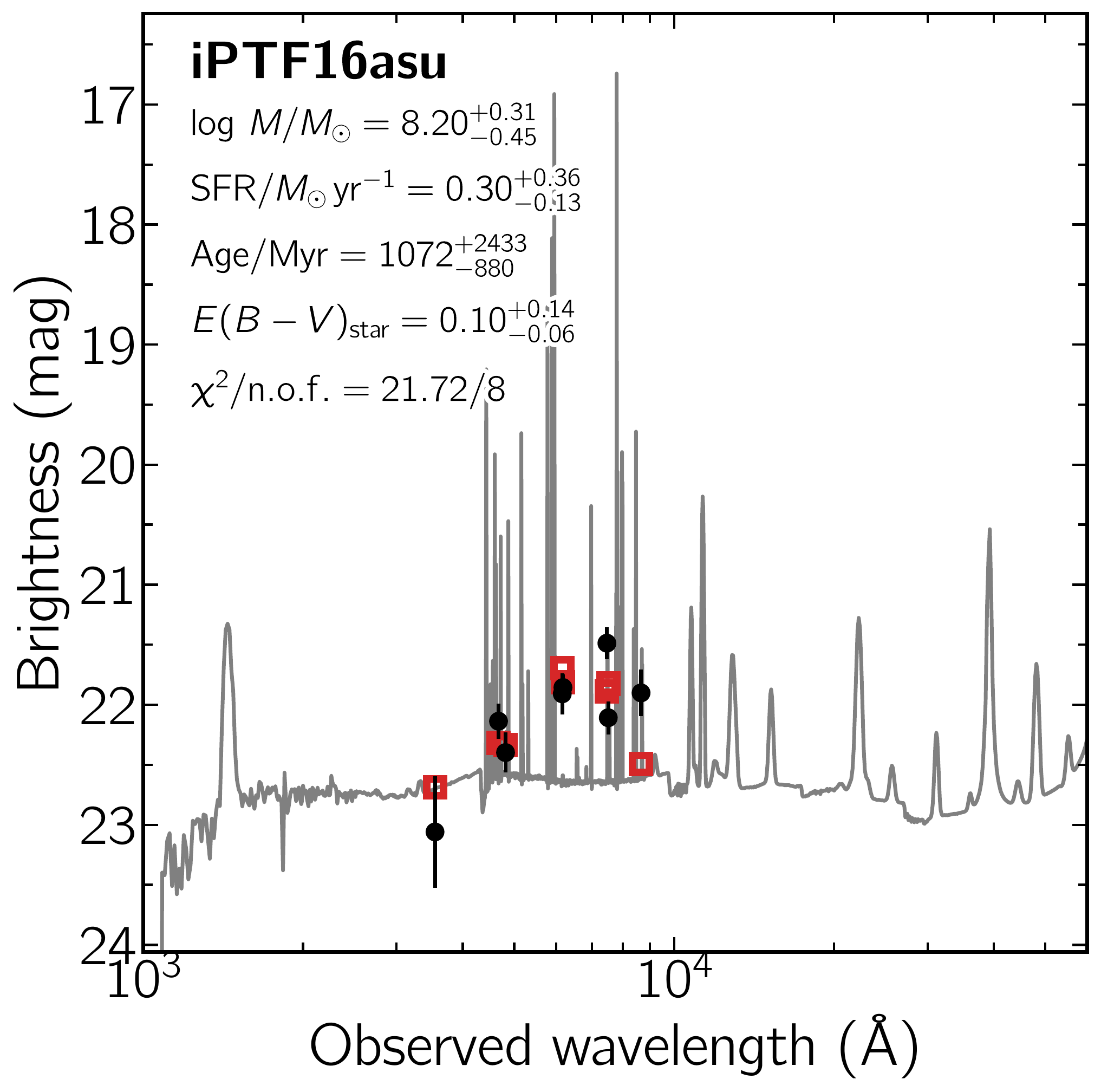}
\caption{Selection of SEDs of SN host galaxies from 1000 to 60,000 \AA\ (detections $\bullet$; upper limits $\blacktriangledown$). The solid line displays the best-fitting model of the SED. The squares represent the model-predicted magnitudes. The fitting parameters are shown in the upper-left corner. The abbreviation ``n.o.f.'' stands for numbers of filters.
}
\label{fig:sedfit}
\end{figure}

Figure~\ref{fig:sedfit} shows examples of the observed SEDs and their fits with \package{Prospector}. The average galaxy SED is observed in 13 bands from the far-ultraviolet (FUV) to the mid-infrared (MIR). After accounting for similarities between SDSS and PS1 filters, each SED has, on average, nine measurements. Our assumed model in \package{Prospector} provides an adequate description of most SEDs (e.g., Fig.~\ref{fig:sedfit}). The median $\chi^2$ divided by the number of filters (n.o.f.) is 0.9. A minority of $\sim3\%$ have a reduced $\chi^2$ between 3 and 12. Nonetheless, these fits are still useful. The large reduced $\chi^2$ is driven by differences between SDSS and PS1 photometry of extended galaxies, small measurement errors, or individual data points.

The derived physical parameters, such as galaxy mass ($M_\star$), star-formation rate (SFR), and specific star-formation rate (sSFSR = SFR / $M_\star$), summarized in Table~\ref{tab:host_prospector}, are comparable to those of other galaxy samples and broadly consistent with results from the literature \citep{Leloudas2015a, Schulze2018a, Modjaz2019a, Taggart2019a}. The galaxy masses and the SFRs are 0.1--0.2 dex smaller compared to the values reported by \citet{Schulze2018a}, \citet{Modjaz2019a}, and \citet{Taggart2019a}, even if identical datasets are used. The bias-corrected root-mean square (r.m.s.) of the galaxy mass, SFR, and sSFR vary between 0.3 and 1.3 dex. The uncertainties of the r.m.s. values reach up to 0.8 dex, making most of these differences statistically not significant. Differences are also expected; they are due to the assumptions of the SED model (e.g., the nebular emission module, SFHs, stellar-population synthesis models) and assumptions inherent to the SED modeling software packages.

\begin{table*}
\movetabledown=6cm
\caption{PTF CCSN sample}\label{tab:general}
\tiny
\hspace{-2cm}
\begin{tabular}{cccccccccccc}
\toprule
PTF	& 	IAU Name	& Type		& Redshift	    & Method    & R.A. (SN)		& Decl. (SN)	& R.A. (Host)	& Decl. (Host)	& Offset	& Offset & $E_{\rm MW}(B-V)$\\ 
			& 				& 			& 			&           & (hh:mm:ss)	& (dd:mm:ss) 	& (hh:mm:ss)	& (dd:mm:ss)	& ($''$)	& (kpc)	 & (mag)\\ 
\midrule
09as & 2009cb	& SLSN-I & 0.1866   & 3& 12:59:15.862    & +27:16:40.80 & 12:59:15.870 & +27:16:40.71 & $0.15\pm0.18$ & $0.48\pm0.57$ & 0.01\\ 
09atu & \nodata	& SLSN-I & 0.5010   & 3& 16:30:24.537    & +23:38:25.59 & 16:30:24.544 & +23:38:25.52 & $0.11\pm0.16$ & $0.69\pm0.99$ & 0.04\\ 
09awk & \nodata	& SN Ib & 0.0616    & 3& 13:37:56.358     & +22:55:04.78 & 13:37:56.359 & +22:55:04.78 & $0.02\pm0.17$ & $0.02\pm0.20$ & 0.02\\ 
09axi & \nodata	& SN II & 0.0640    & 3& 14:12:40.850     & +31:04:03.89 & 14:12:40.943 & +31:04:03.40 & $1.29\pm0.11$ & $1.64\pm0.14$ & 0.01\\ 
09bce & \nodata	& SN II & 0.0234    & 3& 16:35:17.680     & +55:37:59.33 & 16:35:17.657 & +55:38:01.54 & $2.21\pm0.11$ & $1.08\pm0.05$ & 0.01\\ 
09bcl & \nodata	& SN II & 0.0600    & 4& 18:06:26.809     & +17:51:43.15 & 18:06:26.442 & +17:51:42.67 & $5.26\pm0.06$ & $6.30\pm0.07$ & 0.08\\ 
09be & \nodata	& SN II & 0.1020    & 4& 14:10:18.538     & +16:53:38.71 & 14:10:18.493 & +16:53:38.53 & $0.66\pm0.25$ & $1.28\pm0.48$ & 0.02\\ 
09bgf & \nodata	& SN II & 0.0318    & 3& 14:41:38.329     & +19:21:43.80 & 14:41:38.351 & +19:21:43.19 & $0.68\pm0.14$ & $0.45\pm0.10$ & 0.02\\ 
09bw & 2009cw	& SN Ib & 0.1470    & 3& 15:05:01.990     & +48:40:03.49 & 15:05:02.035 & +48:40:03.23 & $0.52\pm0.12$ & $1.37\pm0.33$ & 0.02\\ 
09cjq & \nodata	& SN II & 0.0193    & 1& 21:16:28.502     & +00:49:39.70 & 21:16:27.606 & +00:49:34.77 & $14.32\pm0.14$ & $5.79\pm0.05$ & 0.06\\ 
09cnd & \nodata	& SLSN-I & 0.2583   & 3& 16:12:08.838    & +51:29:16.02 & 16:12:08.838 & +51:29:15.49 & $0.53\pm0.44$ & $2.19\pm1.82$ & 0.02\\ 
09ct & 2009cv	& SN IIn & 0.1560   & 3& 11:42:13.827    & +10:38:54.19 & 11:42:13.843 & +10:38:53.87 & $0.40\pm0.40$ & $1.11\pm1.13$ & 0.03\\ 
09cu & 2009ct	& SN II & 0.0569    & 1& 13:15:23.135     & +46:25:09.16 & 13:15:23.897 & +46:25:13.47 & $8.98\pm0.22$ & $10.24\pm0.25$ & 0.01\\ 
09cvi & \nodata	& SN II & 0.0360    & 4& 21:47:09.947     & +08:18:35.35 & 21:47:09.925 & +08:18:35.55 & $0.38\pm0.20$ & $0.28\pm0.15$ & 0.07\\ 
09cwl & 2009jh	& SLSN-I & 0.3500   & 3& 14:49:10.108    & +29:25:11.68 & 14:49:10.177 & +29:25:12.78 & $1.42\pm0.20$ & $7.21\pm1.04$ & 0.01\\ 
09dah & \nodata	& SN IIb & 0.0238   & 3& 22:45:17.094    & +21:49:15.27 & 22:45:17.102 & +21:49:15.29 & $0.12\pm0.07$ & $0.06\pm0.03$ & 0.05\\ 
09dfk & \nodata	& SN Ib & 0.0158    & 3& 23:09:13.427     & +07:48:15.31 & 23:09:13.483 & +07:48:16.58 & $1.52\pm0.13$ & $0.51\pm0.04$ & 0.05\\ 
09dh & 2009dr	& SN Ic & 0.0770    & 4& 14:44:42.072     & +49:43:45.17 & 14:44:42.105 & +49:43:45.94 & $0.84\pm0.08$ & $1.26\pm0.12$ & 0.02\\ 
09dra & \nodata	& SN II & 0.0766    & 1& 15:48:11.483     & +41:13:28.68 & 15:48:11.318 & +41:13:31.54 & $3.42\pm0.17$ & $5.12\pm0.25$ & 0.01\\ 
\bottomrule
\end{tabular}
\tablecomments{The IAU names were retrieved from the Transient Naming Server (\href{https://wis-tns.weizmann.ac.il}{https://wis-tns.weizmann.ac.il}). The redshifts were obtained either from SDSS (method = 1), the NASA Extragalactic Database (method = 2), galaxy lines in SN spectra (method = 3), or SN-template matching (method = 4). The SN coordinates were measured after aligning SN and host images (for details see Sects. \ref{method:host_id}, \ref{result:host_id}). The coordinates are reported in a  conventional celestial reference system in the J2000.0 system. The full table is available online in a machine-readable form.}
\end{table*}

\begin{table*}
\caption{Results from the host galaxy SED modelling with \package{Prospector}}\label{tab:host_prospector}
\begin{center}
\begin{tiny}
\begin{tabular}{lcccrrrrrrrr}
\toprule
PTF	& 	Type	& Redshift	& $\chi^2/{\rm n.o.f.}$	& \multicolumn{1}{c}{$E_{\rm host}(B-V)$}	&\multicolumn{1}{c}{$M_{\rm FUV}$}	& \multicolumn{1}{c}{$M_{\rm B}$}	& \multicolumn{1}{c}{$ M_{\rm Ks}$} 	& \multicolumn{1}{c}{$\log~{\rm SFR}$}			& \multicolumn{1}{c}{$\log~M$}		&\multicolumn{1}{c}{$\log~{\rm sSFR}$}	&\multicolumn{1}{c}{$\log~{\rm Age}$}\\ 
			& 			& 			& 							& \multicolumn{1}{c}{(mag)}		& \multicolumn{1}{c}{(mag)}			& \multicolumn{1}{c}{(mag)}	 		& \multicolumn{1}{c}{(mag)}				&\multicolumn{1}{c}{$\left(M_\odot\,{\rm yr}^{-1}\right)$}	&\multicolumn{1}{c}{$\left(M_\odot\right)$}	&\multicolumn{1}{c}{$\left({\rm yr}^{-1}\right)$}		&\multicolumn{1}{c}{$\left({\rm Gyr}\right)$} \\ 
\midrule
09as  & SLSN-I & 0.1866 & $8.21/9$ & $0.20\pm0.05$ & $-16.20^{+0.16}_{-0.10}$ & $-17.62^{+0.11}_{-0.07}$ & $-17.96^{+0.31}_{-0.28}$ & $-0.22^{+0.28}_{-0.23}$ & $8.19^{+0.36}_{-0.57}$ & $-8.40^{+0.82}_{-0.55}$ & $0.58^{+1.57}_{-0.50}$ \\ 
09atu & SLSN-I & 0.5010 & $2.94/7$ & $0.23\pm0.14$ & $-14.48^{+0.72}_{-0.58}$ & $-15.70^{+0.27}_{-0.19}$ & $-15.82^{+0.69}_{-0.64}$ & $-0.59^{+0.72}_{-0.63}$ & $6.84^{+0.70}_{-0.49}$ & $-7.43^{+1.03}_{-1.15}$ & $0.06^{+0.81}_{-0.05}$ \\ 
09awk & SN Ib  & 0.0616 & $7.13/15$ & $0.19^{+0.15}_{-0.06}$ & $-17.11^{+0.40}_{-0.14}$ & $-19.07^{+0.12}_{-0.04}$ & $-19.73^{+0.09}_{-0.16}$ & $-0.14^{+0.51}_{-0.23}$ & $9.56^{+0.12}_{-0.27}$ & $-9.69^{+0.92}_{-0.27}$ & $3.98^{+4.45}_{-2.86}$ \\ 
09axi & SN II  & 0.0640 & $7.90/10$ & $0.13^{+0.23}_{-0.09}$ & $-16.04^{+0.68}_{-0.53}$ & $-18.47^{+0.22}_{-0.04}$ & $-18.96^{+0.16}_{-0.43}$ & $-0.82^{+1.01}_{-0.71}$ & $9.17^{+0.12}_{-0.17}$ & $-9.98^{+1.27}_{-0.71}$ & $1.86^{+2.25}_{-0.99}$ \\ 
09bce & SN II  & 0.0234 & \nodata & \nodata & \nodata & \nodata & \nodata & \nodata & \nodata & \nodata & \nodata \\ 
09bcl & SN IIn & 0.0600 & $1.97/10$ & $0.20^{+0.14}_{-0.12}$ & $-13.16^{+1.04}_{-3.99}$ & $-21.19^{+0.39}_{-0.28}$ & $-23.17^{+0.17}_{-0.07}$ & $-2.62^{+3.12}_{-8.82}$ & $11.09^{+0.18}_{-0.34}$ & $-13.67^{+3.75}_{-8.90}$ & $5.81^{+4.39}_{-3.48}$ \\ 
09be  & SN II  & 0.1020 & $0.23/3$ & $0.33^{+0.17}_{-0.16}$ & $-12.75^{+1.03}_{-0.76}$ & $-14.45^{+0.29}_{-0.22}$ & $-14.95^{+0.58}_{-0.66}$ & $-0.91^{+0.71}_{-0.65}$ & $6.51^{+0.66}_{-0.42}$ & $-7.39^{+0.92}_{-1.15}$ & $0.05^{+0.74}_{-0.05}$ \\ 
09bgf & SN II  & 0.0318 & $11.37/14$ & $0.04^{+0.08}_{-0.03}$ & $-16.18^{+0.28}_{-0.13}$ & $-17.49^{+0.15}_{-0.06}$ & $-17.76^{+0.20}_{-0.13}$ & $-1.01^{+0.21}_{-0.11}$ & $8.67^{+0.14}_{-0.26}$ & $-9.68^{+0.52}_{-0.19}$ & $4.68^{+5.48}_{-3.49}$ \\ 
09bw  & SN II  & 0.1470 & $4.76/11$ & $0.17^{+0.09}_{-0.05}$ & $-16.63^{+0.80}_{-0.29}$ & $-18.54^{+0.09}_{-0.06}$ & $-19.46^{+0.26}_{-0.13}$ & $-0.24^{+0.28}_{-0.29}$ & $9.31^{+0.14}_{-0.43}$ & $-9.56^{+0.78}_{-0.34}$ & $4.81^{+5.20}_{-3.90}$ \\ 
09cjq & SN II  & 0.0193 & $25.25/17$ & $0.19^{+0.10}_{-0.05}$ & $-17.98^{+0.44}_{-0.12}$ & $-20.93^{+0.12}_{-0.04}$ & $-22.47^{+0.13}_{-0.06}$ & $0.30^{+0.24}_{-0.16}$ & $10.81^{+0.09}_{-0.29}$ & $-10.51^{+0.69}_{-0.16}$ & $8.59^{+3.55}_{-5.72}$ \\ 
09cnd & SLSN-I & 0.2583 & $5.99/6$ & $0.14^{+0.14}_{-0.10}$ & $-16.43^{+0.47}_{-0.30}$ & $-17.28^{+0.10}_{-0.07}$ & $-17.13^{+0.39}_{-0.49}$ & $-0.46^{+0.59}_{-0.36}$ & $7.91^{+0.37}_{-0.62}$ & $-8.33^{+0.99}_{-0.63}$ & $0.49^{+1.77}_{-0.45}$ \\ 
09ct  & SN IIn & 0.1560 & $5.98/12$ & $0.22^{+0.11}_{-0.10}$ & $-17.06^{+0.82}_{-0.37}$ & $-19.39^{+0.08}_{-0.05}$ & $-20.85^{+0.26}_{-0.11}$ & $0.11^{+0.29}_{-0.27}$ & $9.92^{+0.13}_{-0.28}$ & $-9.81^{+0.66}_{-0.32}$ & $5.34^{+5.04}_{-4.00}$ \\ 
09cu  & SN II  & 0.0569 & $16.71/15$ & $0.23^{+0.24}_{-0.06}$ & $-18.29^{+0.70}_{-0.11}$ & $-20.74^{+0.23}_{-0.04}$ & $-21.87^{+0.08}_{-0.24}$ & $0.52^{+0.77}_{-0.24}$ & $10.54^{+0.10}_{-0.54}$ & $-10.01^{+1.43}_{-0.28}$ & $7.70^{+4.13}_{-6.96}$ \\ 
09cvi & SN II  & 0.0360 & $7.46/4$ & $0.76^{+0.20}_{-0.24}$ & $-9.64^{+1.33}_{-1.38}$ & $-13.85^{+0.18}_{-0.17}$ & $-16.19^{+0.41}_{-0.58}$ & $-0.31^{+0.59}_{-0.66}$ & $6.94^{+0.40}_{-0.23}$ & $-7.21^{+0.62}_{-0.96}$ & $0.04^{+0.32}_{-0.03}$ \\ 
09cwl & SLSN-I & 0.3500 & $4.18/7$ & $0.38^{+0.08}_{-0.14}$ & $-13.00^{+0.65}_{-0.69}$ & $-14.96^{+0.19}_{-0.18}$ & $-15.79^{+0.35}_{-0.22}$ & $-0.55^{+0.46}_{-0.63}$ & $6.78^{+0.69}_{-0.40}$ & $-7.36^{+0.88}_{-1.19}$ & $0.05^{+0.77}_{-0.04}$ \\ 
09dah & SN IIb & 0.0238 & $55.67/17$ & $0.08^{+0.06}_{-0.02}$ & $-16.30^{+0.30}_{-0.11}$ & $-17.84^{+0.17}_{-0.04}$ & $-18.17\pm0.09$ & $-0.88^{+0.23}_{-0.10}$ & $9.01^{+0.11}_{-0.32}$ & $-9.89^{+0.64}_{-0.14}$ & $6.80^{+4.44}_{-4.89}$ \\ 
09dfk & SN Ib  & 0.0158 & $31.60/14$ & $0.19^{+0.05}_{-0.03}$ & $-14.46^{+0.46}_{-0.24}$ & $-16.78^{+0.10}_{-0.04}$ & $-17.57^{+0.07}_{-0.06}$ & $-1.21^{+0.19}_{-0.18}$ & $8.76^{+0.12}_{-0.20}$ & $-9.99^{+0.45}_{-0.17}$ & $4.28^{+4.54}_{-2.76}$ \\ 
09dh  & SN Ic  & 0.0770 & $1.67/3$ & $0.54^{+0.39}_{-0.31}$ & $-9.77^{+2.49}_{-1.61}$ & $-12.88^{+0.52}_{-0.25}$ & $-14.57^{+1.40}_{-1.39}$ & $-1.15^{+1.10}_{-0.92}$ & $6.54\pm0.73$ & $-7.58^{+1.20}_{-1.29}$ & $0.09^{+1.40}_{-0.08}$ \\ 
09dra & SN II  & 0.0766 & $16.39/17$ & $0.21^{+0.05}_{-0.04}$ & $-18.55^{+0.23}_{-0.13}$ & $-20.83^{+0.12}_{-0.04}$ & $-21.83^{+0.08}_{-0.10}$ & $0.55^{+0.18}_{-0.12}$ & $10.43^{+0.14}_{-0.27}$ & $-9.89^{+0.58}_{-0.17}$ & $5.38^{+4.75}_{-3.73}$ \\ 
\bottomrule
\end{tabular}
\end{tiny}
\end{center}
\tablecomments{The absolute magnitudes are not corrected for host reddening. The SFRs are corrected for host reddening. The abbreviation `n.o.f.' stands for number of filters. The `age' in the last column refers to the age of the stellar population. For details of the SED modeling, see Sect. \ref{method:sed}. The full table is available online in a machine-readable form. We omitted modelling the SED of PTF09bce's host because the host galaxy is severely blended with another galaxy and deblending them is impossible.}
\end{table*}

\subsubsection{Impact of Wavelength Coverage}\label{method:impact}

\begin{figure*}
\centering
\includegraphics[width=1\textwidth]{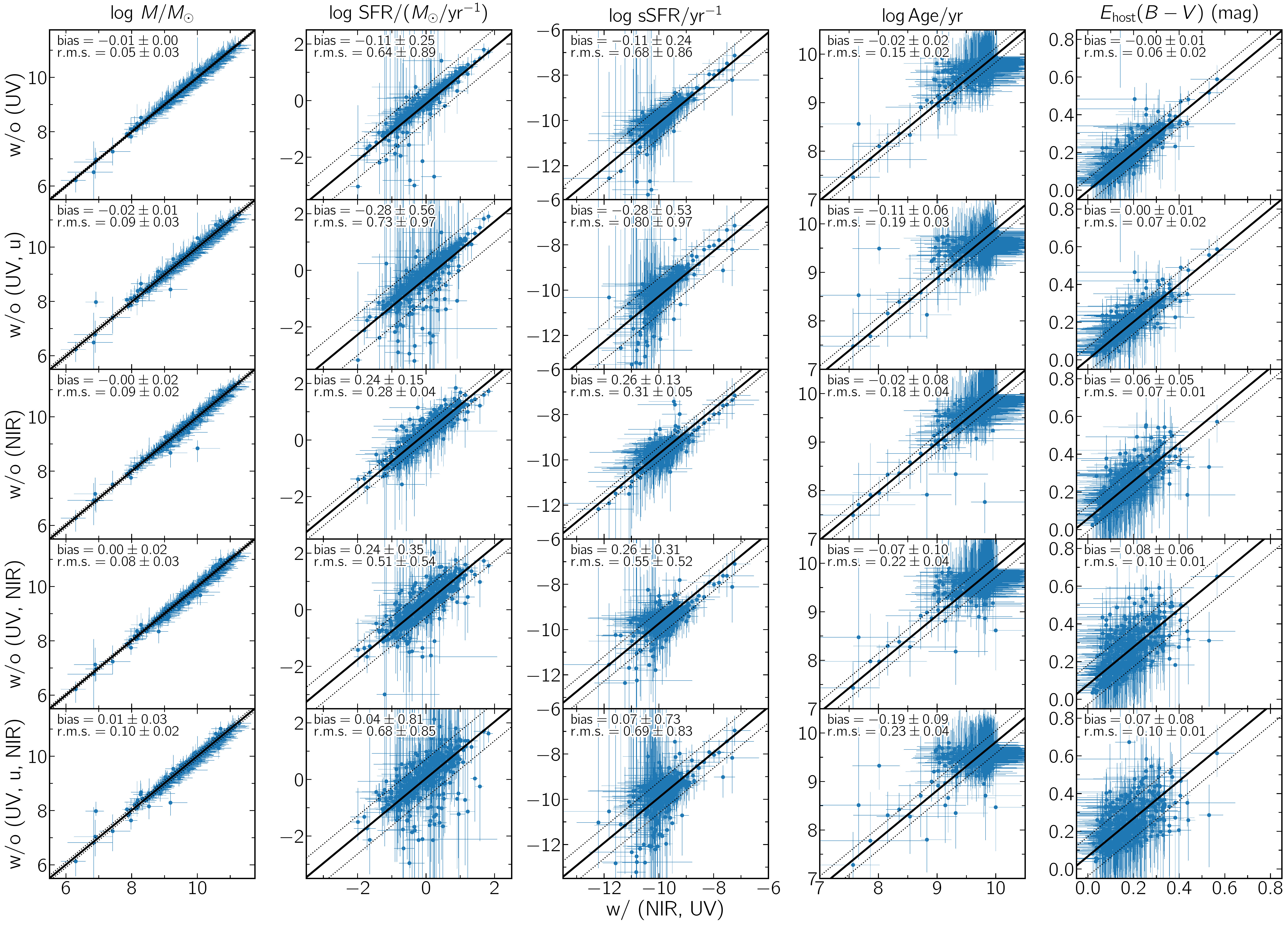}
\caption{Comparison of various host properties derived from galaxy SEDs with detections in the UV, $U$ band, optical, and NIR (all rest frame) and after removing one or more spectral regime. The galaxy mass is the most robustly measured property (negligible bias and an r.m.s. of $<0.1$~dex). In contrast to that, the other parameters are sensitive to the wavelength coverage of the SEDs.
}
\label{fig:sed_sys}
\end{figure*}

Among the 876 identified SN host galaxies, 532 hosts are detected in the rest-frame UV ($<3000$~\AA), 701 in the rest-frame near-infrared (NIR; $>10,000$~\AA), and 70 only in the optical (4000--10000~\AA).\footnote{We declare a host as detected in a given band if the measurement error is $<0.44$ mag, i.e., a $2\sigma$ measurement.} To quantify the systematic uncertainties in the galaxy mass, SFR, sSFR, ages, and attenuation measurements due to the absence of individual spectral bands, we define a subsample of 475 host galaxies with detections in the UV, in the range 3000--4000~\AA, and the NIR (all rest frame). Then, we removed one or more of these spectral bands and repeated the fit. The scatter plots in Fig.~\ref{fig:sed_sys} show how the galaxy mass, SFR, sSFR, age of the stellar population, and attenuation vary if only partial datasets are available.

The galaxy mass (first column in Fig. \ref{fig:sed_sys}) is the most robustly measured quantity and shows no dependence on the availability of NIR data. The bias levels between the measurements are negligible and the r.m.s. values are $<0.1$~dex. In comparison, the median error of the mass measurements is a factor of two larger if the full dataset is used. This confirms findings by \citet[][and references therein]{Conroy2013a}. These authors also pointed out that NIR data are only constraining if dust reddening is significant and larger than in our CCSN sample. Furthermore, galaxies with ages between 0.1 and 1 Gyr and in particular starbursts have very uncertain mass-to-light ratios owing to the difficulty in constraining their SFH. As we show in Sec. \ref{results:sfr}, $4.5\%$ of the entire PTF SN sample is found in starburst galaxies. Their galaxy mass estimates can have larger systematic uncertainties than those reported here if rest-frame NIR data are not available.

In contrast to the robustness of the mass estimates, SFR measurements (second column in Fig. \ref{fig:sed_sys}) are on average overestimated by 0.2~dex, if SEDs consist of only rest-frame UV+optical data, and underestimated by 0.3~dex, if SEDs only consist of rest-frame optical+NIR data. SFR measurements are also more uncertain, which is illustrated by their large r.m.s. of 0.5--0.7 dex and the uncertainties of the r.m.s. values if UV and/or U-band data are lacking (Fig. \ref{fig:sed_sys}). The systematic offset of the SFRs by $\sim0.2$~dex of UV+optical is in agreement with \citet{Conroy2009a}. SFRs are more challenging to measure from SEDs because of the age-dust-metallicity degeneracy and the assumed SFHs. In addition, prominent emission lines in low-mass galaxies add a source of uncertainty to measuring SFRs accurately. Consequently, the uncertainties of the sSFRs increase (third column in Fig. \ref{fig:sed_sys}).

The fourth column in Fig. \ref{fig:sed_sys} presents the dependence of the inferred ages on the wavelength coverage. The absence of wavelength regimes leads to an underestimation of the age of the stellar population. The ages derived from sole optical SEDs are skewed by 0.2 dex toward younger ages. This additional systematic error is smaller than the total error of the age measurements with full wavelength coverage. Although individual age measurements are notoriously difficult to measure accurately and precisely, there is a strong linear correlation between the ages derived from SEDs with complete and partial datasets. This means that we can compare the average ages of the SN host populations and use that to conclude whether a SN class is characterized by a particularly young or old stellar population.

The attenuation measurements (fifth column in Fig. \ref{fig:sed_sys}) are systematically overestimated if incomplete SEDs are used. Pure optical SEDs are affected most, and these attenuation measurements are overestimated by 0.08 mag. The r.m.s. is also of the same order. In contrast to the ages, the bias levels of the attenuation measurements are comparable to individual measurement errors. Furthermore, the median attenuation of SEDs with maximal wavelength coverage is 0.16 mag. This makes these measurements an unsuitable diagnostic to distinguish between different SN host populations.

\section{Results}
\subsection{Supernova Classifications}\label{result:sn_classification}
\begin{figure*}[t!]
\centering
\includegraphics[width=1\textwidth]{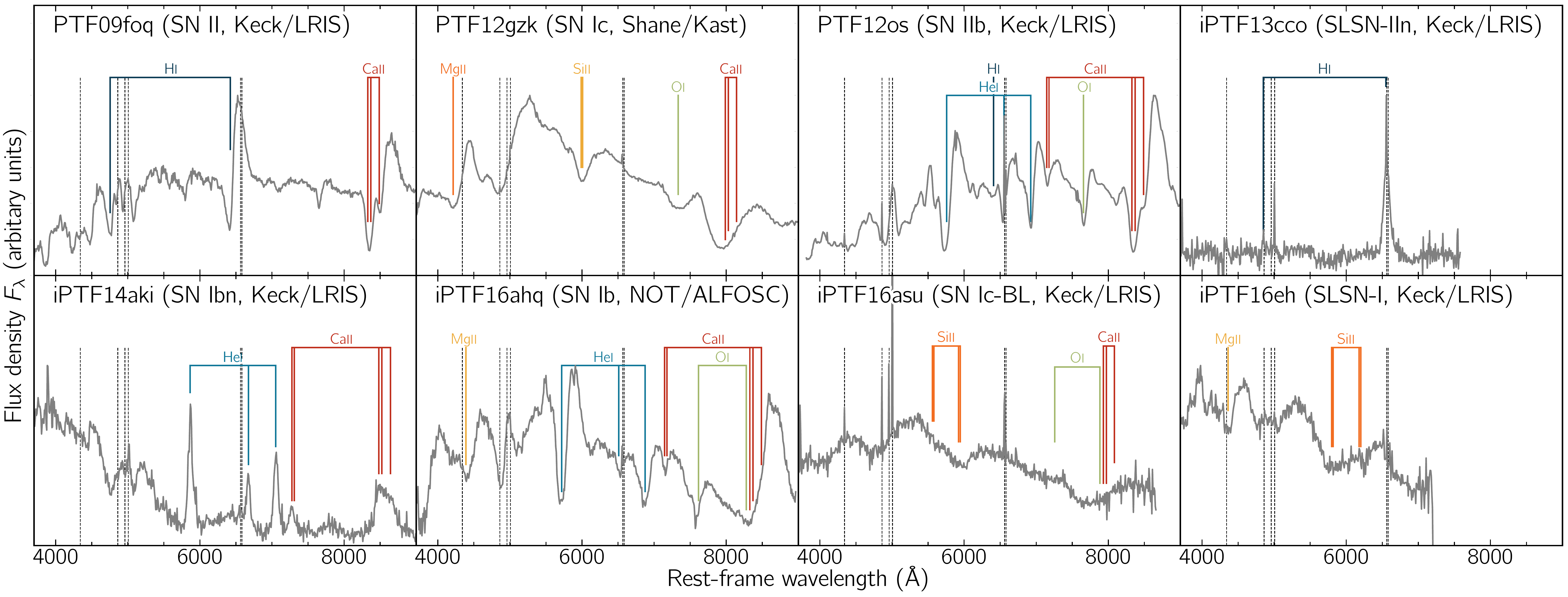}
\caption{A collage of SN classification spectra. Overlaid are absorption and emission lines that are characteristic of each class. The black vertical dashed lines indicate the locations of expected emission lines from the underlying \ion{H}{2} regions in the host galaxies. For presentation purposes, all spectra were rebinned to bin sizes of 8~\AA.
}
\label{fig:classification}
\end{figure*}

\begin{figure*}
\centering
\includegraphics[width=1\textwidth, clip=true]{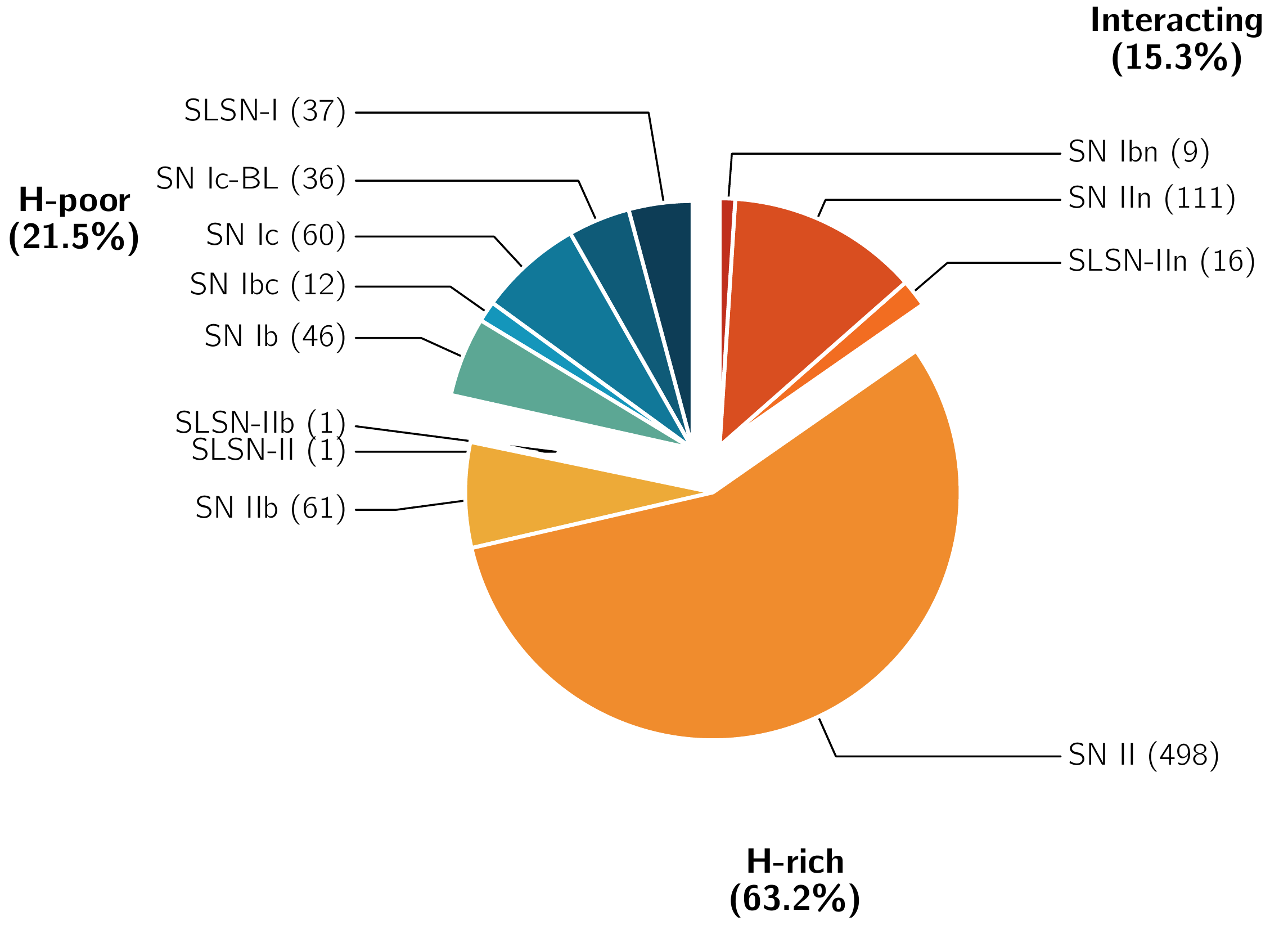}
\caption{The PTF CCSN sample consists of 888 objects divided into three families and 12 classes. The size of each subclass and family is shown in the legend.}
\label{fig:piechart}
\end{figure*}

Our input sample includes over 900 transients that were flagged as CCSNe on the PTF Marshal, a target and observation manager system. Among those, we robustly classify 888 objects using the methods described in Sec. \ref{method:classification}. We divide the sample into three SN families and 12 SN classes: H-poor SNe --- SLSN-I, SN Ic-BL, SN Ic, SN Ibc, SN Ib; H-rich SNe --- SLSN-IIb, SLSN-II, SN IIb, SN II; interacting SNe --- SLSN-IIn\footnote{We use the classifications from Leloudas et al. (in prep.) and added other luminous IIn supernovae to this class. The origin of this class and its relationship to regular Type IIn SNe is highly debated \citep{Gal-Yam2012a, Moriya2019a, GalYam2019a, Jerkstrand2020a, Nicholl2020a}.}, SN IIn, SN Ibn. Figure~\ref{fig:classification} shows examples of the supernova spectra and the characteristic features.
Table~\ref{tab:general} summarizes the classifications and properties of the 888 CCSNe.\footnote{Parallel to this paper, Hangard et al. (in prep.) classified all SNe that were flagged as a SN Ia on the PTF Marshal.}

In addition, we make use of the classifications of SLSNe reported in \citet{Yan2015a}, \citet{Perley2016a}, \citet{Yan2017a}, \citet{Quimby2018a}, \citet{deCia2018a},  \citet{Lunnan2018a} and Leloudas et al. (in prep), Ic-BL SNe by \citet{Taddia2019a} and \citet{Modjaz2019a}, Ibc SNe by \citet{Fremling2018a} and
Barbarino et al. (submitted), IIn SNe by \citet{Nyholm2019a}, CCSNe in general by \citet{Arcavi2010a}, and tidal disruption events (TDEs) by \citet{Arcavi2014a}. In most cases, our classifications are identical to those reported in the papers mentioned above. For a few objects, we prefer a different classification. Information about those objects is provided in Appendix~\ref{app:classification}.

Figure~\ref{fig:piechart} displays the break-down of our sample. It comprises of $\sim63\%$ H-rich and $\sim22\%$ H-poor SNe. The remaining $\sim15\%$  exhibit signatures of strong interaction between the SN ejecta and circumstellar matter. The largest individual SN class in the PTF sample is the class of Type II SNe with 56\% due to their high volumetric rate \citep{Li2011a} and their long-lasting plateaus, which are less demanding for the spectroscopic follow-up. On the other extreme, the Type Ibn SN sample contains only nine objects. Their light curves reach maxima within $\lesssim10$ days and afterward decline by $\sim0.1$~mag/day \citep{Hosseinzadeh2017a}. Therefore, in order to classify them, spectroscopic observations within a few days after discovery are essential.

\subsection{Host Recovery Rate}\label{result:host_id}

\begin{figure}
\centering
\includegraphics[width=0.49\columnwidth]{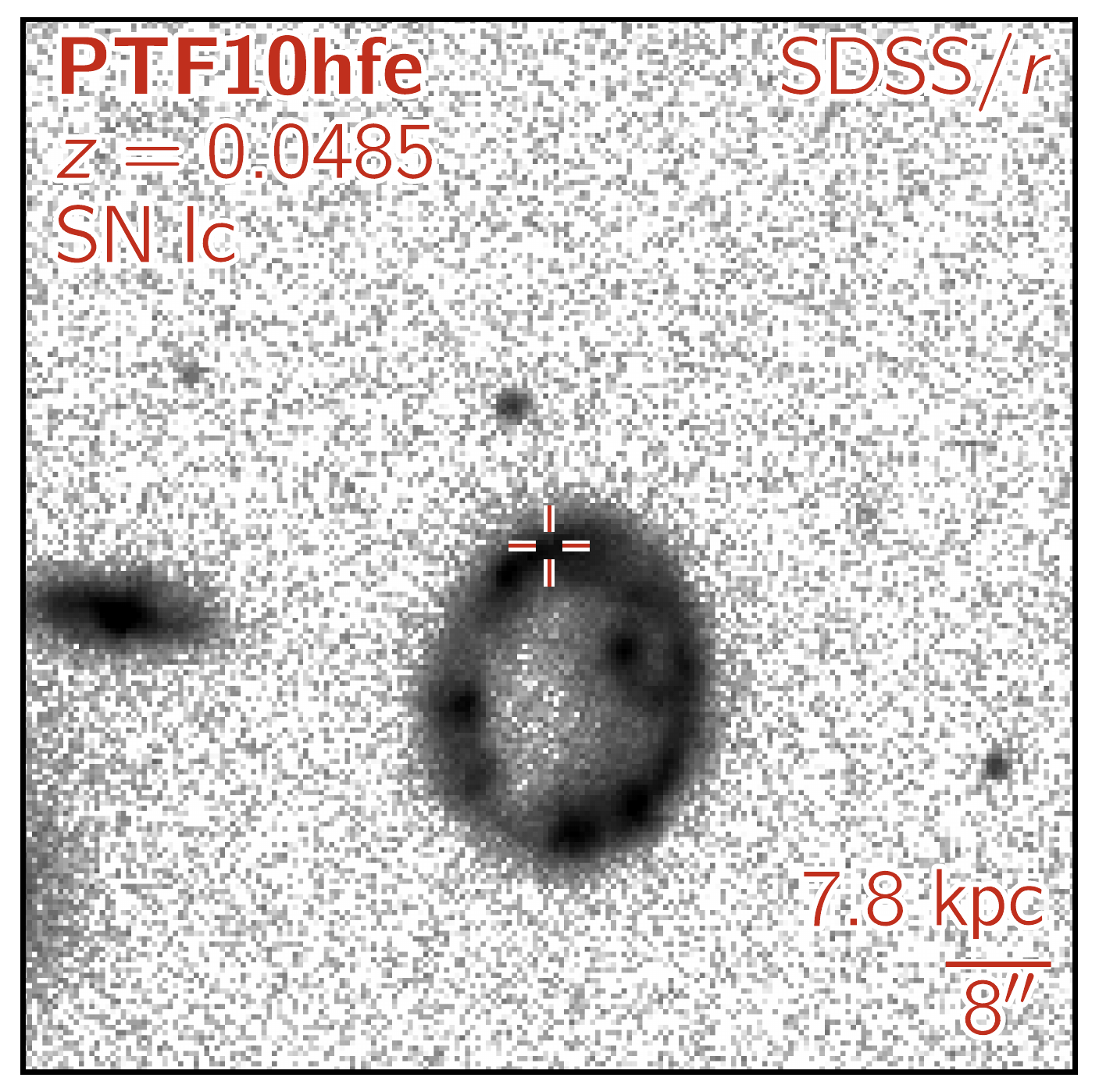}
\includegraphics[width=0.49\columnwidth]{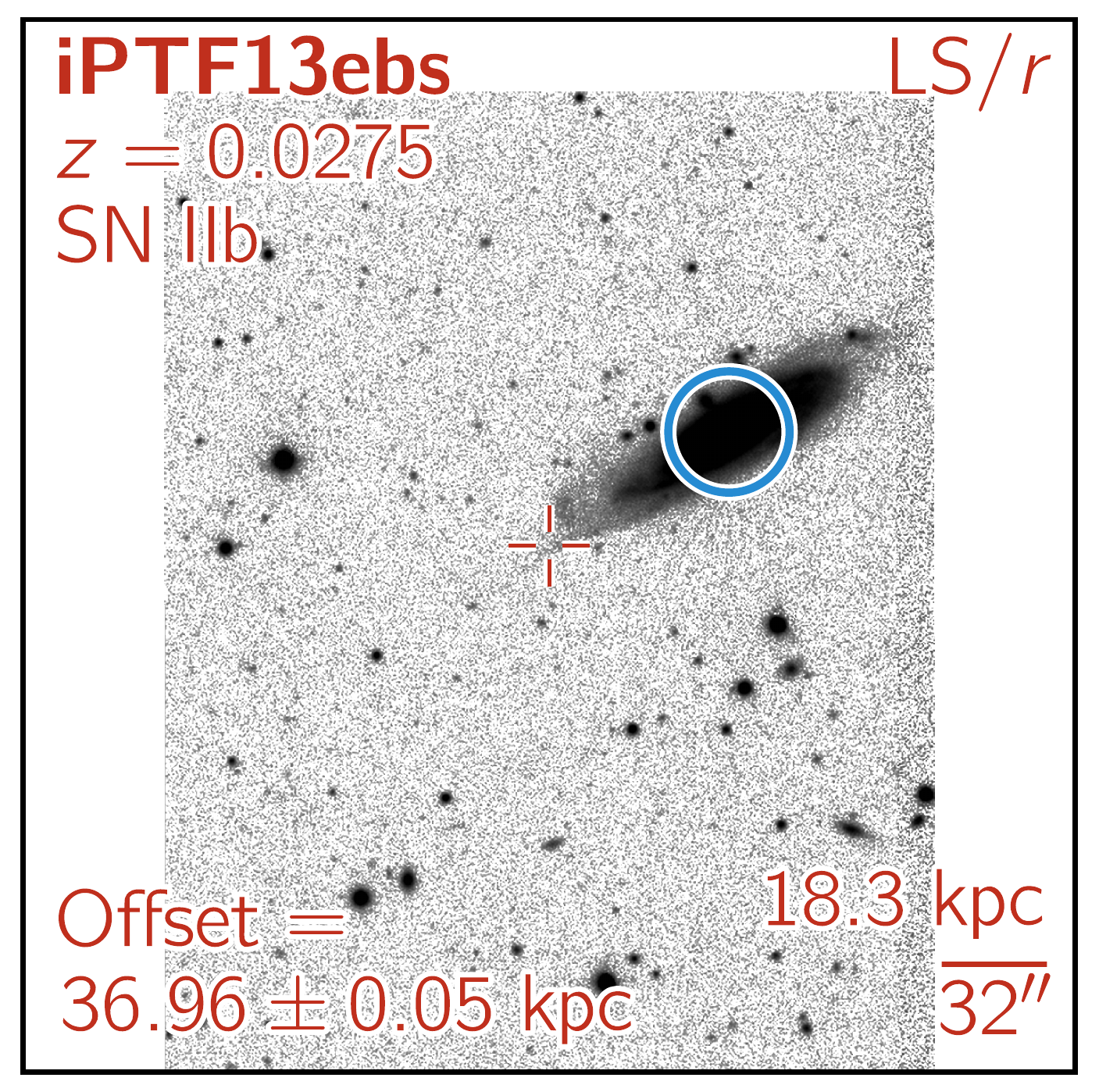}
\includegraphics[width=0.49\columnwidth]{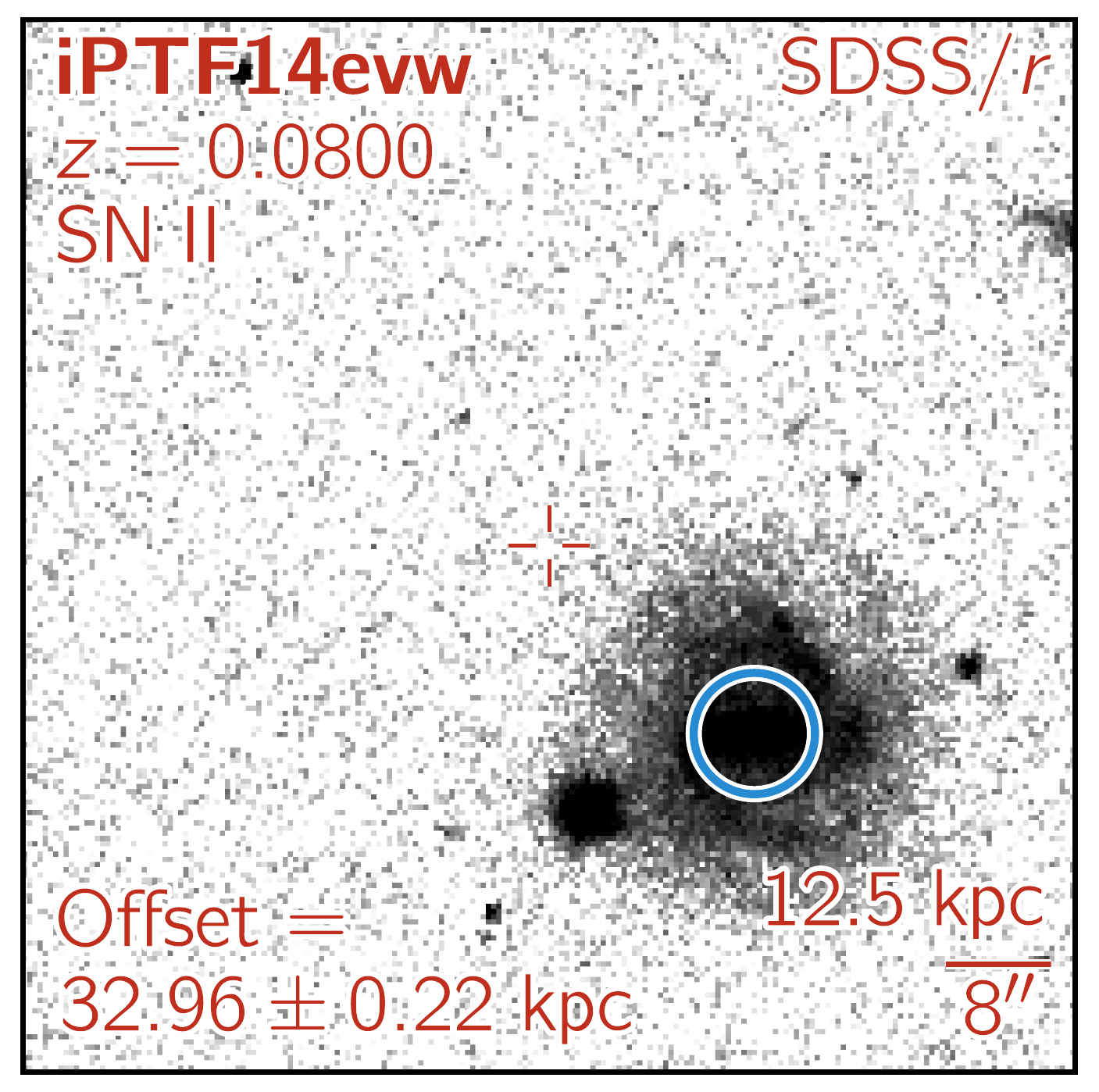}
\includegraphics[width=0.49\columnwidth]{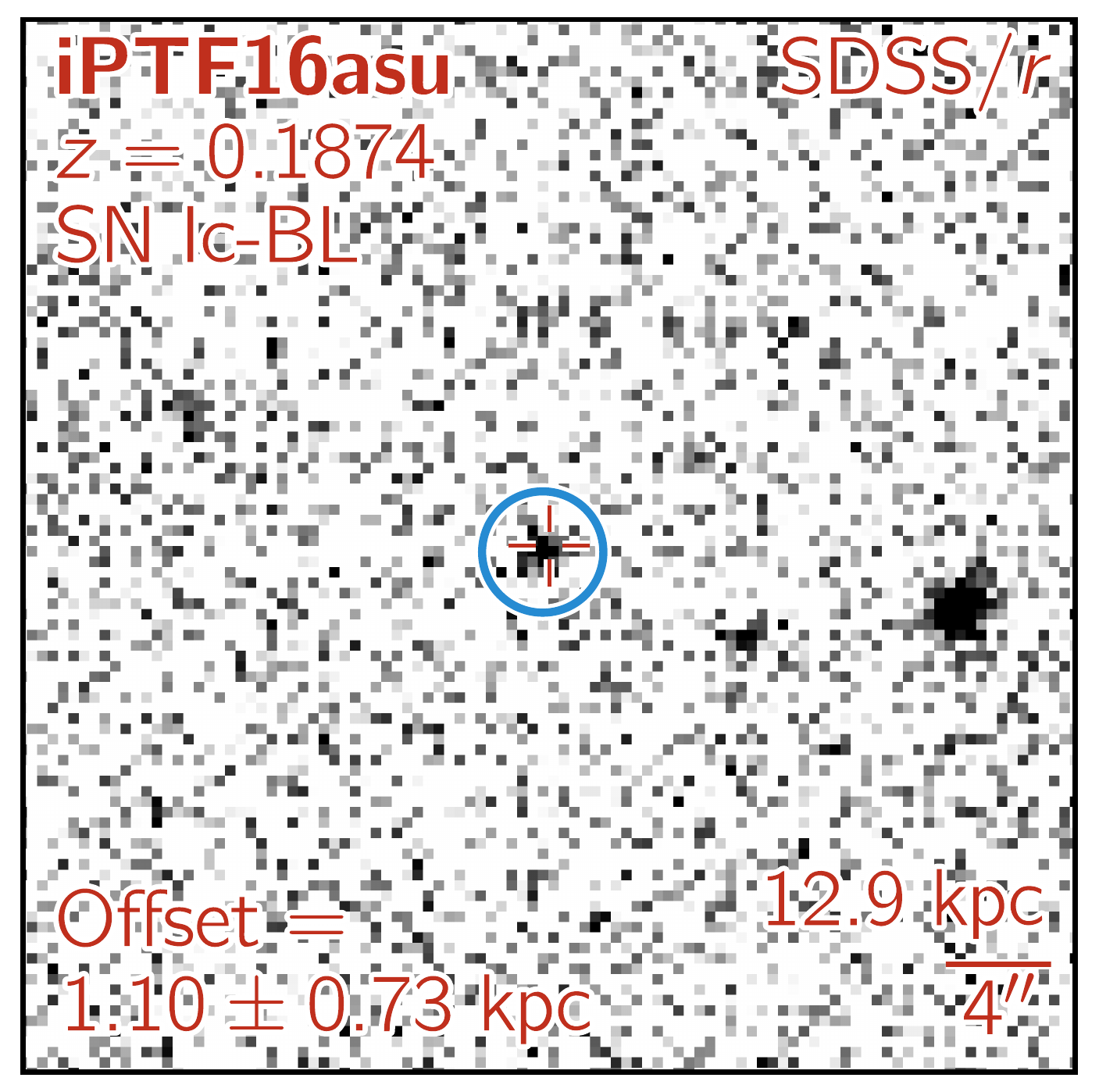}
\caption{
Selection of postage stamps of PTF CCSN host galaxies. The size of the cutouts varies between $40''$ and $320''$. In each panel, North is up and East is left. The crosshair marks the SN position, and the blue circle (arbitrary radius) marks the center of the host galaxy. The projected distance (in kpc) between a SN and the center of its host galaxy is reported in the lower-left corner. A scale is shown in the lower right corner. We report no distance between PTF10hfe and its host because the host galaxy has a ring morphology without a center.
}
\label{fig:postage_stamps}
\end{figure}

\begin{figure*}
\centering
\includegraphics[width=1\textwidth]{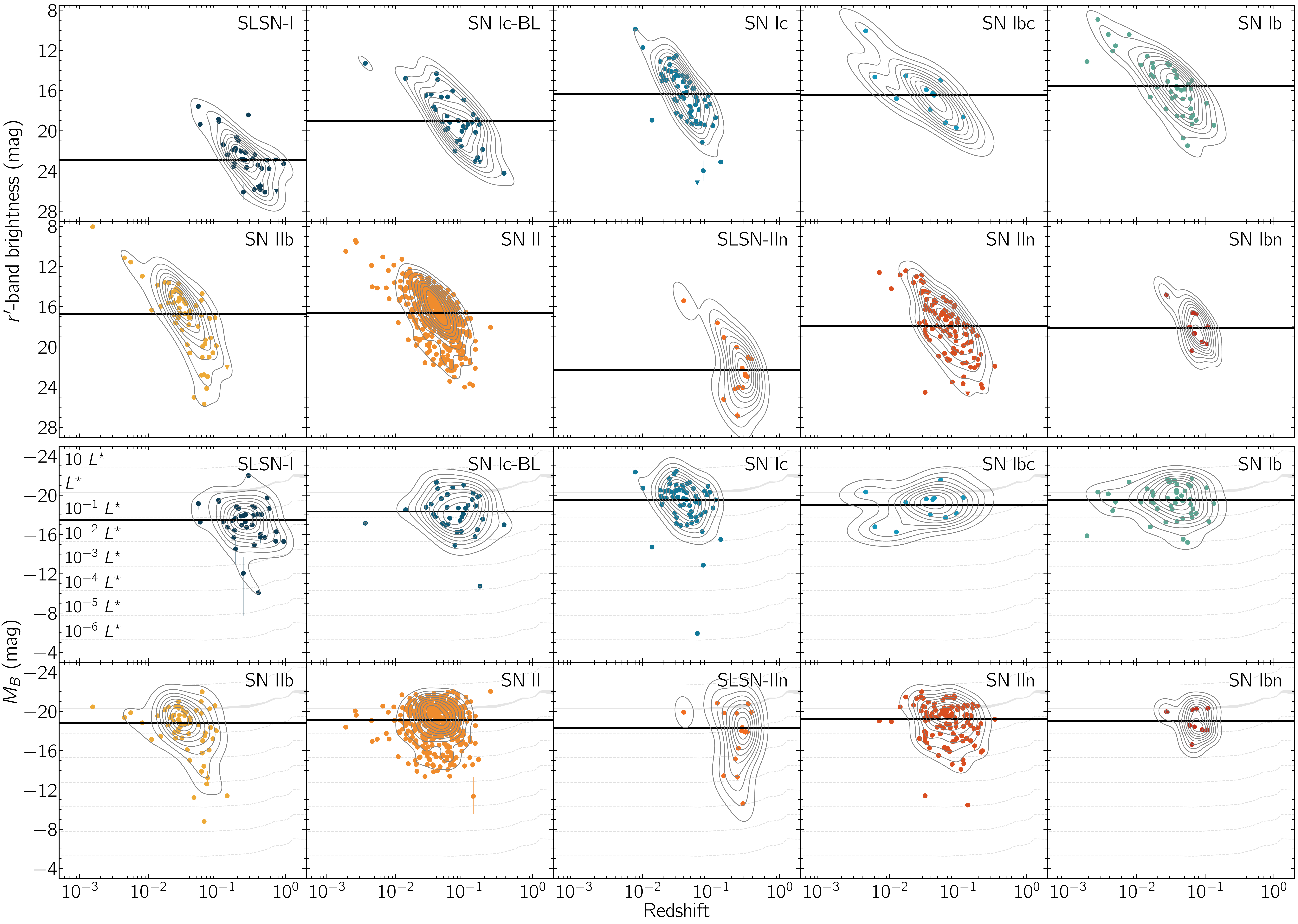}
\caption{The evolution of the apparent $r'$ magnitude (top) and absolute $B$ magnitude (bottom) of the host-galaxy populations as a function of redshift and SN type. The average brightness (indicated by the black horizontal line) varies significantly with each SN type. These dissimilarities are primarily due to differences in the redshift distributions. After accounting for redshift, SLSNe-I are found in galaxies that are still a factor of 4.5 less luminous than SNe II. To guide the eye, contours from 10\% to 90\% are overlaid in both figures. The characteristic luminosity, $L^\star$, of the $B$-band luminosity function of star-forming galaxies presented in \citet{Faber2007a} and multiplies of it are displayed in gray.
}
\label{fig:mag_r}
\end{figure*}

We detect the host galaxies of almost all CCSNe.\footnote{We declare a host as detected if the measurement error in any image is better than 0.44 mag, i.e., a $2\sigma$ measurement.} Only ten hosts evaded detection; this set includes four SLSNe-I and one candidate SLSN-IIn, two SNe IIb, and one Type Ic-BL, Ic and IIn SN each.\footnote{SLSNe-I: iPTF13bdl, 14tb, 15ii, 16bt; candidate SLSN-IIn: PTF10eoo; SNe IIb: PTF09dsj, 13dzy; SN Ic-BL: PTF12grr; SN Ic: iPTF14gqr; SN IIn: PTF10weh.} Three of the five SLSNe were found between redshift $z=0.4$ and $z\approx1$ and are expectably beyond the reach of the Legacy Surveys
\citep[Fig.~\ref{fig:mag_r}; Table~\ref{tab:surveys};][]{Lunnan2014a, Perley2016a, Schulze2018a, Angus2019a}. The fields of the five regular CCSNe were also covered by the Legacy Surveys, except for PTF12grr. Their non-detections in data from the Legacy Surveys, PanSTARRS, and SDSS imply luminosities of $M_R\gtrsim-13.6$~mag. As we show in Sec. \ref{result:host_properties}, these host galaxies are among the faintest galaxies in our sample.

Figure~\ref{fig:postage_stamps} shows postage stamps of a selection of fields in our sample.  The blue circles mark the host galaxies' centers, and red crosshairs indicate the SN positions. The average total alignment error is 0\farcs16 and only exceeds 0\farcs5 in six cases. The coordinates of the SNe, host galaxies, and their projected distances to each other are summarized in Table~\ref{tab:general}.

We remark that the host identification of two SNe has some ambiguity. iPTF14gqr exploded in the outskirts of tidally interacting galaxies. As proposed in \citet{De2018a}, the tidal interaction could trigger star-formation in collisional debris. We assume that the progenitor of iPTF14gqr was formed in such a debris.

PTF12mja is located between two galaxies of a compact galaxy group \citep{McConnachie2009a}. Spectroscopic information would be required to obtain the redshifts of the two galaxies and to identify the actual host.
The host PTF09bce is severely blended by another galaxy so that measuring the host flux alone is not possible. Owing to that, we do not include PTF09bce and 12mja in the discussion of the host properties.

\subsection{Redshift and Distance Measurements}\label{result:redshifts}

About 45\% of the PTF CCSN host galaxies have redshift information listed in the SDSS catalog or the NASA/IPAC Extragalactic Database\footnote{\href{http://ned.ipac.caltech.edu}{http://ned.ipac.caltech.edu}} \citep[NED;][]{Helou1991a}. Spectra of 333 additional SNe ($\sim38\%$) show absorption or emission lines from their host galaxies.  The remaining 18\% (158 SNe) have no redshift information. For those, we use the redshifts inferred by SN-template matching. These redshifts are typically accurate to within a few  hundredths in redshift space \citep{Blondin2007a, Fremling2019a}. Table~\ref{tab:general} summarizes all redshifts and how they are obtained.

\begin{figure*}
\centering
\includegraphics[width=1\textwidth]{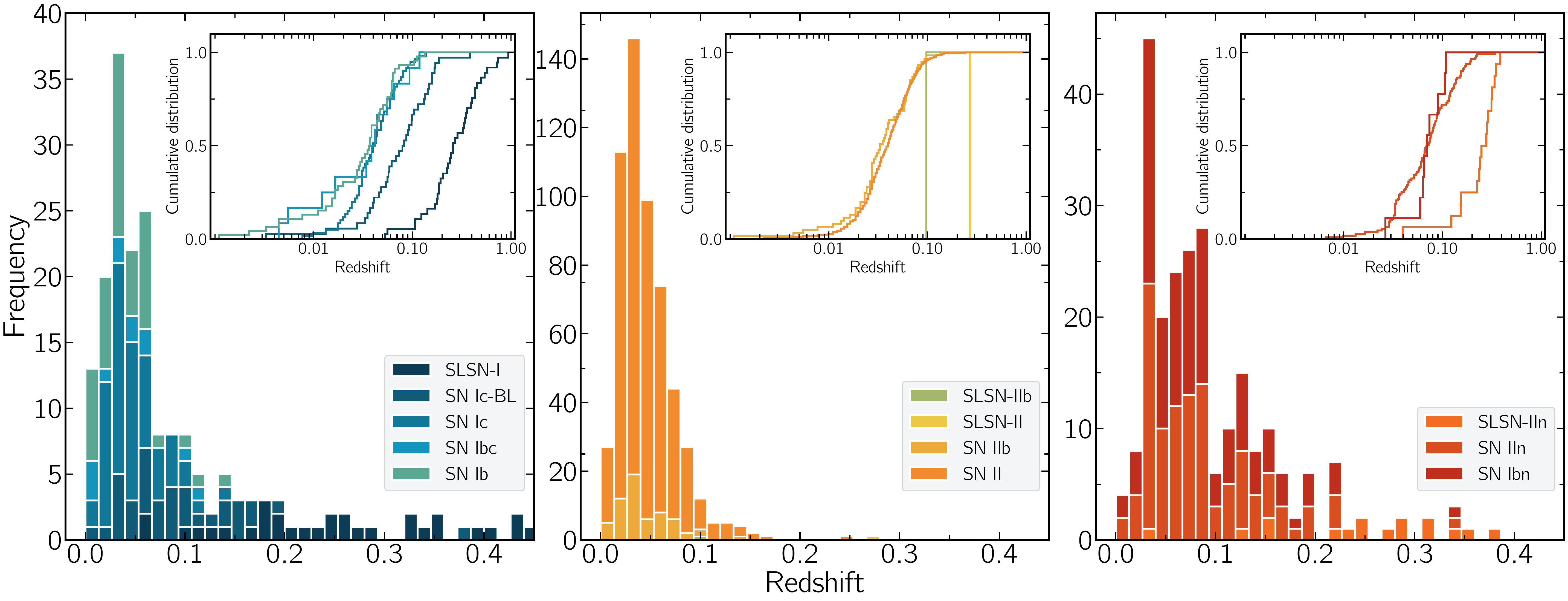}
\caption{The redshift distribution of H-poor (left), H-rich (center), and interaction-powered (right) SNe. The stacked histograms are truncated at $z=0.5$ for presentation purposes, whereas the full distributions are shown as cumulative plots in the insets. The highest redshift SN in the PTF sample is the H-poor SLSN iPTF14tb at $z=0.942$.
}
\label{fig:redshift_distribution}
\end{figure*}

Figure~\ref{fig:redshift_distribution} displays the redshift distribution of each SN class (see also Table~\ref{tab:host_stat} for their median values). The median redshifts reflect the average luminosity of each class. Type Ibc, II, and IIb SNe have the lowest peak luminosities and are found only at low redshifts (median $z\approx0.04$; Table~\ref{tab:host_stat}). In contrast, SLSNe are detected at a median $z\approx0.26$ (Table~\ref{tab:host_stat}). The most distant CCSN in our sample iPTF14tb is, in fact, a H-poor SLSN at $z=0.942$ (Table~\ref{tab:general}). The measured redshift distributions are confirm with previous work by \citet{Perley2019a}, \citet{Nyholm2019a} and \citet{Modjaz2019a}
that are based on subsets of the PTF SLSN, SN IIn and SN Ic-BL host galaxy samples.

At very low redshift, peculiar velocities of galaxies can be significant and can hinder using redshifts as distance measurements \citep[e.g.,][]{Davis2011a}. To quantify the impact of the issue, we compare the distance moduli inferred from the Tully-Fisher (TF) relation and $\Lambda$CDM cosmology without correction for peculiar motion. The NED database has a record of the distance moduli from the TF relation for 24 of 27 hosts at $z<0.01$.\footnote{We limit the comparison to $z<0.01$ because the completeness level of hosts with TF distance moduli plunges from $\sim89\%$ to 30\% (23/76 hosts) as we go from $z<0.01$ to $0.01\leq z < 0.02$.} The distance moduli from $\Lambda$CDM cosmology and the TF relation differ by $0.18^{+0.71}_{-0.68}$~mag on average. This is smaller than the accuracy of the TF relation \citep[0.3--0.4~mag;][]{Freedman2010a}. Given the general consistency between the distances inferred from $\Lambda$CDM cosmology and from the TF relation, we assume that all hosts are in the Hubble flow and that redshifts are a reliable distance measurement for all objects in this paper.\footnote{The differences in the distance moduli of the hosts of iPTF14jku, 14ur, 14va, 15eqv, 16hgm, and 16tu differ by 1 to 2~mag. The TF distance moduli of iPTF15eqv and 16tu can be reconciled with their large measurement uncertainties. Understanding whether the differences of the other objects are significant requires knowledge of how the TF distances were obtained. That is beyond the scope of this paper.}

\begin{figure*}
\centering
\includegraphics[width=1\textwidth]{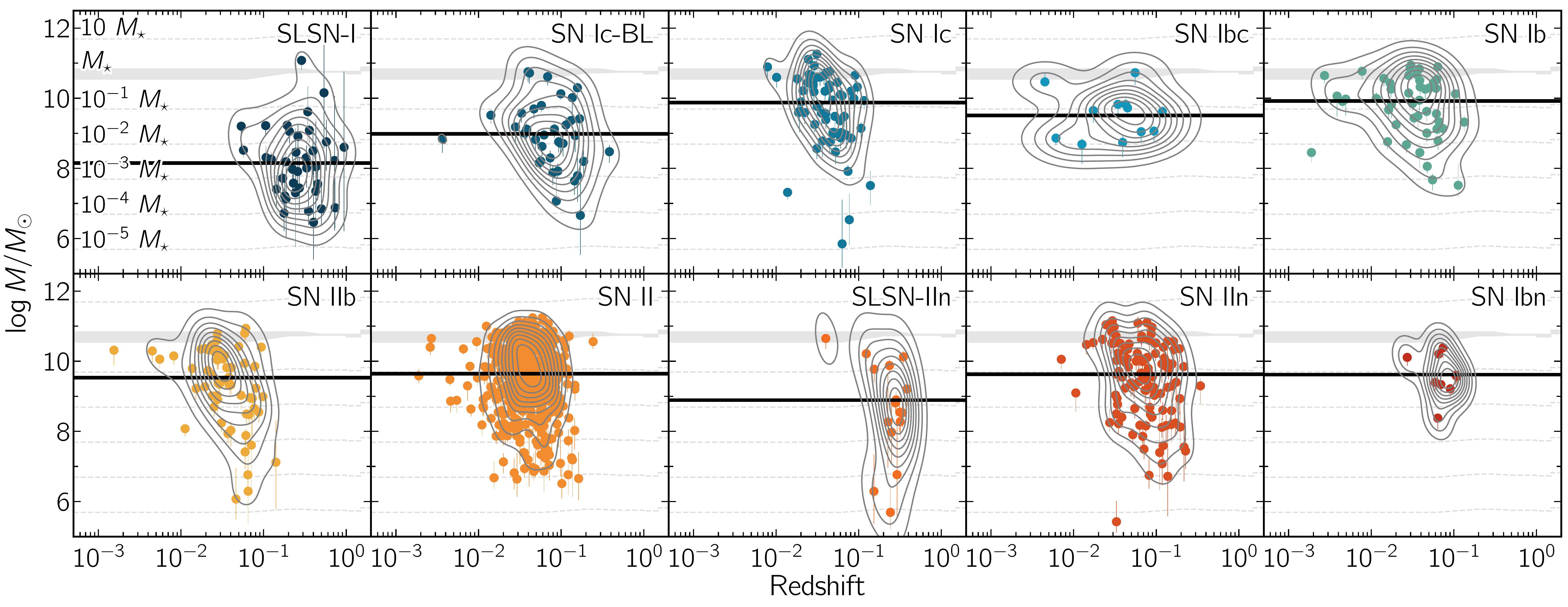}
\caption{Host galaxy mass versus redshift. We overlay the characteristic mass $M_\star$ of the mass function from the GAMA \citep{Baldry2012a} and UltraVISTA \citep{Muzzin2013a} surveys in grey, and several mass tracks. To guide the eye contours from 10\% to 90\% are overlaid to the data. The thick horizontal black line displays the median value of each sample. Note the difference of a factor of 5--25 between the masses of SLSNe and SNe-Ib/c/II/IIb hosts.
}
\label{fig:mass_vs_redshift}
\end{figure*}

\subsection{Brightness and Luminosity}\label{result:host_properties}

The top panel of Fig.~\ref{fig:mag_r} shows the distribution of the hosts' observed $r'$-band brightness as a function of redshift and SN class. The distributions cover a wide range from $r'\approx8$~mag to $r'\approx26$~mag. Clear differences between the different SN classes are visible. Type Ib, Ic, IIb and IIb SNe are found in galaxies with a median brightness of $r'\approx16.6$~mag (horizontal line in the top panel of Fig.~\ref{fig:mag_r}; Table~\ref{tab:host_stat}), whereas Type IIn/Ibn SNe are found in galaxies that are on average 1.5 mag fainter and SLSNe in galaxies that are even $\sim6.3$~mag fainter (Table~\ref{tab:host_stat}).

A significant part of the differences can be attributed to the redshift intervals probed by the different SN populations. To corroborate that, we display the $B$-band luminosities in the bottom panels of Fig.~\ref{fig:mag_r}. Similar to the top panels, the horizontal lines display the median values of the probability distribution functions reported in Sec. \ref{sec:kde} and summarized in Table~\ref{tab:host_stat}. The k-corrected absolute magnitudes of SNe IIn/Ibn hosts are almost identical to the hosts of Type Ibc, IIb and II SNe.
SLSNe and SNe Ic-BL exhibit a preference for low-luminosity galaxies, even after accounting for the evolution of the $B$-band luminosity function in agreement with \citet{Lunnan2014a}, \citet{Leloudas2015a}, \citet{Perley2016a}, \citet{Chen2017a}, \citet{Schulze2018a} and \citet{Modjaz2019a}. Their median luminosities are between $M_B=-17.6$ and $-18.3$~mag and a factor of $\sim2.4$--5 lower than those of regular CCSN host galaxies (Table~\ref{tab:host_stat}). The median luminosity of H-poor SLSN host galaxies is comparable to the LMC at $z\approx0$ \citep{deVaucouleurs1960a,McConnachie2012a}.

The luminosity distribution of detected hosts extends from $M_B\approx-11$ to $-23$~mag. This interval covers the range from $10^{-4}\,L_\star$ to $10\,L_\star$, where $\,L_\star$ is the knee of the $B$-band luminosity functions reported in \citet{Ilbert2005a}. The least-luminous detected hosts in our sample are of PTF09gyp (IIb, $M_B=-11.2^{+0.6}_{-0.4}$~mag; Table~\ref{tab:host_prospector}) and iPTF14ajx (IIn, $M_B=-11.4^{+0.3}_{-0.2}$~mag; Table~\ref{tab:host_prospector}). Including the SN hosts that evaded detection extends the distribution to galaxies fainter than $-11$~mag. This regime is comparable to the least-luminous star-forming galaxies in the Local Group \citep{Mateo1998a,McConnachie2012a}.

\subsection{Galaxy Masses}\label{result:mass}

Spectral energy distribution modeling gives access to the physical properties of the host galaxies. The primary properties we are interested in are the galaxy mass of the stellar component and the star-formation rate. These measurements are summarized in Table~\ref{tab:host_prospector}.

Figure~\ref{fig:mass_vs_redshift} shows the host masses as a function of redshift and SN class. The entire sample spans a range from $10^{5.4}$ to $10^{11.3}~M_\odot$ (Table~\ref{tab:host_prospector}), corresponding to $10^{-5}$ to 10~$M_\star$, where $M_\star$ is the knee of Schechter-type galaxy mass functions \citep[e.g.,][]{Baldry2012a, Muzzin2013a}. About 29\% and 11\% of the sample are found in galaxies that are less massive than $10^9~M_\odot$ and the $10^8~M_\odot$\footnote{The masses of the SMC and the LMC are $10^{8.7}~M_\odot$ and $10^{9.2}~M_\odot$, respectively \citep{McConnachie2012a}.}, respectively. These values are consistent with \citet{Taggart2019a} who studied a more unbiased but significantly smaller SN sample. However, not all SN classes probe this parameter space in the same way. On average, H-poor and H-rich SLSNe and Ic-BL SNe have the least-massive galaxy populations. Their hosts are a factor of 5 to 25 less massive than the hosts of any other CCSN class (Table~\ref{tab:host_stat}). Moreover, these samples exhibit a clear dearth of galaxies above $10^{10}~M_\odot$ (Fig. \ref{fig:mass_vs_redshift}).

To examine these results further, we compare the median mass of a general population of star-forming galaxies to each SN sample. The CANDELS \citep{Grogin2011a, Koekemoer2011a} and COSMOS \citep{Scoville2007a} surveys are the deepest galaxy surveys probing a sufficiently large cosmic volume and have a high level of completeness down to $10^8~M_\odot$ at $z<0.5$. This is still 2--3 orders of magnitude larger than the least massive host in our sample. However, mass functions at $z<0.5$ are well constrained and show no signs of plummeting at the low-mass end. We assume that the mass function parameters are also valid for the range spanned by our host galaxies.

Under the working assumption that massive stars are the progenitors of all CCSN classes, we expect that their stellar mass functions should sample the mass function of star-forming galaxies weighted by their SFR. Using the mass-function parameterization from \citet[][]{Tomczak2014a} and the parameterization of the fundamental correlation between the galaxy stellar mass and SFR from \citet{Lee2015a}, we estimate the SFR-weighted average galaxy mass in the mass interval probed by each of the SN samples is between $10^{9.5}~M_\odot$ and $10^{9.9}~M_\odot$.

These values match well the median host masses of most CCSN classes. Only the hosts of H-poor and H-rich SLSNe and Type Ic-BL SNe show a dearth of massive hosts compared to this model.  Their median masses are $\log M/M_\odot \approx 8.15$ (SLSNe-I) and $\sim8.9$ (SNe Ic-BL and SLSN-IIn;  Table~\ref{tab:host_stat}), but the expected SFR-weighted average galaxy mass would be $10^{9.5}~M_\odot$. The dearth of massive hosts is consistent with results reported in \citet{Perley2016a}, \citet{Chen2017a}, \citet{Schulze2018a} and \citet{Modjaz2019a}. In Sec. \ref{disc:Z_bias} we discuss this further.

Recently, \citet{Modjaz2019a} studied a subsample of Type Ic and Ic-BL SNe and their from the PTF survey. The median mass of their SN Ic-BL sample is consistent with that of the entire PTF sample. However, the hosts of their SN~Ic sample have masses that are 0.5 dex larger.

\begin{figure*}[t!]
\centering
\includegraphics[width=1\textwidth]{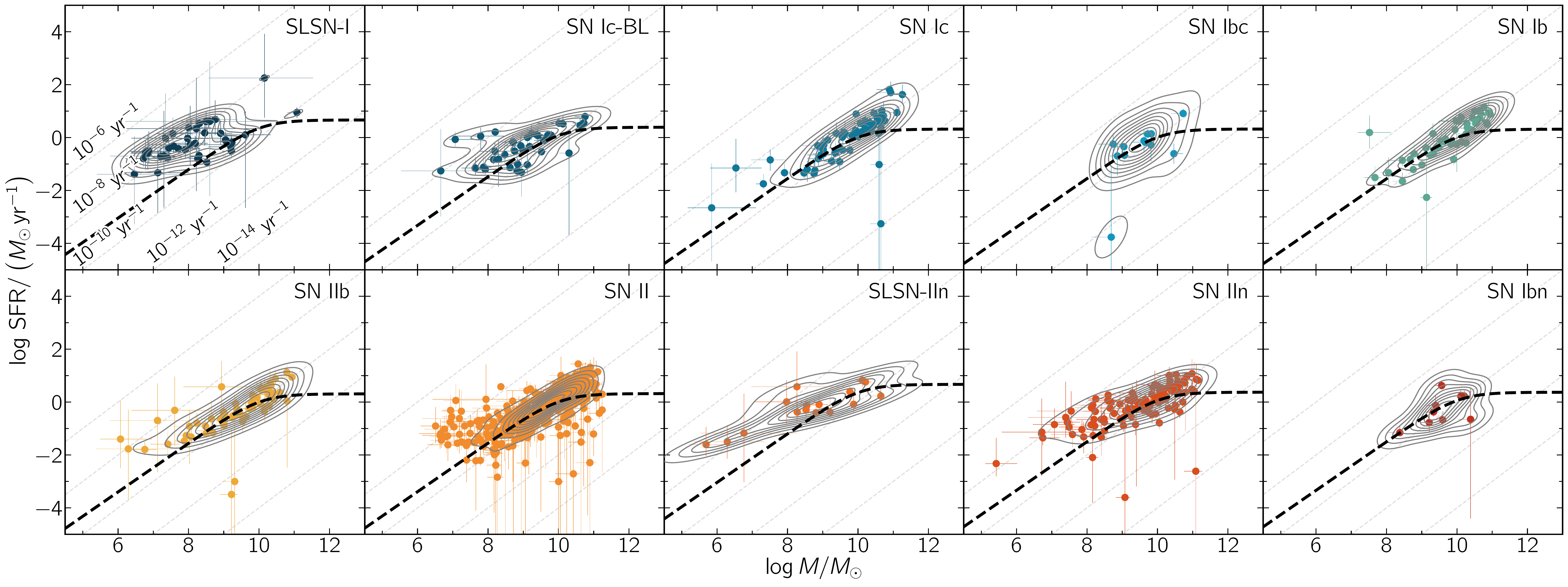}
\caption{The host populations in the mass-SFR plane. All SNe exploded in star-forming galaxies. This is illustrated by their location with respect to the main sequence of star-forming galaxies (black-dashed curve). The grey-dotted diagonal lines display lines of constant specific star-formation rate (SFR normalized by stellar mass). To guide the eye, contours from 10\% to 90\% are overlaid.
}
\label{fig:sfr_mass}
\end{figure*}

\subsection{Star-Formation Rates}\label{results:sfr}

To put the mass measurements in the context of star-forming galaxies, we present in Fig.~\ref{fig:sfr_mass} the galaxy mass as a function of star-formation rate and SN type. The vast majority of hosts occupy a narrow region of the SFR-$M_\star$ parameter space between a specific star-formation rate (sSFR = SFR/$M_\star$) from $10^{-10}$ to $10^{-9}$~yr$^{-1}$. That part of the parameter space is also known as the locus of the galaxy main-sequence of star-forming galaxies (thick dashed line in Fig.~\ref{fig:sfr_mass}), indicating that most SN host galaxies are normal star-forming galaxies. SLSNe are found in more vigorously star-forming galaxies with average sSFR's of $\sim10^{-8.3}~{\rm yr}^{-1}$, consistent with the values reported in \citep{Perley2016a, Schulze2018a, Taggart2019a}. This places them in the regime of starbursting galaxies \citep{Daddi2007a, Elbaz2007a, Noeske2007a, Sargent2012a}.

Starbursts are not exclusive to SLSN hosts. The most vigerously star-forming host galaxies of regular CCSN reach sSFR's of $>10^{-8}\,{\rm yr}^{-1}$ (e.g., iPTF15eqq -- SNII, $>10^{-7}\,{\rm yr}^{-1}$, PTF12eci -- SN Ic-BL, $>10^{-7.1}\,{\rm yr}^{-1}$, PTF09cvi -- SN Ic-BL, $>10^{-7.2}\,{\rm yr}^{-1}$; Table \ref{tab:host_prospector}). More quantitatively, $\sim11\%$ of all regular CCSNe are found in galaxies with sSFR's of $>10^{-9}\,{\rm yr}^{-1}$ (just above the galaxy main sequence) and 3\% are found in extreme starbursts with sSFRs of $>10^{-8}\,{\rm yr}^{-1}$. The frequency of extreme starburst galaxies hosting regular CCSNe is consistent with \citet{Taggart2019a}. In stark contrast to the low starburst fraction of regular CCSNe, $68^{+11}_{-8}\%$ and $38^{+11}_{-8}\%$ of all H-poor SLSNe exploded in galaxies with sSFR's of  $>10^{-9}\,{\rm yr}^{-1}$ and $>10^{-8}\,{\rm yr}^{-1}$, respectively.

On the other extreme of the SFR spectrum, there is also a population of hosts of regular CCSNe with SFR's of $\lesssim0.01~M_\odot\,{\rm yr}^{-1}$ and sSFR's between $10^{-14}~{\rm yr}^{-1}$ and $10^{-11}~{\rm yr}^{-1}$, albeit with very large uncertainties. Whether these galaxies are barely star-forming or, maybe, not star-forming at all requires additional data, such as far-infrared data to assess dust-obscured star-formation and integral-field spectroscopy to search for star-forming regions at SN sites.

\subsection{Projected Distances between SNe and Hosts}

The projected distances (summarised in Table~\ref{tab:general}) between the SNe and the centers of host galaxy centers extends from $0.02\pm0.20$~kpc (PTF09awk, Type II; Fig. \ref{fig:postage_stamps}) to $37.00\pm0.05$~kpc (iPTF13ebs; Type IIb)\footnote{In addition to PTF09bce PTF12mja and the ten ``hostless" SNe, we removed two additional hosts in the host-offset analysis.  The host of PTF10hfe is a ring galaxy without a center (Fig.~\ref{fig:postage_stamps}). PTF11aun exploded in a dwarf galaxy; however, the host is severely blended with a foreground star, and the galaxy center can not be reliably measured.}$^{,\thinspace}$\footnote{The offset measurements of PTF10cd and 17bsi are based on the SN coordinates on the PTF Marshal due to absence of publicly available SN images.}. At the same time, the masses of the host galaxies vary from $10^5$ to $10^{11.5}~M_\odot$, and therefore the sizes of the host galaxy are expected to vary considerably.

\begin{figure*}
\centering
\includegraphics[width=1\textwidth]{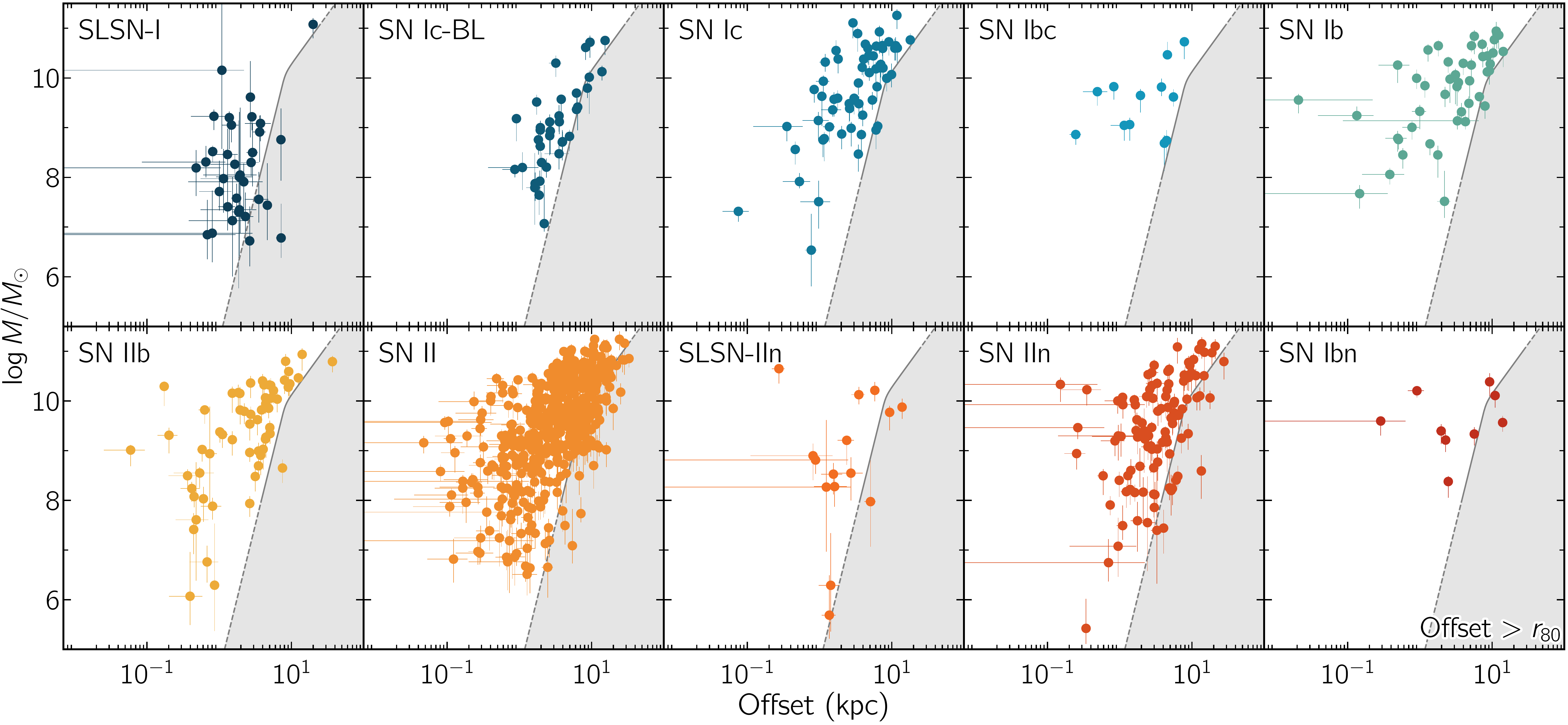}
\caption{Galaxy mass versus supernova-host offset. Most SNe are located within the 80\% light radius of their host galaxies. Up to $15\%$ of each SN class are found at larger galacto-centric distances. This is expected because the offsets are compared to the 80\% light curves.}
\label{fig:offset}
\end{figure*}

To better understand this behavior and whether certain SN classes are found in peculiar host locations, we present the offsets as a function of host-galaxy mass and SN type in Fig.~\ref{fig:offset}. We also overlay the relation between the 80\% light-radius, $r_{80}$, and galaxy mass which is thought to trace the stellar mass content of galaxies independent of whether they are star-forming or not \citep{Miller2019a, Mowla2019a}. The overwhelming majority of CCSNe is found within $r_{80}$. Less than 15\% of each SN class are found at distances larger than $r_{80}$. After propagating uncertainties, the fraction of SNe at large distances decreases to less than 8\% ($3\sigma$ confidence level). Either way, this percentage of SNe with large offsets is expected because we compare their distance to the 80\% light radius. This result reassures that we reliably identified the host galaxies for most CCSNe in our sample.

\section{Discussion}
\subsection {Distribution Functions}\label{sec:kde}

In the previous sections, we focused on a general description of the host galaxy population. In this section, we construct and examine the distribution functions.
The observed distribution functions suffer from the combined effect of the measurement errors and selection bias. Taking selection effects into account requires detailed knowledge of how the survey was operated, how transients were identified, how objects were selected for classification, and how successful were these classification efforts  \citep[e.g.,][]{Frohmaier2017a, Feindt2019a}. This is beyond the scope of this paper. To account for measurement errors, we perform a Monte-Carlo simulation (30000 samples per host property and SN class) and bootstrap each sample. In the Monte-Carlo simulation, each data point is re-sampled as follows:
\begin{itemize}[leftmargin=*, noitemsep]
   \item Redshifts have very small statistical and systematic errors and these errors are assumed to be negligible.
   \item SN-host offsets are represented by the Rice distribution because the offsets are never negative and Gaussian noise superimposed on a vector results in a non-Gaussian probability distribution \citep{Rice1945a}.
   \item The $r'$-band magnitudes are represented by a normal distribution. In the case of a non-detection, a measurement is represented by a uniform distribution where the bright and faint bounds are set to the $3\sigma$ limiting magnitude and the faintest host in the sample (dimmed by additional 0.5 mag), respectively.
   \item For the age of the stellar population, attenuation, galaxy mass, $M_B$, SFR, and sSFR, we use the marginalized posteriors from the SED modeling.
\end{itemize}

\begin{figure*}[ht!]
\centering
\includegraphics[width=1\textwidth]{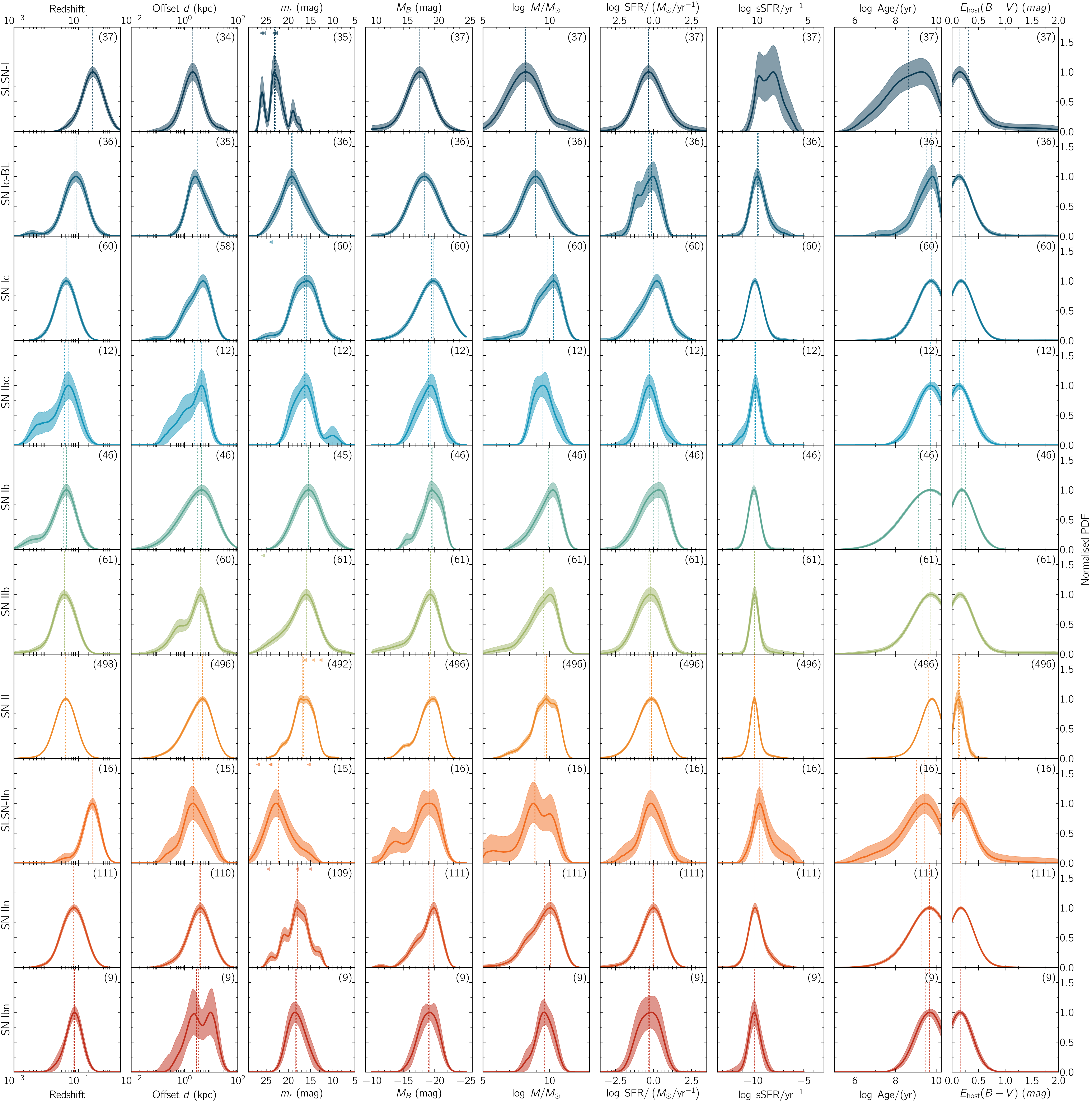}
\caption{Kernel density estimates of SN and host properties. The shaded regions display the point-wise $1\sigma$ confidence intervals. The vertical dashed and dotted lines indicate the mode and the median, respectively. The numbers in parenthesis quote the sample sizes. To guide the eye, the KDEs are normalized to have the same peak height. Upper limits in the $r$-band brightness are indicated by left-pointing  triangles in the upper half of a panel.
}
\label{fig:kde}
\end{figure*}

We estimate the probability distribution functions using the kernel density  (KDE) techniques. The critical parameter of the KDEs is the bandwidth, i.e., the smoothing parameter. For each sample, we estimate an adequate value using the leave-one-out cross-correlation method. For samples with less than 20 objects, e.g., SNe Ibn and SLSNe-IIn, we set the bandwidth parameter to the median value of the other host samples. To compute the point-wise $1\sigma$ confidence intervals, we compute the 68\% confidence interval around the median KDE of each distribution function.

Figure~\ref{fig:kde} shows the KDEs and the point-wise $1\sigma$ confidence intervals of all host properties. Table~\ref{tab:host_stat} summarizes the median, mode and full-width at half-maximum of each KDE. The distribution functions can exhibit complex shapes; they are uni-modal, but they can display asymmetries and pronounced wings primarily towards the faint end, independent of sample size and SN/host property. The asymmetries and the wings reflect, in part, the intrinsic shape of the luminosity and mass functions. Moreover, the existence of pronounced tails in the KDEs limits the effectiveness of median values and widths to identify hosts with outstanding properties. As an auxiliary data product of this paper, we release tabulated versions KDEs to identify singular hosts more easily.

We remark that the multiple peaks seen in the distribution functions are most likely artifacts due to the small sample sizes. As a sanity check, we generated fake samples from symmetric and asymmetric unimodal distributions, where we vary the sample sizes between 30 and 500. Multiple peaks and shoulders are generally observed in the distribution functions with $\lesssim$100 objects which cause deviations from the actual underlying distributions.

\movetabledown=5.5cm
\begin{table*}
\begin{rotatetable*}
\caption{Statistical properties of the SN host galaxy samples}\label{tab:host_stat}
\tiny
\hspace{-1.5cm}
\begin{tabular}{lllllllllll}
\toprule
SN class	& & 	$\log\,z$	& log Offset    & $m_{r}$	&$ M_B	$& $\log~M$                    	& $\log~{\rm SFR}$      	            & $\log~{\rm sSFR}$             & $\log~{\rm Age}$          & $E_{\rm host}(B-V)$\\ 
	    	& & 				& (kpc)         & (mag)		&(mag)	 & $\left(M_\odot\right)$       &$\left(M_\odot\,{\rm yr}^{-1}\right)$  & $\left({\rm yr}^{-1}\right)$  & $\left({\rm yr}\right)$   & (mag)\\ 
\midrule
\multicolumn{11}{c}{\textbf{H-poor SNe}}\\
\midrule
& Median &$-0.570^{+0.045}_{-0.045} $     &$0.283^{+0.068}_{-0.073} $     &$22.90^{+0.33}_{-0.35} $     &$-17.51^{+0.30}_{-0.28} $     &$8.15^{+0.23}_{-0.24} $     &$-0.23^{+0.16}_{-0.15} $     &$-8.34^{+0.30}_{-0.32} $     &$8.63^{+0.24}_{-0.27} $     &$0.31^{+0.05}_{-0.04} $  \\ 
 SLSN-I & Mode &$-0.557^{+0.060}_{-0.055} $  (37)    &$0.313^{+0.090}_{-0.105} $  (34)    &$23.14^{+0.53}_{-0.84} $  (35)    &$-17.63^{+0.33}_{-0.33} $  (37)    &$8.20^{+0.43}_{-0.50} $  (37)    &$-0.32^{+0.17}_{-0.14} $  (37)    &$-8.33^{+0.69}_{-0.99} $  (37)    &$9.05^{+0.49}_{-0.81} $  (37)    &$0.15^{+0.03}_{-0.03} $  (37) \\ 
 \vspace{2mm}  & FWHM &$0.680^{+0.114}_{-0.106} $     &$0.827^{+0.228}_{-0.181} $     &$4.89^{+1.73}_{-1.73} $     &$4.34^{+0.83}_{-0.70} $     &$2.58^{+0.60}_{-0.52} $     &$1.97^{+0.48}_{-0.38} $     &$2.36^{+0.61}_{-0.58} $     &$2.21^{+0.49}_{-0.49} $     &$0.70^{+0.46}_{-0.22} $  \\ 
 & Median &$-1.112^{+0.052}_{-0.054} $     &$0.479^{+0.062}_{-0.057} $     &$19.03^{+0.44}_{-0.46} $     &$-18.34^{+0.29}_{-0.29} $     &$8.98^{+0.16}_{-0.16} $     &$-0.33^{+0.15}_{-0.16} $     &$-9.46^{+0.14}_{-0.13} $     &$9.51^{+0.11}_{-0.13} $     &$0.23^{+0.02}_{-0.01} $  \\ 
 SN Ic-BL & Mode &$-1.077^{+0.080}_{-0.095} $  (36)    &$0.398^{+0.080}_{-0.055} $  (35)    &$19.27^{+0.38}_{-0.46} $  (36)    &$-18.33^{+0.38}_{-0.39} $  (36)    &$8.95^{+0.21}_{-0.21} $  (36)    &$-0.13^{+0.26}_{-0.49} $  (36)    &$-9.57^{+0.18}_{-0.16} $  (36)    &$9.78^{+0.14}_{-0.32} $  (36)    &$0.14^{+0.02}_{-0.01} $  (36) \\ 
 \vspace{2mm}  & FWHM &$0.725^{+0.129}_{-0.115} $     &$0.779^{+0.134}_{-0.134} $     &$6.02^{+1.18}_{-1.17} $     &$4.62^{+0.60}_{-0.61} $     &$2.23^{+0.42}_{-0.39} $     &$1.48^{+0.29}_{-0.30} $     &$1.65^{+0.46}_{-0.36} $     &$1.12^{+0.32}_{-0.26} $     &$0.38^{+0.06}_{-0.04} $  \\ 
 & Median &$-1.389^{+0.033}_{-0.032} $     &$0.540^{+0.059}_{-0.066} $     &$16.36^{+0.39}_{-0.38} $     &$-19.49^{+0.22}_{-0.21} $     &$9.88^{+0.14}_{-0.14} $     &$0.02^{+0.11}_{-0.12} $     &$-9.79^{+0.07}_{-0.07} $     &$9.50^{+0.04}_{-0.05} $     &$0.25^{+0.01}_{-0.01} $  \\ 
 SN Ic & Mode &$-1.382^{+0.045}_{-0.045} $  (60)    &$0.694^{+0.065}_{-0.080} $  (58)    &$15.88^{+1.42}_{-0.77} $  (60)    &$-19.76^{+0.23}_{-0.21} $  (60)    &$10.30^{+0.15}_{-0.35} $  (60)    &$0.23^{+0.11}_{-0.14} $  (60)    &$-9.81^{+0.06}_{-0.06} $  (60)    &$9.75^{+0.05}_{-0.06} $  (60)    &$0.17^{+0.02}_{-0.01} $  (60) \\ 
 \vspace{2mm}  & FWHM &$0.648^{+0.075}_{-0.073} $     &$0.957^{+0.134}_{-0.132} $     &$5.90^{+0.81}_{-0.76} $     &$4.96^{+0.52}_{-0.50} $     &$2.01^{+0.30}_{-0.27} $     &$1.87^{+0.32}_{-0.29} $     &$1.57^{+0.19}_{-0.16} $     &$1.16^{+0.12}_{-0.10} $     &$0.39^{+0.04}_{-0.03} $  \\ 
 & Median &$-1.436^{+0.105}_{-0.153} $     &$0.382^{+0.142}_{-0.199} $     &$16.43^{+0.71}_{-0.65} $     &$-19.00^{+0.51}_{-0.43} $     &$9.51^{+0.21}_{-0.22} $     &$-0.27^{+0.16}_{-0.17} $     &$-9.83^{+0.14}_{-0.17} $     &$9.50^{+0.09}_{-0.10} $     &$0.22^{+0.03}_{-0.02} $  \\ 
 SN Ibc & Mode &$-1.322^{+0.105}_{-0.080} $  (12)    &$0.629^{+0.060}_{-0.495} $  (12)    &$16.22^{+1.80}_{-0.77} $  (12)    &$-19.38^{+1.07}_{-0.32} $  (12)    &$9.51^{+0.26}_{-0.51} $  (12)    &$-0.28^{+0.17}_{-0.18} $  (12)    &$-9.75^{+0.13}_{-0.13} $  (12)    &$9.73^{+0.13}_{-0.15} $  (12)    &$0.14^{+0.03}_{-0.03} $  (12) \\ 
 \vspace{2mm}  & FWHM &$0.999^{+0.276}_{-0.360} $     &$1.041^{+0.298}_{-0.314} $     &$5.03^{+2.53}_{-1.49} $     &$3.58^{+0.97}_{-0.99} $     &$1.57^{+0.43}_{-0.46} $     &$1.26^{+0.60}_{-0.39} $     &$1.28^{+0.70}_{-0.42} $     &$1.12^{+0.18}_{-0.19} $     &$0.35^{+0.06}_{-0.04} $  \\ 
 & Median &$-1.469^{+0.048}_{-0.054} $     &$0.483^{+0.079}_{-0.089} $     &$15.54^{+0.43}_{-0.42} $     &$-19.51^{+0.26}_{-0.25} $     &$9.92^{+0.13}_{-0.15} $     &$0.01^{+0.14}_{-0.15} $     &$-9.89^{+0.07}_{-0.07} $     &$9.13^{+0.04}_{-0.05} $     &$0.25^{+0.01}_{-0.01} $  \\ 
 SN Ib & Mode &$-1.377^{+0.045}_{-0.050} $  (46)    &$0.644^{+0.085}_{-0.105} $  (46)    &$15.47^{+0.67}_{-0.67} $  (45)    &$-19.61^{+0.39}_{-0.89} $  (46)    &$10.26^{+0.16}_{-0.22} $  (46)    &$0.31^{+0.24}_{-0.49} $  (46)    &$-9.90^{+0.07}_{-0.07} $  (46)    &$9.71^{+0.06}_{-0.07} $  (46)    &$0.19^{+0.02}_{-0.01} $  (46) \\ 
   & FWHM &$0.859^{+0.208}_{-0.162} $     &$1.363^{+0.191}_{-0.187} $     &$6.30^{+1.01}_{-0.98} $     &$3.47^{+0.62}_{-0.62} $     &$1.92^{+0.32}_{-0.32} $     &$1.82^{+0.29}_{-0.28} $     &$1.19^{+0.20}_{-0.17} $     &$1.83^{+0.12}_{-0.09} $     &$0.38^{+0.03}_{-0.03} $  \\ 
 \midrule
\multicolumn{11}{c}{\textbf{H-rich SNe}}\\
\midrule
 & Median &$-1.447^{+0.037}_{-0.037} $     &$0.432^{+0.065}_{-0.079} $     &$16.71^{+0.41}_{-0.37} $     &$-18.78^{+0.25}_{-0.23} $     &$9.53^{+0.15}_{-0.16} $     &$-0.30^{+0.12}_{-0.12} $     &$-9.80^{+0.07}_{-0.07} $     &$9.35^{+0.05}_{-0.06} $     &$0.26^{+0.02}_{-0.02} $  \\ 
 SN IIb & Mode &$-1.437^{+0.060}_{-0.050} $  (61)    &$0.609^{+0.045}_{-0.050} $  (60)    &$15.98^{+0.36}_{-0.33} $  (61)    &$-19.32^{+0.33}_{-0.27} $  (61)    &$10.04^{+0.15}_{-0.42} $  (61)    &$-0.18^{+0.25}_{-0.29} $  (61)    &$-9.84^{+0.12}_{-0.11} $  (61)    &$9.74^{+0.06}_{-0.06} $  (61)    &$0.15^{+0.02}_{-0.02} $  (61) \\ 
 \vspace{2mm}  & FWHM &$0.709^{+0.100}_{-0.090} $     &$1.162^{+0.163}_{-0.165} $     &$7.42^{+1.18}_{-1.13} $     &$4.35^{+0.92}_{-0.72} $     &$2.25^{+0.37}_{-0.35} $     &$1.78^{+0.30}_{-0.27} $     &$1.08^{+0.33}_{-0.23} $     &$1.49^{+0.20}_{-0.15} $     &$0.46^{+0.10}_{-0.06} $  \\ 
 & Median &$-1.408^{+0.012}_{-0.012} $     &$0.537^{+0.023}_{-0.024} $     &$16.60^{+0.13}_{-0.13} $     &$-19.15^{+0.09}_{-0.09} $     &$9.65^{+0.05}_{-0.05} $     &$-0.24^{+0.04}_{-0.04} $     &$-9.86^{+0.02}_{-0.02} $     &$9.61^{+0.02}_{-0.02} $     &$0.14^{+0.01}_{-0.01} $  \\ 
 SN II & Mode &$-1.392^{+0.015}_{-0.020} $  (498)    &$0.679^{+0.035}_{-0.040} $  (496)    &$16.77^{+0.58}_{-1.10} $  (492)    &$-19.75^{+0.34}_{-0.24} $  (496)    &$9.77^{+0.47}_{-0.15} $  (496)    &$-0.12^{+0.08}_{-0.09} $  (496)    &$-9.86^{+0.04}_{-0.04} $  (496)    &$9.80^{+0.02}_{-0.02} $  (496)    &$0.12^{+0.04}_{-0.03} $  (496) \\ 
  & FWHM &$0.664^{+0.027}_{-0.026} $     &$1.054^{+0.057}_{-0.056} $     &$4.88^{+0.30}_{-0.30} $     &$3.76^{+0.25}_{-0.23} $     &$1.96^{+0.13}_{-0.13} $     &$1.72^{+0.09}_{-0.09} $     &$1.09^{+0.09}_{-0.08} $     &$0.98^{+0.05}_{-0.04} $     &$0.19^{+0.01}_{-0.01} $  \\ 
\midrule
\multicolumn{11}{c}{\textbf{Interaction-powered SNe}}\\
\midrule
 & Median &$-0.616^{+0.039}_{-0.047} $     &$0.343^{+0.136}_{-0.142} $     &$22.26^{+0.69}_{-0.78} $     &$-18.30^{+0.69}_{-0.62} $     &$8.89^{+0.38}_{-0.37} $     &$-0.15^{+0.22}_{-0.21} $     &$-9.10^{+0.33}_{-0.25} $     &$9.03^{+0.20}_{-0.28} $     &$0.28^{+0.06}_{-0.04} $  \\ 
 SLSN-IIn & Mode &$-0.577^{+0.030}_{-0.035} $  (16)    &$0.321^{+0.283}_{-0.168} $  (15)    &$22.78^{+0.89}_{-1.03} $  (15)    &$-19.13^{+1.07}_{-0.87} $  (16)    &$8.91^{+1.14}_{-0.47} $  (16)    &$-0.16^{+0.34}_{-0.22} $  (16)    &$-9.33^{+0.41}_{-0.22} $  (16)    &$9.44^{+0.24}_{-0.32} $  (16)    &$0.16^{+0.04}_{-0.03} $  (16) \\ 
 \vspace{2mm}  & FWHM &$0.559^{+0.157}_{-0.109} $     &$1.027^{+0.464}_{-0.356} $     &$6.17^{+2.20}_{-1.99} $     &$5.79^{+1.46}_{-2.13} $     &$2.90^{+1.25}_{-1.13} $     &$1.68^{+0.81}_{-0.55} $     &$2.19^{+1.10}_{-0.86} $     &$1.97^{+0.80}_{-0.60} $     &$0.54^{+0.52}_{-0.17} $  \\ 
 & Median &$-1.161^{+0.030}_{-0.030} $     &$0.555^{+0.040}_{-0.042} $     &$17.92^{+0.27}_{-0.28} $     &$-19.25^{+0.20}_{-0.18} $     &$9.63^{+0.12}_{-0.12} $     &$-0.09^{+0.07}_{-0.07} $     &$-9.70^{+0.07}_{-0.07} $     &$9.29^{+0.04}_{-0.05} $     &$0.25^{+0.01}_{-0.01} $  \\ 
 SN IIn & Mode &$-1.137^{+0.045}_{-0.055} $  (111)    &$0.594^{+0.065}_{-0.070} $  (110)    &$17.93^{+0.41}_{-1.27} $  (109)    &$-19.85^{+0.24}_{-0.23} $  (111)    &$10.05^{+0.17}_{-0.25} $  (111)    &$0.01^{+0.11}_{-0.11} $  (111)    &$-9.80^{+0.08}_{-0.07} $  (111)    &$9.67^{+0.05}_{-0.05} $  (111)    &$0.17^{+0.01}_{-0.01} $  (111) \\ 
 \vspace{2mm}  & FWHM &$0.722^{+0.062}_{-0.062} $     &$0.919^{+0.111}_{-0.103} $     &$5.67^{+0.73}_{-0.71} $     &$3.92^{+0.44}_{-0.45} $     &$2.35^{+0.26}_{-0.26} $     &$1.59^{+0.19}_{-0.18} $     &$1.69^{+0.26}_{-0.23} $     &$1.53^{+0.12}_{-0.11} $     &$0.39^{+0.03}_{-0.02} $  \\ 
 & Median &$-1.146^{+0.046}_{-0.054} $     &$0.525^{+0.244}_{-0.193} $     &$18.14^{+0.59}_{-0.65} $     &$-19.02^{+0.48}_{-0.47} $     &$9.62^{+0.21}_{-0.20} $     &$-0.32^{+0.27}_{-0.30} $     &$-9.87^{+0.18}_{-0.18} $     &$9.49^{+0.09}_{-0.10} $     &$0.23^{+0.02}_{-0.02} $  \\ 
 SN Ibn & Mode &$-1.122^{+0.045}_{-0.040} $  (9)    &$0.458^{+0.566}_{-0.190} $  (9)    &$18.45^{+0.82}_{-1.03} $  (9)    &$-19.13^{+0.75}_{-0.83} $  (9)    &$9.59^{+0.43}_{-0.18} $  (9)    &$-0.25^{+0.46}_{-0.57} $  (9)    &$-9.87^{+0.19}_{-0.17} $  (9)    &$9.68^{+0.15}_{-0.16} $  (9)    &$0.16^{+0.03}_{-0.03} $  (9) \\ 
  & FWHM &$0.524^{+0.142}_{-0.124} $     &$1.085^{+0.388}_{-0.366} $     &$4.28^{+1.20}_{-1.33} $     &$3.22^{+0.81}_{-0.89} $     &$1.39^{+0.53}_{-0.47} $     &$1.55^{+0.52}_{-0.54} $     &$1.22^{+0.65}_{-0.42} $     &$1.11^{+0.17}_{-0.18} $     &$0.35^{+0.04}_{-0.04} $  \\ 

\bottomrule
\end{tabular}
\tablecomments{The medians, modes and the full-widths at half-maxima as well as their uncertainties were extracted from the KDEs . The values in parenthesis display the number of objects that were used. The $r$-band brightness and the $B$-band luminosity are not corrected for host attenuation. We omit to report the statistics for SLSN-II and SLSN-IIb because these classes consists of only one object each.
}
\end{rotatetable*}
\end{table*}

\subsection{Environmental Effects on the Formation of CCSNe}\label{disc:Z_bias}
\subsubsection{Metallicity-Dependent SN Production Efficiency}

In Section \ref{result:mass}, we found a dearth of galaxies above $10^{10}~M_\odot$ hosting SLSNe-I/IIn and SNe Ic-BL, while other CCSN classes seem to select galaxies solely based on their star-formation activity. This could point to an environment-dependent production efficiency of progenitor systems of some CCSN classes, as demonstrated in \citet{Perley2016a} and \citet{Schulze2018a}. The primary parameter that could regulate the production efficiency is the galaxy mass because it is known to correlate well with the average galaxy metallicity \citep[e.g.,][]{Tremonti2004a, Mannucci2010a, Andrews2013a}. Metallicity, in turn, has a strong effect on the evolution of massive stars through line-driven stellar winds \citep[e.g.,][]{Kudritzki2000a}.

\begin{figure*}
   \centering
   \includegraphics[width=1\textwidth]{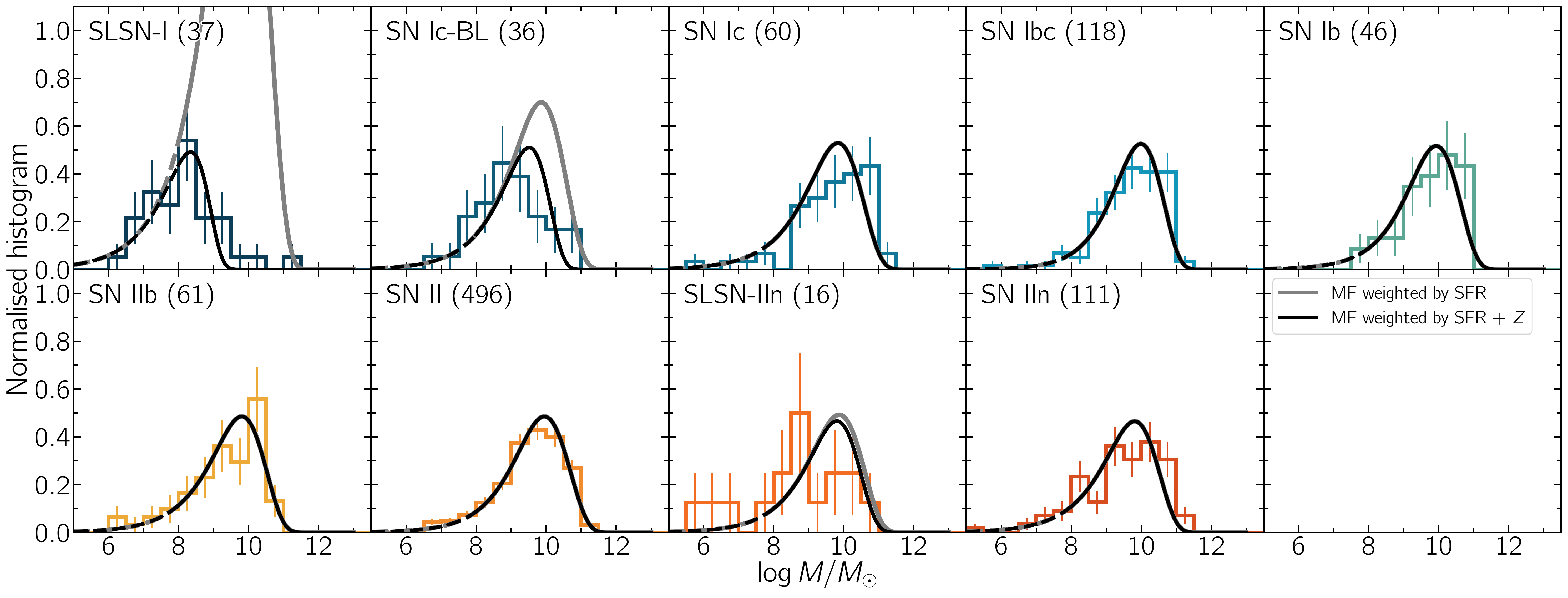}
   \caption{Host mass distributions of each CCSN class. The SFR-weighted mass-function model (grey curves) provides an adequate description of the mass distributions of SN Ib/c, IIb, II as well as IIn host galaxies. The observed mass distributions of SLSN-I and SN Ic-BL hosts show a clear dearth of massive host galaxies. This absence can be accounted for by adding a metallicity-dependent SN production efficiency (black curves). The production efficiencies of SLSNe-I and SNe Ic-BL are stifled in environments with oxygen abundances exceeding $8.26^{+0.22}_{-0.30}$ and $8.66^{+0.20}_{-0.14}$, respectively. The sample of SLSN-IIn hosts is too small to conclude whether SLSNe-IIn require particular environments. The host sample of SNe Ibc includes the Type Ib, Ic and Ibc SNe, to maximize the sample size. We remark that the galaxy mass functions were extrapolated for masses of $<10^8~M_\odot$. This is indicated by the dashed lines. Note, the grey and the black curves are different only for the host-galaxy samples of SLSNe and Ic-BL SNe.
   }
   \label{fig:mass}
   \end{figure*}

To quantify the metallicity-dependent production efficiencies, we apply the method from \citet{Schulze2018a} that we also applied in Sec. \ref{result:mass}. This method goes as follows. We start with the stellar mass function $\Phi(M)$ of star-forming galaxies from CANDELS and use the parameterization of the mass-function for SF galaxies of \citet[][their table 2]{Tomczak2014a}. This yields the number density of galaxies per stellar-mass bin. Next, we weigh each mass bin by its contribution to the cosmic star formation at a given redshift using the fundamental relationship between galaxy mass and star-formation rate and the parameterization of \citet{Lee2015a}, which is valid from $10^{8.5}$ to $10^{11.25}~M_\odot$. The functional form of the SFR-weighted mass function is $w\{{\rm SFR}\left(M\right)\}\times\Phi\{M\}$, where $w$ is the SFR weight.

To find the best-fitting model, we generate 30000 samples of the observed data of each SN class as described in Sec. \ref{sec:kde}. Within each trial, we also vary the galaxy mass function parameters within their uncertainties and the location of the bin centers, and find the best fitting model parameters using least-square-fitting. In the final step, we build distribution functions of the model parameters and extract the median value and its uncertainties of each parameter.

Figure~\ref{fig:mass} shows the observed mass distributions of the eight largest SN classes and in grey the best-fit SFR-weighted mass functions. The predicted mass functions provide excellent matches to the samples of hosts of Type Ibc\footnote{In this analysis and in Sec. \ref{disc:higher_order}, we combined Ib, Ic and Ibc SNe to maximize the sample size.}, II/IIb SNe as well as Type IIn SNe. This means that their occurrence is independent of global galaxy metallicity, to the level we are sensitive, and driven by the star-formation activity of their hosts.

In contrast to the agreement between the observed mass distributions and the SFR-weighted mass functions for the hosts of Ib, Ic, Ibc\,+\,Ib\,+\,Ic, II, IIb, and IIn SNe, the mass distributions of SLSN and SN Ic-BL host galaxies peak 1--2 orders of magnitudes lower than predicted by SFR-weighted mass functions. To account for the lack of massive galaxies, we introduce a function that describes an efficiency $\rho(M)$ of producing SNe from star formation. Similar to \citet{Schulze2018a}, we chose $\rho\left(M\right)$ as an exponential function of the form $\rho\left(M\right) = \exp\left(-M/M_0\right)$, where $M_0$ is a characteristic cut-off mass and therefore a cut-off metallicity. The functional form of the metallicity-dependent SFR-weighted star-formation history is $\rho\{M\} \times w\{{\rm SFR}\left(M\right)\}\times\Phi\{M\}$.

The best fits are shown by the black curves in Fig.~\ref{fig:mass} and fit parameters are summarised in Table~\ref{tab:zbias}. This model adequately describes the observed mass distributions of SLSN-I, SN Ic-BL and SLSN-IIn host galaxies, albeit the sample size of the latter SN class is too small to draw a firm conclusion. We convert these mass cut-offs into a cut-off oxygen abundance using equation 5 in \citet{Mannucci2010a}. The best-fitting models point to stifled production efficiencies at oxygen abundances exceeding $12 + \log {\rm O}/{\rm H} =8.26^{+0.26}_{-0.30}$, $8.65^{+0.20}_{-0.14}$ and $8.75^{+0.33}_{-0.41}$ for SLSNe-I, SNe Ic-BL and SLSNe-IIn, respectively. This translates to cut-off metallicity of $\sim0.4$, $\sim1$ and $\sim1.1$ solar metallicity for SLSNe-I, SNe Ic-BL, SLSNe-IIn, respectively, using the solar oxygen abundance reported in \citet{Asplund2009a}.

The value of H-poor SLSNe is consistent with the values reported in \citet{Perley2016a}, \citet{Chen2017a}, and \citet{Schulze2018a}. Furthermore, the hosts of H-poor SLSNe are characterized by the youngest stellar population. Their median age is $\log {\rm Age}/{\rm yr}\approx8.6$ in contrast to the average age of $\log {\rm Age}/{\rm yr}\approx9.7$ of all regular CCSN host galaxies (Table \ref{tab:host_stat}). Measuring ages is notoriously difficult. However, as we demonstrated in Sec. \ref{method:impact}, the inferred ages are reliable in a comparative sense, i.e., samples of young galaxies remain young and samples of evolved galaxies remain old independent of the wavelength coverage of the SEDs or the assumptions of the SED model. This means that the difference in the age distributions reflects a genuine difference. Therefore, not only a low metallicity but also young age play an important role in the formation of SLSN-I progenitors. This corroborates the conjecture in \citet{Leloudas2015a}, \citet{Thoene2015a} and \citet{Schulze2018a} that SLSNe could be connected with the death of very massive stars \citep[see also][]{Taggart2019a}.

Our result for Ic-BL SNe is consistent with \citet{Modjaz2019a}. These authors analyzed a subsample of the entire PTF SN Ic-BL sample and reported a median oxygen abundance of $12+\log\,{\rm O} / {\rm H} \approx 8.5$. The median mass of the entire SN Ic-BL host sample is $\log\,M/M_\odot\approx8.98$ (Table \ref{tab:host_stat}). This value translates to a median oxygen abundance of $\sim8.4$ using equation 5 in \citet{Mannucci2010a}. Our analysis expands upon \citet{Modjaz2019a} by providing a critical galaxy metallicity ($0.9^{+0.5}_{-0.3}$ solar) above which the production is stifled.

Long-duration gamma-ray bursts are thought to be accompanied by Ic-BL SNe \citep[for a recent review see][but see \citealt{Fynbo2006a, Greiner2015a, Michalowski2018a, Tanga2018a, Kann2019a}]{Cano2017a}. An outstanding question in the SN and GRB fields is how both subpopulations of Ic-BL SNe are connected \citep[e.g.,][]{Modjaz2019a}. GRBs also show a pronounced environment dependence. At $z<1$, their production efficiency is stifled above 0.5--0.9 solar metallicity \citep{Kruehler2015a, Schulze2015a, Vergani2015a, Perley2016a, Vergani2017a, Schulze2018a} consistent with our measurement of Ic-BL SNe. However, the comparison is limited by the large statistical errors.

The samples of SLSNe-IIn is too small to draw a firm conclusion on any environment-dependent production efficiency, similar to Type Ibn SNe which are even rarer in the PTF sample.

\begin{table}
\caption{Summary of the cut-off mass and metallicities of the metallicity-dependent SN production efficiencies}\label{tab:zbias}
{
\hspace{-1cm}
\begin{tabular}{ccccc}
\toprule
SN class         & $\log M_0/M_\odot$        & $12 + \log\,{\rm O/H}$     & $\chi^2/{\rm d.o.f.}$ \\
\midrule
SLSN-I          & $8.64^{+0.46}_{-0.64}$     & $8.26^{+0.26}_{-0.30}$    & 4.6/9\\
SN Ic-BL        & $9.47^{+0.42}_{-0.30}$     & $8.65^{+0.20}_{-0.14}$    & 2.0/8\\
SLSN-IIn        & $9.67^{+1.56}_{-0.87}$     & $8.75^{+0.33}_{-0.41}$    & 1.9/7\\
\bottomrule
\end{tabular}
}
\tablecomments{The abbreviation `d.o.f.' stands for degree of freedom.}
\end{table}

\begin{figure*}[ht!]
\centering
\includegraphics[width=1\textwidth]{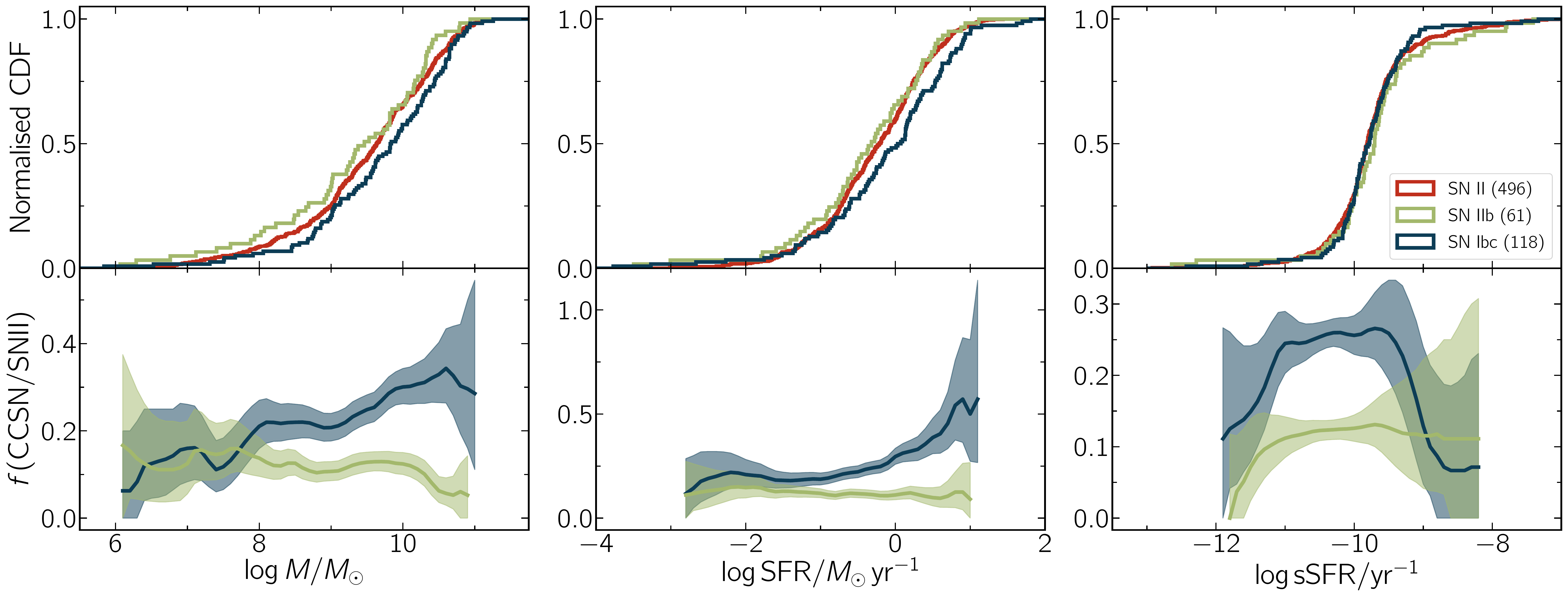}
\caption{Comparison of the distribution functions (top) and number ratios between Type Ibc, IIb and II SNe (bottom) as a function of different galaxy properties. Type Ibc SNe are found in significantly more massive galaxies with higher absolute star-formation rates (but similar specific star-formation rates) than Type II and IIb SNe. In contrast to that, Type IIb SNe explode in galaxies very similar to that of SNe II. The shaded regions indicate the $1\sigma$ confidence interval of the number ratio (see main text for details). The ratios doe not extend to the ends of the distribution because of too low-number statistics.
}
\label{fig:ad}
\end{figure*}

\subsubsection{Differences between SNe Ibc and SNe II/IIb}\label{disc:higher_order}

To identify more subtle differences in the SN environments, we directly compare the mass, SFR and sSFR distribution functions. Critical for this analysis is that the redshift distributions are statistically identical to minimize the impact of the secular evolution of SN and host properties. This limits the comparison to Type Ibc, IIb and II SNe.

The top panels of Fig.~\ref{fig:ad} show the cumulative distribution functions of the galaxy mass, the SFR and the sSFR of the three samples. The galaxy mass, SFR and sSFR distribution functions of all samples span the same range. To quantify dissimilarities between the SN classes, we apply the Anderson-Darling test on 30000 resampled distributions for each SN class and host property (generated as described in Sec. \ref{sec:kde}). Our null-hypothesis, $H_0$, is that the test sample is drawn from the parent sample. We reject $H_0$ if the chance probability $p_{\rm ch}$ is smaller than 5\%.

The mass, SFR and sSFR distribution functions of SN IIb and SN II host galaxies are statistically identical. The chance probabilities to randomly draw a mass, SFR or sSFR distribution from the SN-II host sample, which are as extreme as the SN IIb host sample, are between 12\% and 24\% (Table~\ref{tab:ad}). Similarly, the mass, the SFR and the sSFR distributions of Type Ib and Ic SN hosts are statistically identical ($p_{\rm ch}=19$--35\%; Table~\ref{tab:ad}).

As the SN Ib and SN Ic host populations are shown to be similar, we combine the SN Ib, Ibc and Ic samples. This gives us a larger statistical power to trace differences between the host populations of Type Ibc and Type II SNe, and Type Ibc and IIb SNe. Indeed, there are statistically significant differences between the mass and SFR distributions of SN Ibc and SN IIb as well as SN II hosts. The chance probabilities vary between 0.4 and 4\%, which are below our threshold of rejecting the null hypothesis (Table~\ref{tab:ad}).

\begin{table}
\caption{Summary of the Anderson Darling tests}\label{tab:ad}
{
\hspace{-1.5cm}
\scriptsize
\begin{tabular}{c@{\hskip 2mm}c@{\hskip 2mm}c@{\hskip 2mm}c@{\hskip 2mm}c@{\hskip 2mm}c}
\toprule
Sample              & Sample    & \multicolumn{4}{c}{Chance probability}\\
                    & sizes     & Redshift  & $\log M$      & $\log {\rm SFR}$          & $\log {\rm sSFR}$\\
                    &           &           & $(M_\odot)$   & $(M_\odot{\rm yr}^{-1})$  & $({\rm yr}^{-1})$\\
\midrule
SN Ib vs. SN Ic   & 46/60    & 0.048 & 0.345 & 0.287 & 0.191 \\
SN Ic vs. SN IIb  & 46/61    & 0.143 & 0.025 & 0.040 & 0.258 \\
SN Ib vs. SN IIb  & 46/61    & 0.187 & 0.025 & 0.032 & 0.155 \\
SN Ic vs. SN II   & 60/496   & 0.378 & 0.066 & 0.011 & 0.213 \\
SN Ib vs. SN II   & 46/496   & 0.025 & 0.076 & 0.018 & 0.206 \\
SN Ibc vs. SN IIb & 118/61   & 0.233 & 0.015 & 0.032 & 0.190 \\
SN Ibc vs. SN II  & 118/496  & 0.112 & 0.035 & 0.004 & 0.206 \\
SN IIb vs. SN II  & 61/496   & 0.148 & 0.175 & 0.244 & 0.123 \\

\bottomrule
\end{tabular}
}
\tablecomments{The null hypothesis $H_0$ is that the test distribution from the parent distribution. $H_0$ is rejected if the chance probability is smaller than 5\%.}
\end{table}

To illustrate these results differently, we present in the bottom panels of Fig. \ref{fig:ad} the number ratios between Type Ibc and Type II SNe (blue) and Type IIb and II SNe (green) as a function galaxy mass, SFR and sSFR. We computed this ratio for moving bins (bin width 1 dex and bin stepsize 0.1 dex) for each resampled distribution and, then, extracted the median ratio and its $1\sigma$ confidence interval at each bin step. Type Ibc SNe are found in more massive galaxies with higher absolute star-formation but the same specific star-formation rates than Type II SNe. In contrast to that, the number ratio between Type IIb and II SNe is not changing with host properties.

In summary, any difference in the production efficiencies of regular CCSNe is only mildly dependent on global galaxy properties. Understanding the mapping between SNe and their progenitors may require examining the particular conditions of the explosion sites across the full spectrum of host galaxy properties  \citep[e.g.,][]{Modjaz2008a,Leloudas2011a,Kelly2012a,Sanders2012a,Kuncarayakti2013a,Galbany2018a}.

Previously, \citet{Arcavi2010a} reported the CCSN number ratio based on PTF Year-1 data. This sample included 72 CCSNe of various types. These authors concluded that Type Ic SNe are almost exclusively found in galaxies brighter than $M_r<-19$~mag. In less luminous galaxies, the stripped-envelope SN population is dominated by Type Ib and Ic-BL SNe. Furthermore, these authors found an excess of SNe IIb in low-luminosity galaxies. To explain these results, \citet{Arcavi2010a} hypothesized that metallicity-driven mass-loss leads to reduced stripping of SN Ic progenitors in low-metallicity environments, which allows their progenitors to retain some hydrogen and helium.

Using the full PTF sample, we cannot recover several of these suggested trends. Type II, IIb, Ib, and Ic SNe are found in galaxies from $10^6$ to $10^{12}~M_\odot$ (Sect.~\ref{result:mass}), and they select their hosts according to their star-formation activity (Sect.~\ref{disc:Z_bias}). We also do not recover the over-abundance of SNe IIb in low-mass galaxies in our complete sample. However, Type Ic \textit{and} Ib SNe are found in slightly more massive galaxies with higher absolute star-formation rates than Type II and IIb SNe (Fig.~\ref{fig:ad}; Table~\ref{tab:ad}), similar to what was seen in \citet{Arcavi2010a}.

The interpretation suggested by \citet{Arcavi2010a}, and in particular, that similar progenitors produce SNe Ib at high metallicity (more massive galaxies) and SNe IIb at lower metallicity, is not evident in our data. Localized metallicity studies in SN explosion sites, or direct progenitor metallicity studies of individual SN progenitors possible through rapid \hst\ UV spectroscopy of infant SNe, could further illuminate this issue.

\citet{Graur2017a, Graur2017b} used the volume-limited Lick Observatory Supernova Survey that primarily targeted massive galaxies in the nearby Universe. These authors reported that Type Ibc SNe are a factor of 2.5--3 less frequent in galaxies between $10^{9.5}$ and $10^{10.7}~M_\odot$, and  that the SNe IIb / SN II number ratio is the same in low- and high-mass galaxies. These authors also found a flattening of the SN Ibc / SNII rate in galaxies with $M>10^{10}~M_\odot$. Our data set does not support the rapid increase of the SN Ibc / SN II number ratio with galaxy mass claimed. We find a moderate increase of the SN Ibc / SN II number ratio by 25\% from $\sim10^{9.5}~M_\odot$ to $\sim10^{10.7}~M_\odot$ (bottom left panel in Fig.~\ref{fig:ad}).
We do confirm that that number ratio between Type IIb and Type II SN does not evolve with galaxy properties on a statistically significant level. Our data is inconclusive about whether the SN Ibc / SN II number ratio is flattening in galaxies more massive than $\sim10^{10}~M_\odot$.

\section{Conclusions}

In this paper, we have presented all core-collapse supernovae detected by the Palomar Transient Factory between 2009 and mid-2017, and their host galaxies. This sample includes 888 objects from 12 distinct classes out to $z\approx1$. We measured the brightness of the hosts from the FUV to the MIR and modeled their spectral energy distributions with stellar-population-synthesis models to extract physical properties, such as galaxy masses and star-formation rates, as well as phenomenological properties, such as absolute magnitudes. Our main conclusions are:

\begin{itemize}[leftmargin=*]

\item The PTF CCSN sample probes the complete spectrum of star-forming galaxies from $10^{5.4}$ to $10^{11.3}~M_\odot$, including galaxies comparable to the least massive least-massive star-forming galaxies in the Local Group. About 29\% and 11\% of the entire CCSN sample are found in galaxies less massive than $10^9$ and $10^8~M_\odot$, respectively. About 3\% of all regular CCSNe are found in starbursting galaxies with specific star-formation rates of $>10^{-8}~{\rm yr}^{-1}$. If SLSNe are included, the starburst frequency increases to $4.5\%$.

\item Regular CCSNe (Type Ib/c, IIb, II, IIn) are direct tracers of star-formation. Their mass functions are consistent with those of the general population of star-forming galaxies weighted by their star-formation activity. The production efficiencies of their progenitor systems are close to independent of the \textit{host-integrated} metallicity and sSFR. Explosion site studies are needed to identify the true environmental differences between these SN classes.

\item The mass, SFR and sSFR distribution functions of Type Ib, Ic, IIb and II SN host galaxies span the same ranges. However, the mass and SFR distribution functions of SN Ib+Ic+Ibc host galaxies (as individual classes and combined) are skewed towards galaxies with slightly higher masses and higher star-formation rates. These differences are less pronounced than in previous studies that were based on smaller and/or heterogeneous samples.

\item H-poor SLSNe, as well as SNe Ic-BL, are biased tracers of star-formation. This bias can be corrected for by introducing a metallicity-dependent production efficiency. The occurrence of H-poor SLSNe and SNe Ic-BL is stifled above an oxygen abundance of $12+\log\,{\rm O/H} = 8.26^{+0.26}_{-0.30}$ ($\sim0.4~Z_\odot$) and $8.65^{+0.20}_{-0.14}$ ($\sim1~Z_\odot$), respectively, confirming previous studies. In addition, H-poor SLSNe are found in galaxies with younger stellar-populations ($10^{8.3}$ vs. $10^{9.7}$~yr) and higher specific star-formation rates $>10^{-8}~{\rm yr}^{-1}$ (68\% vs. 3\%) than all other SN classes discussed in this paper. This lends further support to the notion that low-metallicity and young age play an important in the formation of SLSN progenitors.

\item The samples of H-rich SLSNe and Type Ibn SNe are still too small to conclude on whether their progenitors require special galaxy environments, e.g., low-metallicity galaxies.

\item The distribution functions of the projected distances of SNe to the center of their host galaxies extends to 37~kpc. In most cases, the projected distances are smaller than the expected 80\% light radii of their host galaxies. Less than 14\% of all CCSNe (of all types) are found at larger distances, but in most cases, still within the extent of the diffuse galaxy light.
\end{itemize}

\noindent On a more technical note, we conclude that

\begin{itemize}[leftmargin=*]

\item Galaxy surveys with limiting magnitudes of $\sim24.5$~mag, such as the DESI Legacy Imaging Surveys, are sufficient to detect the host galaxies of almost every CCSN in transient surveys with limiting magnitudes of $\sim21$~mag. The host recovery-rate of SLSNe is lower due to their larger redshifts and their preference for low-luminosity galaxies.

\item The probability distributions of the host properties are characterized by a Gaussian core with different levels of asymmetry and pronounced wings, in particular towards the faint end. These shapes are reminiscent of the underlying luminosity and mass functions of star-forming galaxies and need to be taken into account to identify singular CCSN host galaxies.

\end{itemize}

We presented the host galaxies of the most common CCSN classes and indirectly constrained their progenitor populations. However, rare classes, such as Type Ibn SNe and various flavors of H-rich SLSNe (SLSN-II, SLSN-IIb, SLSN-IIn), are still too scarce to constrain their host properties and quantify environment-dependent production efficiencies. The Zwicky Transient Facility (ZTF; \citealt{Bellm2019a, Graham2019a}) will allow building up larger and more homogeneous samples in a shorter period of time (\citealt{Fremling2019a}; Perley et al., to be submitted). It uses the P48 telescope like PTF but a new camera with a 6-times larger field of view. Furthermore, the implementation of a sophisticated alert distribution system \citep{Patterson2019a} allows defining unbiased surveys with reproducible selection functions \citep[such as the public ZTF Bright Transient Survey; ][]{Fremling2019a} which can build samples containing thousands of CCSNe within a mere of a few years. The absence of an alert distribution system inhibits us in quantifying the selection effects of the PTF survey.

Large and well-defined SN samples offer novel techniques to address open questions in galaxy science. The PTF CCSN sample revealed extreme environments of star-formation, such as early-type galaxies \citep{Irani2019a}, extremely low-mass galaxies \citep{De2018a}, and starburst galaxies \citep{Perley2016a,Leloudas2015a,Schulze2018a}. Galaxies with such extreme properties are also rare in an absolute sense. In Schulze (in prep), we examine these peculiar environments in detail and show how SNe can be used as probes to identify these extreme environments in real-time.

\begin{acknowledgements}

We thank Nino Cucchiara, Thomas de Jaeger, Harald Ebeling, David Levitan, Bruce Margon, Jon Mauerhan, Jacob Rex, David Sand, Jeffrey M. Silverman, Vicky Toy, and Brad Tucker for performing some of the observations and Ido Irani and Maryam Modjaz for valuable discussions.

I. Arcavi is a CIFAR Azrieli Global Scholar in the Gravity and the Extreme Universe Program and acknowledges support from that program, from the European Research Council (ERC) under the European Union’s Horizon 2020 research and innovation program (grant agreement number 852097), from the Israel Science Foundation (grant numbers 2108/18 and 2752/19), from the United States - Israel Binational Science Foundation (BSF), and from the Israeli Council for Higher Education Alon Fellowship.
J.~S. Bloom was partially supported by a Gordon and Betty Moore Foundation Data-Driven Discovery grant.
The UCSC team is supported in part by NASA grant NNG17PX03C, the Gordon \& Betty Moore Foundation, the Heising-Simons Foundation, and by a fellowship from the David and Lucile Packard Foundation to R.~J. Foley.
A. V. Filippenko acknowledges support from the U.S. National Science Foundation, the Christopher R. Redlich Fund, the TABASGO Foundation, and the Miller Institute for Basic Research in Science (U.C. Berkeley).
M. Fraser is supported by a Royal Society - Science Foundation Ireland University Research Fellowship.
A. Gal-Yam's research is supported by the EU via ERC grant 725161, the ISF GW excellence center, an IMOS space infrastructure grant and BSF/Transformative and GIF grants, as well as The Benoziyo Endowment Fund for the Advancement of Science, the Deloro Institute for Advanced Research in Space and Optics, The Veronika A. Rabl Physics Discretionary Fund, Paul and Tina Gardner, Yeda-Sela and the WIS-CIT joint research grant;  A. Gal-Yam is the recipient of the Helen and Martin Kimmel Award for Innovative Investigation.
A.~Y.~Q. Ho was supported by a National Science Foundation Graduate Research Fellowship under grant DGE‐1144469, and by the GROWTH project funded by the National Science Foundation under PIRE grant 1545949.
D.~A. Howell, G. Hosseinzadeh, and C. McCully were supported by NSF grant AST-1313484.
M.~M. Kasliwal acknowledges support by the GROWTH (Global Relay of Observatories Watching Transients Happen) project funded by the National Science Foundation under PIRE Grant No 1545949.
S. Kne\v{z}evi\'c was supported by the Ministry of Education, Science and Technological Development of the Republic of Serbia through contract 451-03-68/2020/14/20002 made with the Astronomical Observatory of Belgrade.
G. Leloudas is supported by a research grant (19054) from VILLUM FONDEN.
R. Lunnan is supported by a Marie Sk\l{}odowska-Curie Individual Fellowship within the Horizon 2020 European Union (EU) Framework Programme for Research and Innovation (H2020-MSCA-IF-2017-794467).
K. Maguire acknowledges funding from EU H2020 ERC grant 758638
T. Petrushevska acknowledges the financial support from the Slovenian Research Agency (grants I0- 0033, P1-0031, J1-8136 and Z1-1853).
S. Schulze gratefully acknowledges support provided by the Feinberg Graduate School at the Weizmann Institute, Israel.
A.~H. Wright is supported by an European Research Council Consolidator Grant (770935).

The Palomar Transient Factory project is a scientific collaboration among the California Institute of Technology, Los Alamos National Laboratory, the University of Wisconsin, Milwaukee, the Oskar Klein Center, the Weizmann Institute of Science, the TANGO Program of the University System of Taiwan, and the Kavli Institute for the Physics and Mathematics of the Universe. LANL participation in iPTF is supported by the US Department of Energy as a part of the Laboratory Directed Research and Development program.

This work makes use of data from the Las Cumbres Observatory network. Some of the data presented herein were obtained at the W. M. Keck Observatory, which is operated as a scientific partnership among the California Institute of Technology, the University of California, and NASA; the observatory was made possible by the generous financial support of the W. M. Keck Foundation. Research at Lick Observatory is partially supported by a generous gift from Google. We thank the staffs of the various observatories at which data were obtained for their excellent assistance.

Funding for the Sloan Digital Sky Survey IV has been provided by the Alfred P. Sloan Foundation, the U.S. Department of Energy Office of Science, and the Participating Institutions. SDSS-IV acknowledges support and resources from the Center for High-Performance Computing at the University of Utah. The SDSS web site is \href{www.sdss.org}{www.sdss.org}.

SDSS-IV is managed by the Astrophysical Research Consortium for the Participating Institutions of the SDSS Collaboration, including the Brazilian Participation Group, the Carnegie Institution for Science, Carnegie Mellon University, the Chilean Participation Group, the French Participation Group, Harvard-Smithsonian Center for Astrophysics,  Instituto de Astrof\'isica de Canarias, The Johns Hopkins University,  Kavli Institute for the Physics and Mathematics of the Universe (IPMU) / University of Tokyo, Lawrence Berkeley National Laboratory, Leibniz Institut f\"ur Astrophysik Potsdam (AIP),   Max-Planck-Institut f\"ur Astronomie (MPIA Heidelberg),  Max-Planck-Institut f\"ur Astrophysik (MPA Garching),  Max-Planck-Institut f\"ur Extraterrestrische Physik (MPE),  National Astronomical Observatories of China, New Mexico State University, New York University, University of Notre Dame,  Observat\'ario Nacional / MCTI, The Ohio State University,  Pennsylvania State University, Shanghai Astronomical Observatory, United Kingdom Participation Group, Universidad Nacional Aut\'onoma de M\'exico, University of Arizona,  University of Colorado Boulder, University of Oxford, University of Portsmouth, University of Utah, University of Virginia, University of Washington, University of Wisconsin,
Vanderbilt University, and Yale University.

The Pan-STARRS1 Surveys (PS1) have been made possible through contributions of the Institute for Astronomy, the University of Hawaii, the Pan-STARRS Project Office, the Max-Planck Society and its participating institutes, the Max Planck Institute for Astronomy, Heidelberg and the Max Planck Institute for Extraterrestrial Physics, Garching, The Johns Hopkins University, Durham University, the University of Edinburgh, Queen's University Belfast, the Harvard-Smithsonian Center for Astrophysics, the Las Cumbres Observatory Global Telescope Network Incorporated, the National Central University of Taiwan, the Space Telescope Science Institute, the National Aeronautics and Space Administration under Grant NNX08AR22G issued through the Planetary Science Division of the NASA Science Mission Directorate, the National Science Foundation under Grant AST-1238877, the University of Maryland, and Eotvos Lorand University (ELTE).

This publication makes use of data products from the Two Micron All Sky Survey, which is a joint project of the University of Massachusetts and the Infrared Processing and Analysis Center/California Institute of Technology, funded by the National Aeronautics and Space Administration and the National Science Foundation.

This publication makes use of data products from the Wide-field Infrared Survey Explorer, which is a joint project of the University of California, Los Angeles, and the Jet Propulsion Laboratory/California Institute of Technology, funded by the National Aeronautics and Space Administration.

The LBNL Physics Division is supported by the U.S. Department of Energy Office of Science High Energy Physics.

The Cerro Tololo Inter-American Observatory and the National Optical Astronomy Observatory are operated by the Association of Universities for Research in Astronomy (AURA) under cooperative agreement with the National Science Foundation.

The National Energy Research Scientific Computing Center, which is supported by the Office of Science of the U.S. Department of Energy under Contract DE-AC02-05CH11231, provided staff, computational resources, and data storage for this project.

The Computational HEP program in The Department of Energy's Science Office of High Energy Physics provided resources through the ``Cosmology Data Repository" project (Grant \#KA2401022).

The data presented here were obtained in part with ALFOSC, which is provided by the Instituto de Astrof\'isica de Andaluc\'ia (IAA) under a joint agreement with the University of Copenhagen and NOTSA.

\end{acknowledgements}

\software{Astropy version 3.2.3 \citep{Astropy2013a, Astropy2018a},
Flexible Stellar Population Synthesis  \citep[FSPS;][]{Conroy2009a},
High Order Transform of Psf ANd Template Subtraction version 5.1.11 \citep[Hotpants;][]{Becker2015a},
IRAF \citep{Tody1986a},
LAMBDAR \citep{Wright2016a},
Prospector version 0.3 \citep{Leja2017a},
python-fsps \citep{Foreman-Mackey2014a}
scikit-learn version 0.21.2 \citep{Pedregosa2011a},
Software for Calibrating AstroMetry and Photometry \citep[SCAMP;][]{Bertin2006a} version 2.0.4,
Source Extractor version 2.19.5 \citep{Bertin1996a},
Supernova Identification version 5.0 \citep{Blondin2007a},
Superfit version 3.5 \citep{Howell2005a}
}


\appendix

\newpage
\section{Spectroscopic log}\label{appendix:spec}
\begin{table}
\movetabledown=7cm
\caption{Log of the spectroscopic observations}\label{tab:spec}
\hspace{-1.5cm}
\begin{tabular}{lccccc}
\toprule
PTF	    & Date	 & Telescope/		& Reference	\\
            &        & Instrument       &           \\
\midrule
09as     &  2009-03-31     & Keck/LRIS  & 1     \\  
09atu    &  2009-08-25     & Keck/LRIS  & 1     \\ 
09awk    &  2009-07-22     & Keck/LRIS  & 2,3   \\ 
09axi    &  2009-07-22     & Keck/LRIS  & 4     \\ 
09bce    &  2009-07-25     & Lick/KAST  & 4     \\ 
09bcl    &  2009-10-24     & Keck/LRIS  & 5     \\ 
09be     &  2009-04-27     & P200/DBSP  & 4     \\ 
09bgf    &  2009-07-25     & Lick/KAST  & 4     \\ 
09bw     &  2009-03-31     & Keck/LRIS  & 4     \\ 
09cjq    &  2009-10-22     & Keck/LRIS  & 4     \\ 
09cnd    &  2009-08-16     & Keck/LRIS  & 1     \\ 
09ct     &  2009-10-17     & Keck/LRIS  & 4     \\ 
09cu     &  2009-04-27     & P200/DBSP  & 4     \\ 
09cvi    &  2009-10-22     & Keck/LRIS  & 4     \\ 
09cwl    &  2009-08-25     & Keck/LRIS  & 1     \\ 
09dah    &  2009-09-16     & WHT/ISIS   & 2, 3  \\ 
09dfk    &  2009-10-22     & Keck/LRIS  & 2, 3  \\ 
09dh     &  2009-09-23     & Keck/LRIS  & 4     \\ 
09dra    &  2009-08-25     & Keck/LRIS  & 4     \\ 
\bottomrule
\end{tabular}
\tablecomments{All spectra are publicly available on WISeREP (\href{https://wiserep.weizmann.ac.il}{https://wiserep.weizmann.ac.il}). The full table is available online in a machine-readable form.}
\tablerefs{1) \citet{Quimby2018a}; 2) \citet{Fremling2018a}; 3) Fremling (in prep.); 4) This work; 5) \citet{Nyholm2019a}.}
\end{table}

\newpage
\section{Host photometry}\label{appendix:phot}
\begin{table}
\caption{Photometry of the PTF CCSN host galaxies}\label{tab:data}
\begin{tabular}{cccc}
\toprule
PTF & Survey & Filter & Brightness \\
\midrule
09awk & \galex & $FUV$ & $20.24\pm0.24$ \\
09awk & \galex & $NUV$ & $20.03\pm0.03$ \\
09awk & SDSS & $u'$ & $19.2\pm0.04$ \\
09awk & SDSS & $g'$ & $18.16\pm0.0$ \\
09awk & SDSS & $r'$ & $17.79\pm0.01$ \\
09awk & SDSS & $i'$ & $17.51\pm0.01$ \\
09awk & SDSS & $z'$ & $17.43\pm0.03$ \\
09awk & PS1 & $g_{\rm PS1}$ & $18.15\pm0.01$ \\
09awk & PS1 & $r_{\rm PS1}$ & $17.86\pm0.0$ \\
09awk & PS1 & $i_{\rm PS1}$ & $17.52\pm0.01$ \\
09awk & PS1 & $z_{\rm PS1}$ & $17.48\pm0.01$ \\
09awk & PS1 & $y_{\rm PS1}$ & $17.41\pm0.04$ \\
09awk & 2MASS & $J$ & $17.09\pm0.1$ \\
09awk & 2MASS & $H$ & $17.21\pm0.07$ \\
09awk & 2MASS & $K$ & $17.48\pm0.09$ \\
09axi & SDSS & $u'$ & $20.17\pm0.13$ \\
09axi & SDSS & $g'$ & $18.76\pm0.03$ \\
09axi & SDSS & $r'$ & $18.47\pm0.03$ \\
09axi & SDSS & $i'$ & $18.21\pm0.03$ \\
09axi & SDSS & $z'$ & $18.16\pm0.09$ \\
09axi & PS1 & $g_{\rm PS1}$ & $18.88\pm0.05$ \\
09axi & PS1 & $r_{\rm PS1}$ & $18.52\pm0.04$ \\
09axi & PS1 & $i_{\rm PS1}$ & $18.29\pm0.04$ \\
09axi & PS1 & $z_{\rm PS1}$ & $18.26\pm0.08$ \\
09axi & PS1 & $y_{\rm PS1}$ & $18.29\pm0.2$ \\
09bce & \galex & $FUV$ & $18.03\pm0.08$ \\
09bce & \galex & $NUV$ & $17.5\pm0.05$ \\
09bce & SDSS & $u'$ & $16.31\pm0.07$ \\
09bce & SDSS & $g'$ & $14.74\pm0.04$ \\
09bce & SDSS & $r'$ & $14.05\pm0.03$ \\
09bce & SDSS & $i'$ & $13.62\pm0.03$ \\
09bce & SDSS & $z'$ & $13.34\pm0.06$ \\
09bce & PS1 & $g_{\rm PS1}$ & $14.75\pm0.02$ \\
09bce & PS1 & $r_{\rm PS1}$ & $14.08\pm0.02$ \\
09bce & PS1 & $i_{\rm PS1}$ & $13.72\pm0.02$ \\
09bce & PS1 & $z_{\rm PS1}$ & $13.52\pm0.03$ \\
09bce & PS1 & $y_{\rm PS1}$ & $13.37\pm0.04$ \\
09bce & 2MASS & $J$ & $12.98\pm0.05$ \\
09bce & 2MASS & $H$ & $12.76\pm0.05$ \\
09bce & 2MASS & $K$ & $12.96\pm0.05$ \\
09bce & unWISE & $W1$ & $13.47\pm0.01$ \\
09bce & unWISE & $W2$ & $14.1\pm0.02$ \\
\bottomrule
\end{tabular}
\tablecomments{All magnitudes are reported in the AB system and not corrected for reddening. The full table is available online in a machine-readable form.}
\end{table}

\newpage
\section{Supernova classifications}\label{app:classification}

\paragraph{PTF09aux} \citet{Arcavi2010a} classified the SN as a Type Ia/Ic SN. Due to the ambiguity of the classification, we excluded this SN from our sample.

\paragraph{PTF09axc} Initially classified as a Type II SN \citep{Arcavi2010a}, \citet{Arcavi2014a} showed that this transient is in fact a TDE.

\paragraph{PTF09bcl} \citet{Arcavi2010a} classified this transient as a Type II SN. \citet{Nyholm2019a} showed that this transient is a Type IIn SN.

\paragraph{PTF09bfz} Initially classified as a Type Ic-BL SN \citep{Arcavi2010a}, \citet{Quimby2018a} concluded that this is a SLSN using a new library of SN templates that were not available in 2010.

\paragraph{PTF09ct} \citet{Arcavi2010a} classified this transient as a Type II SN. Spectra not available to \citet{Arcavi2010a} showed that this is in fact a Type IIn SN.

\paragraph{PTF09djl} Initially classified as a Type II SN \citep{Arcavi2010a},  \citet{Arcavi2014a} showed that this transient is a TDE.

\paragraph{PTF09ebq} \citet{Arcavi2010a} initially classified the transient as a Type II SN in the center of its host galaxy. A re-inspection of all data revealed variability about 1--2 months before the discovery of the transient. This transient is more likely an active galactic nucleus.

\paragraph{PTF09ejz} \citet{Arcavi2010a} classified the SN as a Type Ia/Ic SN. Due to the ambiguity of the classification, we excluded this SN from our sample.

\paragraph{PTF09ism} A re-inspection of the spectra revealed \ion{He}{1}\,$\lambda$\,6678 in absorption, albeit weaker than what is deemed for a Type IIb SN. We change the classification of this SN from Type II \citep{Arcavi2010a} to Type IIb SN.

\paragraph{PTF09q} \citet{Quimby2018a} concluded that this object could be an H-poor SLSN, instead of being a Type Ic SN \citep{Arcavi2010a}. \citet{Quimby2018a} noted that its massive host galaxy would be at odds with the known population of SLSN hosts, albeit a few SLSNe in massive hosts were reported in the literature, e.g., SN2017egm \citep{Nicholl2017a, Bose2018a}. We follow \citet{Arcavi2010a}, \citet{De2018a} and \citet{Fremling2018a} and classify the SN as Type Ic SN.

\paragraph{PTF09tm} The Lick spectrum from 25 July 2009 shows narrow Balmer emission lines from H$\alpha$ to H$\gamma$, making it a Type IIn SN \citep{Nyholm2019a}, instead of a Type II SN \citep{Arcavi2010a}.

\paragraph{PTF10aaxi}
\citet{Smith2012a} classified PTF10aaxi as a Type II SN. \hst\ images obtained after the transient faded show a resolved source with an absolute magnitude of $-8$~mag at the explosion site. With the current data in hand, it remains ambiguous whether this object is connected with PTF10aaxi or whether it is a companion star. Owing to this ambiguity, we removed PTF10aaxi from our final sample.

\paragraph{PTF10acbu} The spectrum of this stripped-envelope supernova shows no helium in absorption. Therefore, we change the classification from Type Ib \citep{Fremling2018a} to Type Ic.

\paragraph{PTF10bip} We classify this object as a Type Ic SN whereas it was listed as uncertain (Ic/Ic-BL) in \citet{Modjaz2019a}.

\paragraph{PTF10cwx} Initially classified as a Type II SN \citep{Arcavi2010a}, a reassessment of the spectrum shows narrow Balmer emission lines from H$\alpha$ to H$\gamma$ making it a Type IIn SN \citep{Nyholm2019a}.

\paragraph{PTF10cxx} The spectrum shows narrow Balmer emission lines from H$\alpha$ to H$\delta$, making it a Type IIn not Type II SN \citep{Arcavi2010a}.

\paragraph{PTF10gvb} \citet{deCia2018a} reported that the spectra of the candidate SLSN \citep{Quimby2018a, Modjaz2019a} lack the typical features of SLSNe and they are better matched with templates of Ic-BL SNe. We follow this assessment and put PTF10gvb in the sample of Type Ic-BL SNe.

\paragraph{PTF10hgi} We followed \citet{Quimby2018a} and \citet{GalYam2019a} and classify this transient as a SLSN-IIb instead of a SLSN-I \citep{Inserra2013a}.

\paragraph{PTF10izr} \citet{Fremling2018a} classified this transient as a Type Ic SN. We removed this object from our sample because of the poor data quality.

\paragraph{PTF10qwu} \citet{Perley2016a} classified PTF10qwu as a SLSN-IIn. However, a detailed analysis by Leloudas et al. (in prep) raises concerns about this classification. We follow Leloudas et al. (in prep) and put PTF10qwu in the class of Type IIn SNe.

\paragraph{PTF10svt} We classified this object as a Type Ic SN, whereas it was listed as uncertain (Ic/Ic-BL) in \citet{Modjaz2019a}.

\paragraph{PTF10tqv} This SN was classified as a Ic-BL SN in \citet{Taddia2019a} and \citet{Modjaz2019a} and a Ic SN in \citet{Fremling2018a}.  \citet{Taddia2019a} and \citet{Modjaz2019a} showed that the spectra are better matched with Ic-BL templates than templates of normal Type Ic SNe. We follow this analysis and classify this SLSNe as a Ic-BL SN.

\paragraph{PTF10ts} The classification spectrum shows narrow Balmer emission lines from H$\alpha$ to H$\delta$, making it a Type IIn SN instead of a Type II SN \citep{Arcavi2010a}.

\paragraph{PTF10u} The spectrum shows narrow Balmer emission lines from H$\alpha$ to H$\delta$, making it a Type IIn not Type II SN \citep{Arcavi2010a}.

\paragraph{PTF10wg} The classification of this transient is not free of ambiguity \citep{Arcavi2010a, Fremling2018a}; therefore, we exclude it.

\paragraph{PTF10ysd} \citet{Modjaz2019a} concluded that the SN-Ic templates provide a better match for this SN than Ic-BL templates \citep{Taddia2019a}.

\paragraph{PTF10yyc} \citet{Perley2016a} classified PTF10yyc as a SLSN-IIn. However, a detailed analysis by Leloudas et al. (in prep) raised concerns about this classification. We follow Leloudas et al. (in prep) and put PTF10yyc in the class of Type IIn SNe.

\paragraph{PTF11gcj} \citet{Fremling2018a} classified this transient as a Type Ic SN. \citet{Modjaz2019a} used a newer SN template library and showed that SN Ic-BL templates describe the spectra of PTF11gcj better.

\paragraph{PTF11mnb} \citet{Quimby2018a} pointed out that the spectra are similar to the H-poor SLSN SN1999as, however it only reached a peak luminosity of $M_r\approx-18.9$~mag. We follow \citet{deCia2018a} and \citet{Taddia2019a} and classify the transient as a Type Ic SN.

\paragraph{PTF11pnq} The SN spectra are better matched with SN-Ib templates than with SN-Ibc templates \citep{Fremling2019a}.

\paragraph{PTF12epg} \citet{Perley2016a} classified PTF12epg as a SLSN-IIn. However, a detailed analysis by Leloudas et al. (in prep) raised doubts about the nature of this transient. We follow Leloudas et al. (in prep) and put PTF12epg in the class of Type IIn SNe.

\paragraph{PTF12gty} \citet{Fremling2018a} and Barbarino et al. (submitted) classified this transient as a Type Ic SN. \citet{Quimby2018a} used a larger template database and showed that this is an H-poor SLSNe.

\paragraph{PTF12hni} \citet{Fremling2018a} classified this transient as a Type Ic SN. \citet{Quimby2018a} used a larger template database and showed that this is an H-poor SLSNe.

\paragraph{iPTF13doq} \citet{Fremling2018a} classified this transient as a Type Ic SN. We removed this object from our sample because of the low contrast between the SN and the host galaxy in the transient spectrum.

\paragraph{iPTF14jhf} \citet{Fremling2018a} classified this transient as a Type Ic SN and Barbarino et al. (submitted) as a Type Ibc SN. We removed this object from our sample due to the poor data quality.

\paragraph{iPTF15eov}
\citet{Taddia2019a} classified iPTF15eov as a broad-lined Ic supernova. The authors also pointed out several peculiarities. The light curve is too broad, and the peak luminosity of $M_r=-21.8$~mag is 3.2 mag more luminous than the average SN Ic-BL in their sample and even 1~mag more luminous than the second most luminous Ic-BL SN in their sample. The authors also concluded that the light curve could not be powered by the radioactive decay of nickel. The peak luminosity and the broad-light curve are rather characteristic of a superluminous supernova. Furthermore, these authors pointed out that the spectra of iPTF15eov are similar to the H-poor SLSN PTF11rks. We agree with this assessment and change the classification of iPTF15eov from SN Ic-BL to SLSN-I.

\paragraph{iPTF16flq} \citep{Fremling2019a} and Barbarino et al. (submitted) classified the SN as a Type Ibc SN. We see no He in absorption in any of the spectra and hence classify the SN as a Type Ic SN.

\end{document}